\documentclass[
aps,
pre,
reprint,
superscriptaddress,
amsmath,
amssymb,
nofootinbib
]{revtex4-2}

\usepackage[utf8]{inputenc}
\usepackage[english]{babel}

\usepackage{amsmath,amssymb,amsfonts,latexsym}
\usepackage{bm}
\usepackage{braket}
\usepackage{physics}
\usepackage{stmaryrd}

\usepackage{graphicx}
\usepackage{adjustbox}
\usepackage{microtype}
\usepackage{dcolumn}
\usepackage{booktabs}
\usepackage{xcolor}
\usepackage{soul}
\usepackage{subfigure}
\usepackage{placeins}

\usepackage{tikz}
\usetikzlibrary{arrows}
\usetikzlibrary{arrows.meta,positioning,calc,decorations.pathmorphing}

\usepackage{marvosym}

\usepackage{chngcntr}

\usepackage{hyperref}

\newcommand{\be}{\begin{equation}\begin{adjustbox}{max width=\columnwidth}$\displaystyle}
\newcommand{\en}{$\end{adjustbox}\end{equation}}
\newcommand{\bea}{\begin{equation}\begin{adjustbox}{max width=\columnwidth}$\displaystyle\begin{aligned}}
\newcommand{\ena}{\end{aligned}$\end{adjustbox}\end{equation}}
\newcommand{\beano}{\begin{equation*}\begin{adjustbox}{max width=\columnwidth}$\displaystyle\begin{aligned}}
\newcommand{\enano}{\end{aligned}$\end{adjustbox}\end{equation*}}
\newcommand{\bee}{\begin{enumerate}}
\newcommand{\ene}{\end{enumerate}}

\begin{document}

% Two-column text safeguards: these reduce overfull text lines without changing the APS layout.
\sloppy
\emergencystretch=3em
\tolerance=2000
\hbadness=10000

\title{Turing mechanisms in a multimode open quantum system}

\author{Giorgia Comparato}
\email{giorgia.comparato@unipa.it}
\affiliation{Dipartimento di Ingegneria, Universit\`a di Palermo, 90128 Palermo, Italy}

\author{Francesco Gargano}
\email{francesco.gargano@unipa.it}
\affiliation{Dipartimento di Ingegneria, Universit\`a di Palermo, 90128 Palermo, Italy}

\author{Rosario {Lo Franco}}
\email{rosario.lofranco@unipa.it}
\affiliation{Dipartimento di Ingegneria, Universit\`a di Palermo, 90128 Palermo, Italy}

\begin{abstract}
We investigate pattern formation in a finite chain of bosonic modes whose dynamics is governed by a Gorini-Kossakowski-Sudarshan-Lindblad (GKSL) master equation. The model combines local parametric driving and nonlinear damping with nonlocal dissipative couplings between modes that work on different discrete spatial scales. In the classical limit, these mechanisms generate a reaction–diffusion-like dynamics, allowing the emergence of Turing-type instabilities.
The key aspect of the analysis is the coexistence and competition of different unstable spatial modes. Depending on the range of parameters, the system may select different stationary nonuniform configurations, oscillatory wave-like states, or regimes in which multiple modes interact before a dominant pattern is established, thus providing a mechanism for pattern selection.
We compare the deterministic bifurcation scenario, generated by a reaction-diffusion-like system derived from semiclassical drift dynamics, with the quantum dynamics, derived via the GKSL master equation, using phase-space methods and reduced Wigner functions. The results show how Turing instabilities, mode competition, and pattern selection can be extended to multimode open quantum systems, providing a bridge between nonlinear dynamical systems, dissipative quantum mechanics, and spatial self-organization.
\end{abstract}


\maketitle

\section{Introduction}
In classical systems, the Turing instability provides one of the paradigmatic mechanisms for pattern formation \cite{Cross1993PatternFormation}. It arises when a spatially uniform state, stable in the absence of diffusion, becomes unstable once the latter phenomenon is introduced. This mechanism is commonly associated with activator-inhibitor antagonism, where the inhibitor diffuses faster than the activator \cite{Turing1952,wang2026,NakaoMeravigliao}. A complementary and particularly intuitive interpretation is provided by the Gierer-Meinhardt framework \cite{GiererMeinhardt1972}, according to which patterns emerge from the interplay between short-range facilitation and long-range competition. This perspective extends the Turing mechanism especially relevant beyond its original reaction-diffusion setting, identifying a general principle for the spontaneous emergence of spatial organization.

Despite its central role in classical nonlinear dynamics, the quantum counterpart of the Turing mechanism remains only partially understood in genuinely quantum systems. In recent years, the term ``Turing'', in connection with pattern formation and symmetry-breaking instabilities, has appeared with increasing frequency in quantum-mechanical contexts \cite{KoseskaVolkovKurths2013,Ban20,BandyopadhyayKhatunBanerjee2021,Kato2022,Paul2024}. Starting from the classical transition from amplitude death to oscillation death in coupled oscillators, interpreted as a Turing bifurcation \cite{KoseskaVolkovKurths2013}, a quantum version was investigated in coupled van der Pol oscillators \cite{Ban20}. A transition from homogeneous to inhomogeneous steady states via Turing instabilities was also reported \cite{BandyopadhyayKhatunBanerjee2021}. Furthermore, it was shown that a Turing instability can occur in a pair of quantum bosonic units, where it may generate both non-uniformity and entanglement \cite{Kato2022}. Related symmetry-breaking transitions have also been studied in coupled quantum Stuart-Landau oscillators, where a transition from quantum limit-cycle oscillations to quantum oscillation death was found \cite{Paul2024}.

From a methodological perspective, these studies share a common framework based on open quantum systems, driven-dissipative bosonic dynamics, phase-space representations, and semiclassical or stochastic descriptions of nonlinear quantum systems \cite{RobertsClerk2020,Chia2023,LimEtAl2024}. This naturally leads to an analysis based on quantum master equations, Wigner phase-space methods \cite{Polkovnikov2010PhaseSpace}, Fokker-Planck and stochastic semiclassical approximations, and classical bifurcation theory, which have become standard tools for connecting microscopic open quantum dynamics with emergent nonlinear behavior in driven-dissipative systems \cite{CarusottoCiuti2013,Sieberer2016,Minganti2018,McDonaldClerk2023,Sieberer2025,Dutta2025}. 
Within this setting, the quadratures of a quantum bosonic mode may play a role analogous to that of activator-inhibitor variables in classical systems. Suitable combinations of coherent evolution and engineered dissipation can then destabilize homogeneous configurations and promote spatial non-uniformity, whose signatures can be detected, among other observables, in the structure of the Wigner function.

Existing quantum studies, however, have mostly focused on single units or on pairs of coupled units. In such minimal settings, spatial structure is necessarily restricted: the selection of a pattern wavelength, which is a key aspect of classical Turing phenomena, becomes essentially trivial, and the role of boundary conditions or the competition among distinct spatial modes cannot be systematically addressed. The present work aims to go beyond this limitation by introducing a multimode bosonic lattice in which spatial mode selection, boundary effects, and facilitative-competitive mechanisms arise directly from microscopic open quantum dynamics. In particular, our construction follows the Gierer-Meinhardt interpretation of pattern formation \cite{GiererMeinhardt1972}, implementing quantum dissipative processes that act as counterparts of short-range facilitation or excitation and long-range competition or suppression.

The system considered here is a one-dimensional three-site bosonic chain with no-flux boundary conditions. The local coherent dynamics includes squeezing and detuning, while the coupling to the environment gives rise to several dissipative mechanisms: local single-photon losses, nonlinear two-photon losses, a nearest-neighbor incoherent pumping channel acting on antisymmetric combinations of consecutive modes, and a longer-range dissipative channel damping antisymmetric combinations over three neighboring sites. All these processes are described microscopically by Lindblad operators, and the full dynamics is governed by a Gorini-Kossakowski-Sudarshan-Lindblad (GKSL) master equation for the density operator.

Starting from this microscopic model, we derive a reaction-diffusion-like mean-field dynamics for the complex field amplitudes and analyze the corresponding deterministic bifurcation structure. This semiclassical drift dynamics provides a reference framework for interpreting the full quantum evolution generated by the GKSL master equation. We identify two main regimes. In the squeezing-dominant regime, homogeneous stationary states can lose stability and generate stationary non-uniform branches: the corresponding quantum steady states display spatially structured reduced Wigner functions. In the detuning-dominant regime, the instability is oscillatory and leads to wave-like time-dependent patterns: in the quantum description, these are associated with ring-like structures in the reduced Wigner functions.

We also compare weak and strong quantum regimes. In the weak quantum regime, where nonlinear damping is small, the semiclassical stochastic description agrees well with the full quantum dynamics, as expected from the truncated phase-space approach. In the strong quantum regime, this agreement is progressively reduced: quantum fluctuations become more relevant and tend to smooth or mask the deterministic signatures of the Turing mechanism. Nevertheless, the reduced Wigner functions still retain traces of the underlying spatial organization, showing that the deterministic analysis remains a useful guide to the quantum behavior.

The paper is organized as follows. In Section~\ref{sec:Nsites} we introduce the $N$-site open quantum model, specifying the local coherent dynamics and the dissipative channels responsible for spatial coupling. In Section~\ref{sec_Turing}, we derive the mean-field equations and perform the linear stability analysis of the homogeneous equilibria. Section~\ref{sec:QuantumTuring} is devoted to the quantum counterpart of the Turing mechanisms in the minimal three-site chain. Finally, in Section~\ref{sec:conclusions}, we summarize the main results and discuss possible extensions. The Appendices collect the derivation of the Wigner equation, the mean-field limit, and the classical noisy equations.

\section{The $N$-site model}\label{sec:Nsites}

We consider a one-dimensional chain of $N\ge 3$ bosonic spatial modes, labeled by $j=1,\dots,N$, with associated annihilation and creation operators $\hat a_j$ and $\hat a_j^\dagger$, satisfying the canonical commutation relations
$$
[\hat a_j,\hat a_k^\dagger]=\delta_{jk}.
$$
The dynamics of the system is governed by a GKSL quantum master equation, derived from the combination of coherent local dynamics and the presence of local and nonlocal dissipative processes caused by the interaction with the environment.

The coherent part of the evolution is generated by a local Hamiltonian of degenerate optical parametric oscillator (DOPO) type \cite{Tezak2017}, given by
\bea
\hat H_{\mathrm{loc}}
=
\sum_{j=1}^N
\left[
\Delta\,\hat a_j^\dagger \hat a_j
+
i\eta
\left(
\hat a_j^2 e^{-i\theta}
-
\hat a_j^{\dagger 2} e^{i\theta}
\right)
\right].
\ena
The first term describes the free energy of each mode in the rotating frame, with $\Delta\in\mathbb{R}$ denoting the detuning, while the second term represents a coherent parametric drive of strength $\eta>0$ and phase $\theta$, responsible for pair creation and phase symmetry breaking.

Local dissipation is modeled by two families of Lindblad operators.
Single-photon losses are described by
\bea
\hat L_j^{(sl)}=\sqrt{\gamma_1}\,\hat a_j,
\qquad j=1,\dots,N,
\ena
where $\gamma_1$ is the linear decay rate.
To ensure saturation of the parametric instability, we also include nonlinear two-photon losses
\bea
\hat L_j^{(dl)}=\sqrt{\gamma_2}\,\hat a_j^2,
\qquad j=1,\dots,N,
\ena
where $\gamma_2$ is the nonlinear decay rate, which introduces an intensity-dependent damping mechanism.

Two other nonlocal dissipative channels, acting on different spatial scales along the chain, are also introduced to promote the spatial structure of the system.
The first one is the short-range dissipative coupling acting on neighboring sites, and is described through jump operators acting on antisymmetric combinations of creation operators
\bea
\hat L^{(\lambda)}_{j}
=
\sqrt{\lambda}\,
\left(
\hat a_{j+1}^\dagger-\hat a_j^\dagger
\right),
\qquad j=1,\dots,N-1.
\ena
Each of these operators injects excitations into the antisymmetric modes and, being linear in $\hat a^\dagger$, acts as an incoherent pump with strength $\lambda$.
Since the first and last sites interact with only one neighboring node, the boundary sites enter only one short-range link, implementing effective no-flux boundary conditions for the finite chain.
The second nonlocal mechanism acts as a long-range dissipative channel with strength $\kappa$ by damping antisymmetric modes over three consecutive nodes. The corresponding jump operators are defined as
\beano
\text{First site:}\qquad
\hat L^{(\kappa)}_{1}
&=
\sqrt{\kappa}\,
\big(
\hat a_{2}-\hat a_{1}
\big),
\\[0.3em]
\text{Bulk sites:}\qquad
\hat L^{(\kappa)}_{j}
&=
\sqrt{\kappa}\,
\big(
(\hat a_j-\hat a_{j+1})
+
(\hat a_j-\hat a_{j-1})
\big),
\qquad j=2,\ldots,N-1,
\\[0.3em]
\text{Last site:}\qquad
\hat L^{(\kappa)}_{N}
&=
\sqrt{\kappa}\,
\big(
\hat a_{N-1}-\hat a_{N}
\big).
\enano
The full GKSL master equation thus reads
\bea
\dot{\hat\rho}
=
-i[\hat H_{\mathrm{loc}},\hat\rho]
+
\sum_{j=1}^N
\left(
\mathcal D[\hat L^{(sl)}_{j}]\hat\rho
+
\mathcal D[\hat L^{(dl)}_{j}]\hat\rho
\right)
+
\sum_{j=1}^{N-1}
\mathcal D[\hat L^{(\lambda)}_{j}]\hat\rho
+
\sum_{j=1}^N
\mathcal D[\hat L^{(\kappa)}_j]\hat\rho,\label{QME}
\ena
where we have defined $\mathcal{D}[\hat O]\hat\rho=\hat O \hat \rho \hat O^\dagger- \frac12\{\hat O^\dagger \hat O,\hat \rho\}$ for an operator $\hat O$.
We shall see that the interplay between the $\lambda$ and $\kappa$ channels defines a discrete pattern-selection
mechanism.

\section{Turing instability analysis for the mean-field equations}\label{sec_Turing}

Starting from the quantum master equation previously defined, the dynamics of observables is obtained from the general relation
$\frac{d}{dt}\langle \hat O\rangle=\mathrm{Tr}(\hat O\dot{\hat\rho})$.
In particular, we focus on the evolution of the field amplitudes
$\alpha_j=\langle \hat a_j\rangle=x_j+i p_j, \, j=1,\ldots,N$, and we determine the conditions for the emergence of Turing instabilities. The equations of motion are derived in Appendix \ref{APP:MFE} and generate the following system of reaction-diffusion-like equations:
\bea
\dot{\alpha}_j = (s-i\Delta)\alpha_j - \gamma_2|\alpha_j|^2\alpha_j - 2\eta e^{i\theta}\alpha_j^*
+\\  \frac{\lambda}{2} \sum_k (\nabla^{(2)})_{jk} \alpha_k
- \frac{\kappa}{2} \sum_k (\nabla^{(4)})_{jk} \alpha_k,
\label{alphaeq}
\ena
where $s=(2\gamma_2-\gamma_1)/2$. The reason why these equations describe reaction-diffusion mechanisms relies on the fact that the matrix $\nabla^{(2)}$ is the discrete anti-diffusive Laplacian defined in Eq.~\eqref{matD2} and $\nabla^{(4)}=(\nabla^{(2)})^2$ is the discrete bi-Laplacian.

We search for steady homogeneous equilibria of the system defined by Eq.~\eqref{alphaeq} such that $\alpha_j$ are all equal for every $j=1,\ldots,N$. These equilibria are the starting point for the Turing instability analysis. 
For these homogeneous steady solutions, one has
$$\nabla^{(2)}\boldsymbol{\alpha}^\mathrm{T}=
\nabla^{(4)}\boldsymbol{\alpha}^\mathrm{T}=
\boldsymbol{0},$$ where $\boldsymbol{\alpha}=(\alpha_1,\dots,\alpha_N),$
so that it is sufficient to look for the zeros of the local nonlinear equations.

The first homogeneous equilibrium, always existing, is the trivial (null) solution
\bea
\alpha_j=\alpha^{\text{null}} = 0, \qquad j=1,\dots,N.\label{nulle}
\ena

A second class of nontrivial equilibria exists only in the squeezing-dominated regime,
$\eta\geq|\Delta|/2$.
They are determined by
\bea\alpha_j=\alpha^{\mathrm{hom}}_\pm=r_\pm e^{i\phi},\quad j=1,\dots,N, \label{nontrivial}\ena
where
\bea r_{\pm}^2 = \frac{s\pm\sqrt{4\eta^2-\Delta^2}}{\gamma_2}, \label{condr1}\ena
\bea 2\phi = \theta-\text{arg}\left(s-\gamma_2 r_\pm^2-i\Delta\right)
\quad (\text{mod } 2\pi),\label{condr2} \ena
which are admissible only if $r_{+}^2$ and $r_{-}^2$ are non-negative.
Because of the $\mathbb{Z}_2$ symmetry of the equations, if $\alpha^{\mathrm{hom}}$ is a homogeneous equilibrium, then
$-\alpha^{\mathrm{hom}}$ is a homogeneous equilibrium too.

%We stress that in the detuning-dominated regime,
%$\eta<|\Delta|/2$,
%no stationary homogeneous solutions exist, and only time-periodic
%(limit-cycle) solutions could be possible. As shown in the next subsection this regime is however important to determine the kind of instability leading to non homogeneous states.

In the following subsections, we perform the linear instability analysis around the above equilibria, using perturbations of the form 
\bea\delta \boldsymbol\alpha(t)=\sum_{m=0}^{N-1}c_m(t) \boldsymbol{v}_m \label{perturb}\ena
with $|c_m(0)|\ll1$, where $\boldsymbol{v}_m$ are the eigenvectors of  $\nabla^{(2)}$,
\bea
\nabla^{(2)}\boldsymbol{v}_m =\mu_m\boldsymbol{v}_m,\,\quad m=0,\ldots, N-1\label{eigensystem}
\ena
with $\mu_m=4\sin^2\left(\frac{m\pi}{2N}\right)$ and $\boldsymbol{v}_m=(v_m^{0},\ldots,v_m^{N-1})$ defined in Ref.~\cite{strang99} 
\bea
v_0^{j}=\sqrt{\frac{1}{N}},\,v_{m>0}^{j}=\sqrt{\frac{2}{N}}\cos\left(\frac{m(j+\frac12)\pi}{N}\right),\quad j=0,\ldots,N-1.
\ena
\subsection{Stability of the null equilibrium}\label{TuringNull}

Considering a small perturbation
\bea
\alpha_j(t)=\delta\alpha_j(t),\qquad |\delta\alpha_j(0)|\ll 1,
\ena
after linearization, Eq.~\eqref{alphaeq} reads as
\bea
\delta\dot{\alpha}_j &=& (s-i\Delta)\,\delta\alpha_j - 2\eta e^{i\theta}\,\delta\alpha_j^*
+ \frac{\lambda}{2} \sum_{k=0}^{N-1} (\nabla^{(2)})_{jk} \delta\alpha_k
- \frac{\kappa}{2} \sum_{k=0}^{N-1} (\nabla^{(4)})_{jk} \delta\alpha_k.
\label{eq:lin_null_realspace}
\ena
Substituting the perturbations of Eq.~\eqref{perturb} into Eq.~\eqref{eq:lin_null_realspace}, for each discrete spatial mode $m$ we obtain the linear system
\bea
\frac{d}{dt}
\begin{pmatrix}
	c_m(t)\\
	c_m^*(t)
\end{pmatrix}
&=&
J_m
\begin{pmatrix}
	c_m(t)\\
	c_m^*(t)
\end{pmatrix},
\quad
J_m=
\begin{pmatrix}
	A_m & B\\
	B^* & A_m^*
\end{pmatrix},
\label{eq:block_matrix}
\ena
with
\bea
A_m &=& (s-i\Delta) + \Gamma_m,
\qquad
\Gamma_m = \frac{\lambda}{2}\mu_m - \frac{\kappa}{2}\mu_m^2,
\qquad
B=-2\eta e^{i\theta}.
\label{eq:Am_Bm}
\ena
The  eigenvalue $\sigma^{(m)}$ of the Jacobian $J_m$ with the largest real part is 
\bea
\sigma^{(m)} = s+ \sqrt{4\eta^2-\Delta^2}+\Gamma_m.
\label{eq:eigs_explicit}
\ena
If $\eta\geq |\Delta|/2$, Turing instability of the homogeneous zero solution occurs when the uniform mode ($m=0$) is linearly stable, that is 
\bea
\sigma^{(0)}=s+ \sqrt{4\eta^2-\Delta^2}<0
\label{eq:stab_homogeneous_null},
\ena
and at the same time  there exists at least one non-uniform mode $m\ge 1$ such that 
\bea
\sigma^{(m)}=\sigma^{(0)}+\frac{\lambda}{2}\,\mu_m-
\frac{\kappa}{2}\,\mu_m^2>0.\label{eq:instab_homogeneous_null}
\ena
In the case $\eta<|\Delta|/2$, the eigenvalues of the Jacobian are complex conjugate, and we have the emergence of the so-called \emph{wave instability}  \cite{NemesEtAl2005, JungEtAl1998}. In particular, this arises when the stability condition for the uniform mode ($m=0$) is fulfilled,
\bea
\Re{\sigma^{(0)}}=s<0
\label{eq:wave_homogeneous_null},
\ena
and there exists at least one non-uniform mode $m\ge 1$ such that 
\bea
\Re{\sigma^{(m)}}=s+\frac{\lambda}{2}\,\mu_m-
\frac{\kappa}{2}\,\mu_m^2>0.\label{eq:inwave_homogeneous_null}
\ena

\subsection{Stability of the nontrivial homogeneous equilibrium}
We now consider the linear stability of the nontrivial homogeneous equilibria
$\alpha_j=\alpha^{\mathrm{hom}}_\pm=r_\pm e^{i\phi}$, $j=1,\dots,N$, which exist only in the squeezing-dominated regime $\eta\ge |\Delta|/2$ and for which $r_\pm^2\ge 0$, that is, when the previous null equilibrium is always linearly unstable because of the violation of Eq.~\eqref{eq:stab_homogeneous_null}.
Considering a small perturbation of $\bar\alpha$ which is one of $\alpha^{\mathrm{hom}}_\pm$, we have
\bea
\alpha_j(t)=\bar\alpha+\delta\alpha_j(t),\qquad |\delta\alpha_j(0)|\ll 1,
\ena
and expanding the nonlinear term as
\bea
|\alpha_j|^2\alpha_j &\approx& |\bar\alpha|^2\bar\alpha
+ 2|\bar\alpha|^2\,\delta\alpha_j + \bar\alpha^2\,\delta\alpha_j^*,
\ena
the linearized equations read
\bea
\delta\dot{\alpha}_j = \Bigl[(s-i\Delta)-2\gamma_2|\bar\alpha|^2\Bigr]\delta\alpha_j
-\Bigl[\gamma_2\bar\alpha^2+2\eta e^{i\theta}\Bigr]\delta\alpha_j^*
+\\ +\frac{\lambda}{2}\sum_{k=0}^{N-1}(\nabla^{(2)})_{jk}\,\delta\alpha_k
-\frac{\kappa}{2}\sum_{k=0}^{N-1}(\nabla^{(4)})_{jk}\,\delta\alpha_k.
\label{eq:lin_hom_realspace}
\ena
Using again Eq.~\eqref{perturb}, we obtain for each discrete mode $m$ the linear system
\bea
\frac{d}{dt}
\begin{pmatrix}
	c_m(t)\\
	c_m^*(t)
\end{pmatrix}
&=&
J_m^{(\mathrm{hom})}
\begin{pmatrix}
	c_m(t)\\
	c_m^*(t)
\end{pmatrix},
\qquad
J_m^{(\mathrm{hom})}=
\begin{pmatrix}
	A_m^{(\mathrm{hom})} & B^{(\mathrm{hom})}\\
	\bigl(B^{(\mathrm{hom})}\bigr)^* & \bigl(A_m^{(\mathrm{hom})}\bigr)^*
\end{pmatrix},\\
\label{eq:block_matrix_hom}
\ena
with
\bea
A_m^{(\mathrm{hom})} &= (s-i\Delta)-2\gamma_2|\bar\alpha|^2+\Gamma_m,\\
B^{(\mathrm{hom})} &= -\gamma_2\bar\alpha^2-2\eta e^{i\theta}=-(s-i\Delta)e^{2i\phi},
\label{eq:Am_Bm_hom}
\ena
and $\Gamma_m$ as in Eq.~\eqref{eq:Am_Bm} and where we have used the stationary condition $2\eta e^{i\theta}=\bigl(s-\gamma_2|\bar\alpha|^2-i\Delta\bigr)\,e^{2i\phi}$ obtained from Eq.~\eqref{alphaeq}.
Considering the two possible cases given by $r^2_{\pm}$ in Eq.~\eqref{condr1}, the largest eigenvalues of the Jacobian $J_m^{(\mathrm{hom})}$ are
\bea
\sigma_{\pm,\mathrm{hom}}^{(m)}
&=&  -s+|s|\mp 2\sqrt{4\eta^2-\Delta^2}+\Gamma_m.
\label{eq:eigs_hom_explicit}
\ena
Notice that in the case $r^2_{-}$, when existing, we always have 
\bea
\sigma_{-,\mathrm{hom}}^{(0)}=-s+|s|+ 2\sqrt{4\eta^2-\Delta^2}\geq 0,
\ena
so that the equilibrium related to $r_-^2$ is always (at least marginally) unstable also in the absence of nonlocal interactions. Hence, we focus on the branch $r^2_+$.
As before, a Turing instability occurs when the homogeneous mode $m=0$ is linearly stable, namely
\bea
\sigma_{+,\mathrm{hom}}^{(0)}=-s+|s|- 2\sqrt{4\eta^2-\Delta^2}<0,
\label{eq:stab_homogeneous_nontriv}
\ena
%% and at the same time there exists at least one non-uniform mode $m\ge 1$ such that
while at least one non-uniform mode \(m\ge 1\) becomes linearly unstable, namely
\bea
\sigma_{+,\mathrm{hom}}^{(m)}
&=&  \sigma_{+,\mathrm{hom}}^{(0)}+\frac{\lambda}{2}\,\mu_m-
\frac{\kappa}{2}\,\mu_m^2>0.
\label{eq:turing_nontriv_condition}
\ena

\section{Quantum Turing mechanisms}\label{sec:QuantumTuring}

In this section, we numerically investigate the emergence of non-uniform states, starting from the analytical conditions derived in Section~\ref{TuringNull} and their quantum counterpart.
In order to illustrate the mechanism, it is sufficient to consider the minimal case of $N=3$ sites, which already allows the appearance of distinct spatially non-uniform configurations. All numerical computations for the QME of Eq.~\eqref{QME} have been performed using the QuTiP computational library \cite{qutip5_2026}. Depending on the specific case under study, we employed either the direct computation of the QME steady-state solution or the Monte Carlo wavefunction method \cite{Molmer1993}. We stress here that having a larger number of sites would require, depending on system parameters, especially for large squeezing effects (large $\eta$), a computational effort that is difficult to satisfy, both in terms of execution time and RAM resources. Even with $3$ sites, the simulations required a machine with 128 GB of RAM and several hours of execution time. The local bosonic Hilbert space was truncated at a finite Fock cutoff $n_\textrm{cut}$, chosen separately for each parameter set according to the local steady-state occupations $\langle \hat a_j^\dagger \hat a_j\rangle$ and checked by increasing the cutoff until the reduced Wigner functions were stable within the numerical accuracy.
In this section, we also show some results concerning the classical noisy equation given in Eq.~\eqref{SDE} derived from the semiclassical limit of the QME (SDE). Numerical integration of the SDE relies on an adaptive-step Euler-Maruyama scheme \cite{Lamba2007}, designed to ensure numerical stability. The state-dependent, spatially correlated noise is evaluated at each time step by factorizing the covariance matrix via either Cholesky decomposition or a positive-semidefinite spectral projection.

\subsection{Squeezing-dominant regime $\eta\geq|\Delta|/2$}\label{subsec:squeeze}

In the squeezing-dominant regime $\eta\geq|\Delta|/2$, Turing instabilities may emerge both from the null solution of Eq.~\eqref{nulle} and from the nontrivial solution of Eq.~\eqref{nontrivial}.
For $N=3$, it is useful to write the eigensystem of the discrete Laplacian introduced in Eq.~\eqref{eigensystem},
$
\mu_0 = 0, \, \mu_1=1,\, \mu_2=3,
$
with corresponding normalized eigenvectors
\bea
\boldsymbol{v}_0=(c,c,c), \quad
\boldsymbol{v}_1=(a,0,-a), \quad
\boldsymbol{v}_2=(b,-2b,b),
\ena
where $c=\sqrt{1/3}$, $a=\sqrt{1/2}$ and  $b=\sqrt{1/6}$. 
The mode $\boldsymbol{v}_0$ represents the homogeneous/uniform configuration, whereas $\boldsymbol{v}_1$ and $\boldsymbol{v}_2$ span the subspace of non–uniform configurations associated with the emerging patterns.

\subsubsection{Deterministic pattern roadmap for the null homogeneous solution}
We begin with the instability analysis of the null solution. The deterministic Turing instability conditions derived in Eqs.~\eqref{eq:stab_homogeneous_null}–\eqref{eq:instab_homogeneous_null}
require the homogeneous null equilibrium to be stable with respect to the uniform mode, namely
\bea
s+\sqrt{4\eta^2-\Delta^2}<0.
\ena
Instead, a non–uniform mode $m\geq 1$ becomes unstable when the continuation parameter $\kappa$ is below a certain threshold, namely
\bea
\kappa
<
\kappa_m^{\mathrm{th}}
=
\frac{\lambda\,\mu_m+2\big(s+\sqrt{4\eta^2-\Delta^2}\big)}{\mu_m^2}.
\label{eq:kappath}
\ena

\begin{figure*}[t!]
	\includegraphics[width=0.75\textwidth]{ 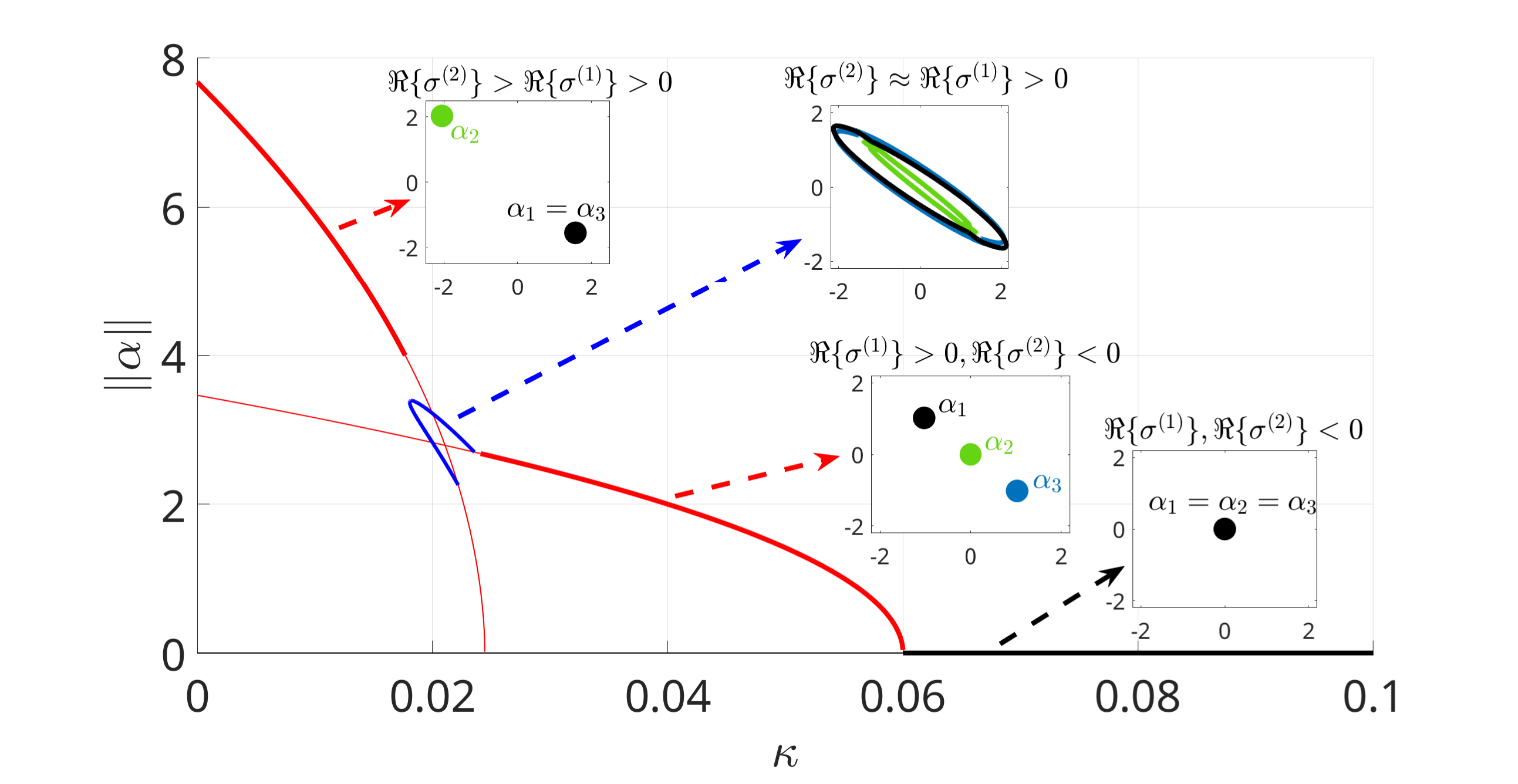}
	\caption{
		Bifurcation diagram of the deterministic model of Eq.~\eqref{alphaeq} for $N=3$ sites, obtained by numerical continuation in the parameter $\kappa$ in the squeezing-dominant regime $\eta\geq|\Delta|/2$. System parameters: $
		\Delta = 0.05,
		\gamma_1 = 0.03,  
		\gamma_2 = 0.005,
		\theta = \pi,  
		\eta = 0.025, 
		\lambda = 0.08.
		$
		The vertical axis represents the norm $\|\boldsymbol{\alpha}\|$ of the stationary solution. 
		Thick (thin) black lines denote stable (unstable) null equilibrium, whereas red curves represent stationary non–homogeneous solutions emerging via Turing instability. 
		The blue branch indicates a time–periodic (oscillatory) spatial pattern arising when two non-uniform eigenmodes have similar positive growth rates. 
		The inset shows representative spatial configurations $(\alpha_1,\alpha_2,\alpha_3)$ corresponding to selected points along the various branches.
	}\label{fig:bifurcation_delta05}
\end{figure*}

We now construct the bifurcation diagram of the deterministic model of Eq.~\eqref{alphaeq} for a parameter set lying at the boundary of the squeezing-dominant domain.
\bea
\Delta = 0.05,
\gamma_1 = 0.03,  
\gamma_2 = 0.005,
\theta = \pi,  
\eta = 0.025, 
\lambda = 0.08,\label{param1}
\ena
using $\kappa$ as a continuation (bifurcation) parameter.
For the chosen parameters, the instability thresholds for the non-uniform modes are \bea
\kappa_1^{\mathrm{th}} \approx 0.06,
\qquad
\kappa_2^{\mathrm{th}} \approx 0.0244.\label{kappa_tresh}
\ena
The diagram has been obtained by means of numerical continuation techniques implemented in \textsc{Julia} \cite{Julia}, and
the branches are represented by plotting the Euclidean norm 
$\|\boldsymbol{\alpha}\|$
as a function of $\kappa$ in Fig.~\ref{fig:bifurcation_delta05}. One clearly observes the instability scenario predicted by the linear analysis.
For $\kappa > \kappa_1^{\mathrm{th}}\approx0.06$, the null homogeneous solution 
$\boldsymbol{\alpha}=(0,0,0)$ is linearly stable (thick black line). 
At $\kappa=\kappa_1^{\mathrm{th}}\approx0.06$, the null equilibrium loses stability via a Turing instability, and a new branch of stationary non-uniform solutions emerges. 
This branch corresponds to the spatial structure dictated by the first Laplacian eigenmode $\boldsymbol{v}_1=(a,0,-a)$, and it is represented in Fig.~\ref{fig:bifurcation_delta05} by the red curve emerging from $\kappa=0.06$. 
A representative configuration of this patterned state is shown in the corresponding inset.
As $\kappa$ is further decreased, at $\kappa=\kappa_2^{\mathrm{th}}\approx 0.0244$, the second mode also becomes unstable, and oscillatory solutions emerge. These oscillating patterns arise when the two growth rates of Eq.~\eqref{eq:eigs_explicit} become comparable,
$
\sigma^{(1)} \approx \sigma^{(2)},
$
so that the dynamics is no longer dominated by a single spatial eigenmode. 
The interaction between the two unstable modes, being also at the boundary of the squeezing/detuning dominant domains as $\eta=|\Delta|/2$, gives rise to a time–periodic spatial pattern, visible in the diagram as the blue branch. This oscillatory behavior is typically explained through an analysis of subharmonic instabilities, as in Ref.~\cite{Gambino2018}, {and is not performed here because it lies outside the primary scope of the present work}.
For smaller values of $\kappa$, below approximately $\kappa\simeq 0.0175$, this oscillatory branch disappears and the system selects a purely stationary non-uniform configuration, corresponding to spatial structure reflecting the geometry of  $\boldsymbol{v}_2=(b,-2b,b)$  (shown in the leftmost inset of Fig.~\ref{fig:bifurcation_delta05}).

\begin{figure}[ht]
	\begin{center}
\includegraphics[width=0.5\columnwidth]{ 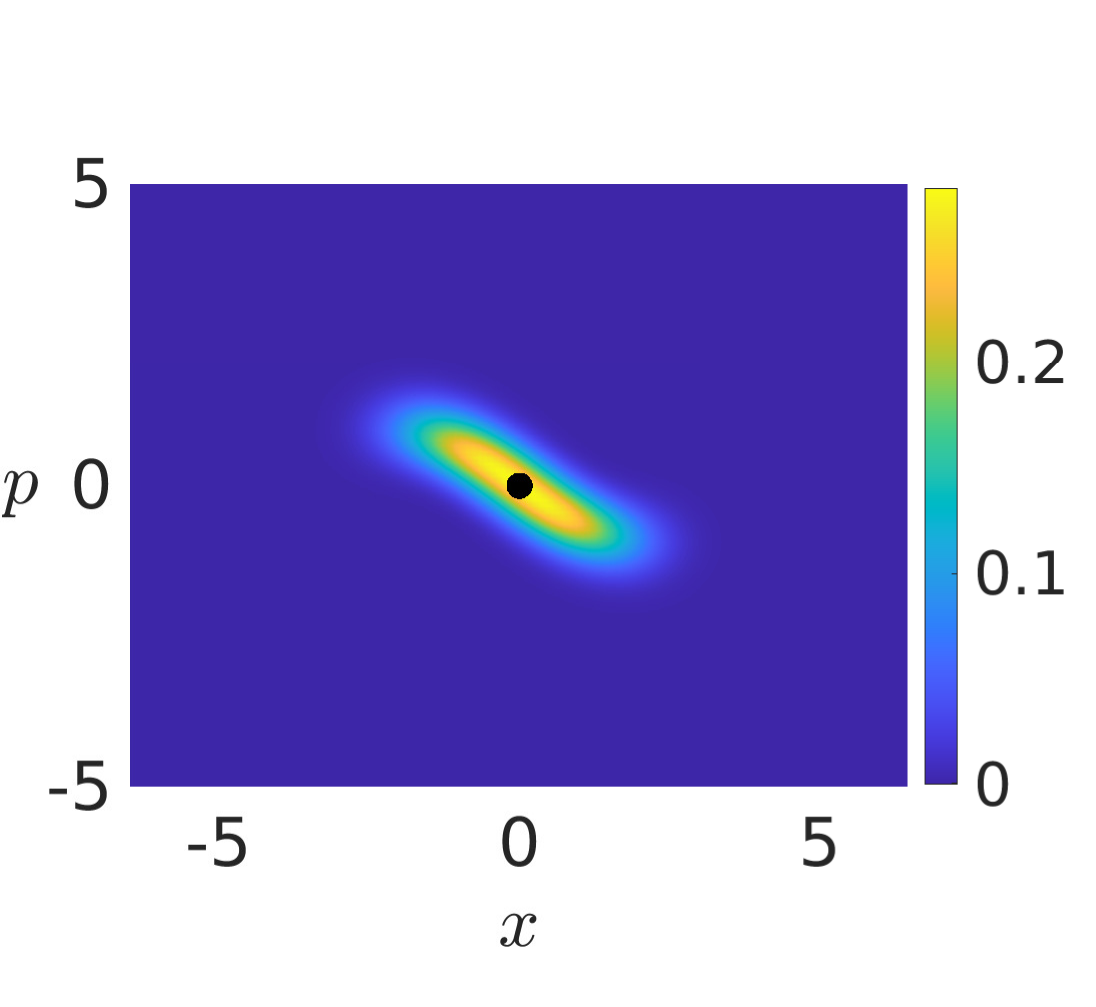}
	\caption{
Reduced Wigner functions in the phase space $(x,p)=(\Re{\alpha},\Im{\alpha})$ for the three–site quantum system in the absence of nonlocal couplings ($\lambda=\kappa=0$). 
Since the sites are dynamically decoupled, the reduced Wigner functions coincide, 
$W_1=W_2=W_3$. The parameters are 
$\Delta=0.05$, $\gamma_1=0.03$, $\gamma_2=0.005$, $\theta=\pi$, and $\eta=0.025$.  The black dot marks the homogeneous null solution for the deterministic model.
	}\label{fig:HomoSOL}
\end{center}
\end{figure}

\paragraph{Quantum counterpart and comparison with the SDE.}
We now consider the quantum counterpart of the full three-site quantum system. 
To characterize the local quantum state at each site, we compute the reduced Wigner functions  
$
W_j(\alpha_j,\alpha_j^\ast,t), j=1,2,3,
$
by performing a partial trace over the remaining degrees of freedom of the composite system, that is by considering the reduced density matrices
\beano
\hat\rho_1=\mathrm{Tr}_{2,3}\{\hat\rho\},\qquad
\hat\rho_2=\mathrm{Tr}_{1,3}\{\hat\rho\},\qquad
\hat\rho_3=\mathrm{Tr}_{1,2}\{\hat\rho\},
\enano
and then applying the Wigner transform of Eq.~\eqref{Wigner} (see Appendix \ref{APP:WIGNER}).  
We first consider the quantum dynamics in the absence of nonlocal interactions, namely, for
$
\lambda = 0,
\kappa = 0,
$
so that the three sites are dynamically decoupled and evolve independently with 
$
W_1 = W_2 = W_3.
$
The steady-state equilibrium configuration is shown in Fig.~\ref{fig:HomoSOL}, and it is represented by a squeezed Gaussian-like structure showing a unimodal behavior centered at the origin of the phase space ($\alpha=0$), that is at the deterministic equilibrium.

\begin{figure}[h!]
	\begin{center}
		\hspace*{-0.41cm}\subfigure[$W_1$]{\includegraphics[width=0.4\columnwidth]{ 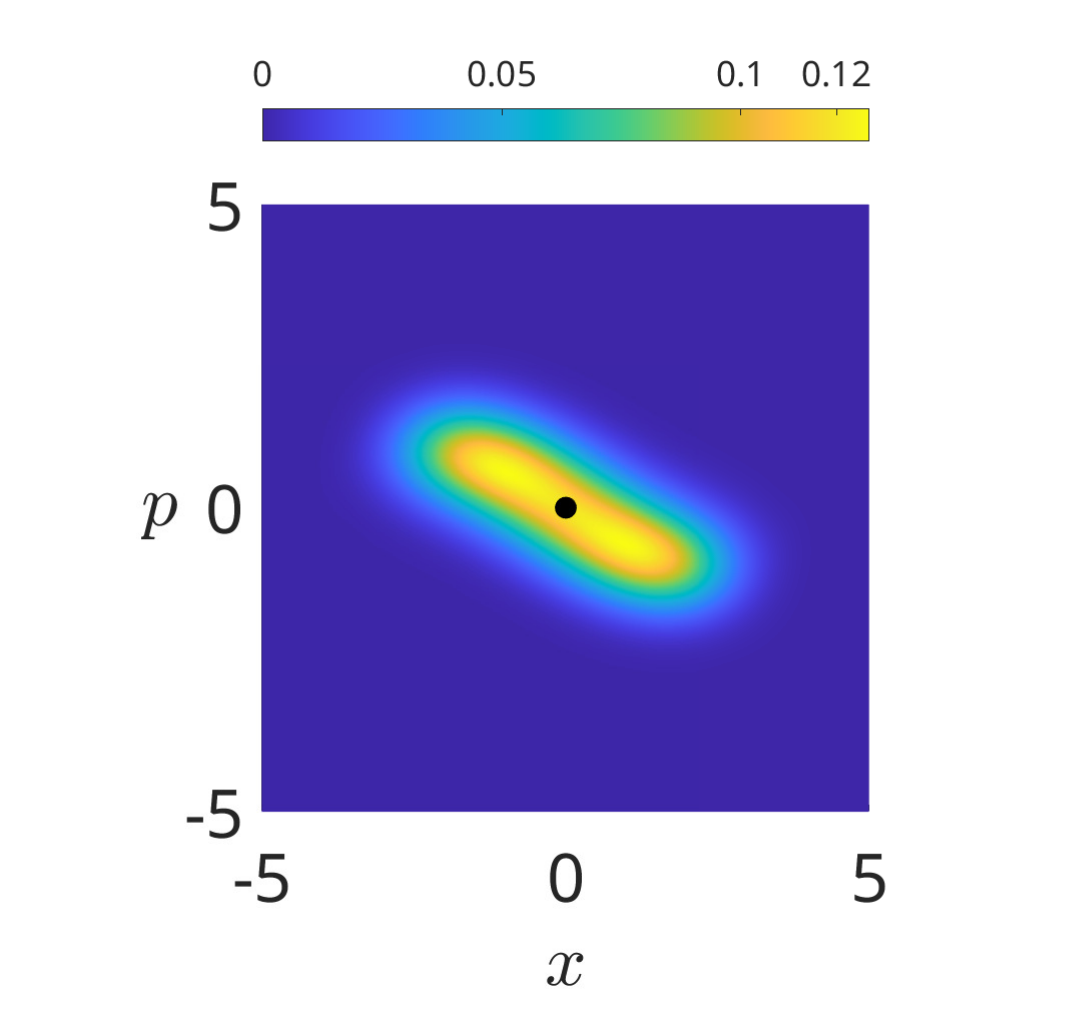}\label{fig:kappa008a}}
		\hspace*{-0.78cm}\subfigure[$W_2$]{\includegraphics[width=0.4\columnwidth]{ 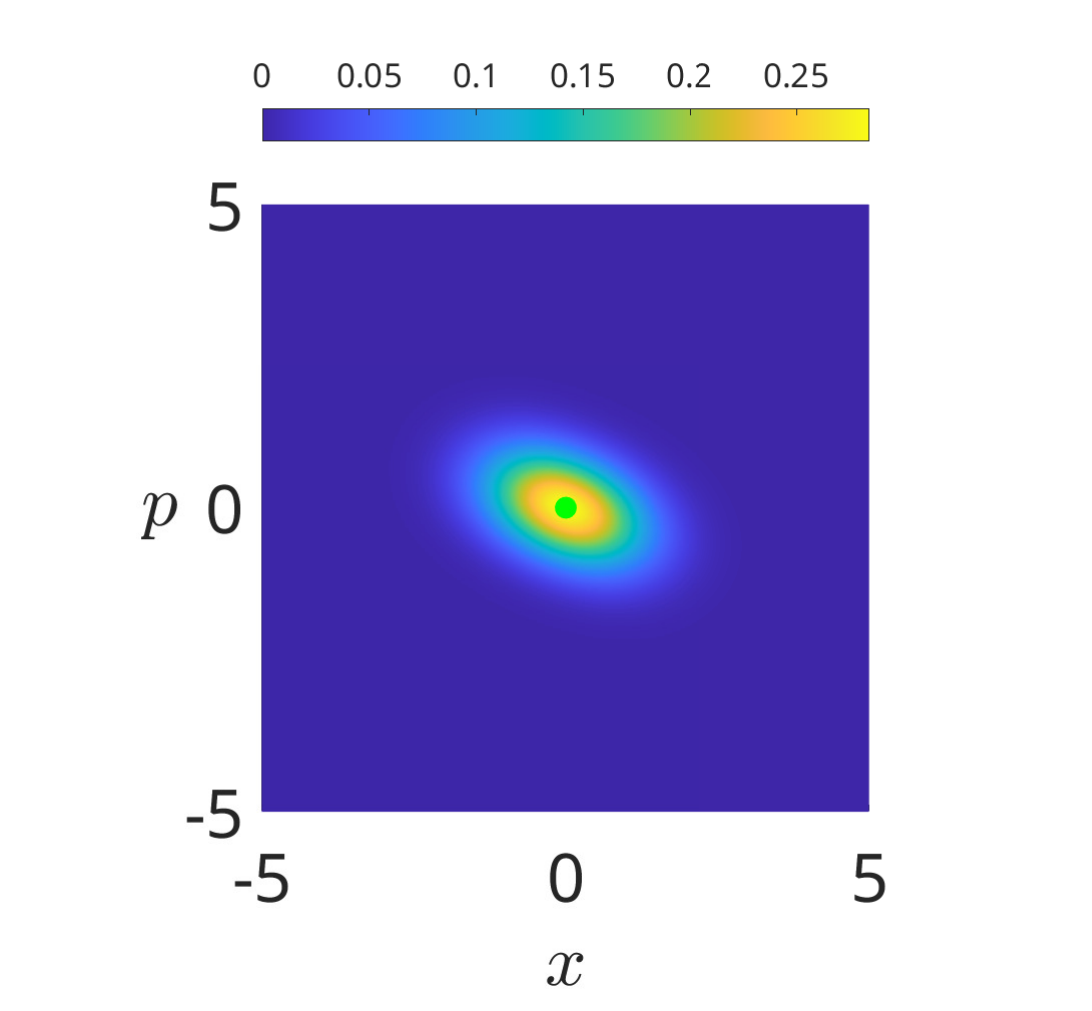}\label{fig:kappa008b}}
		\hspace*{-0.77cm}\subfigure[$W_3$]{\includegraphics[width=0.4\columnwidth]{ 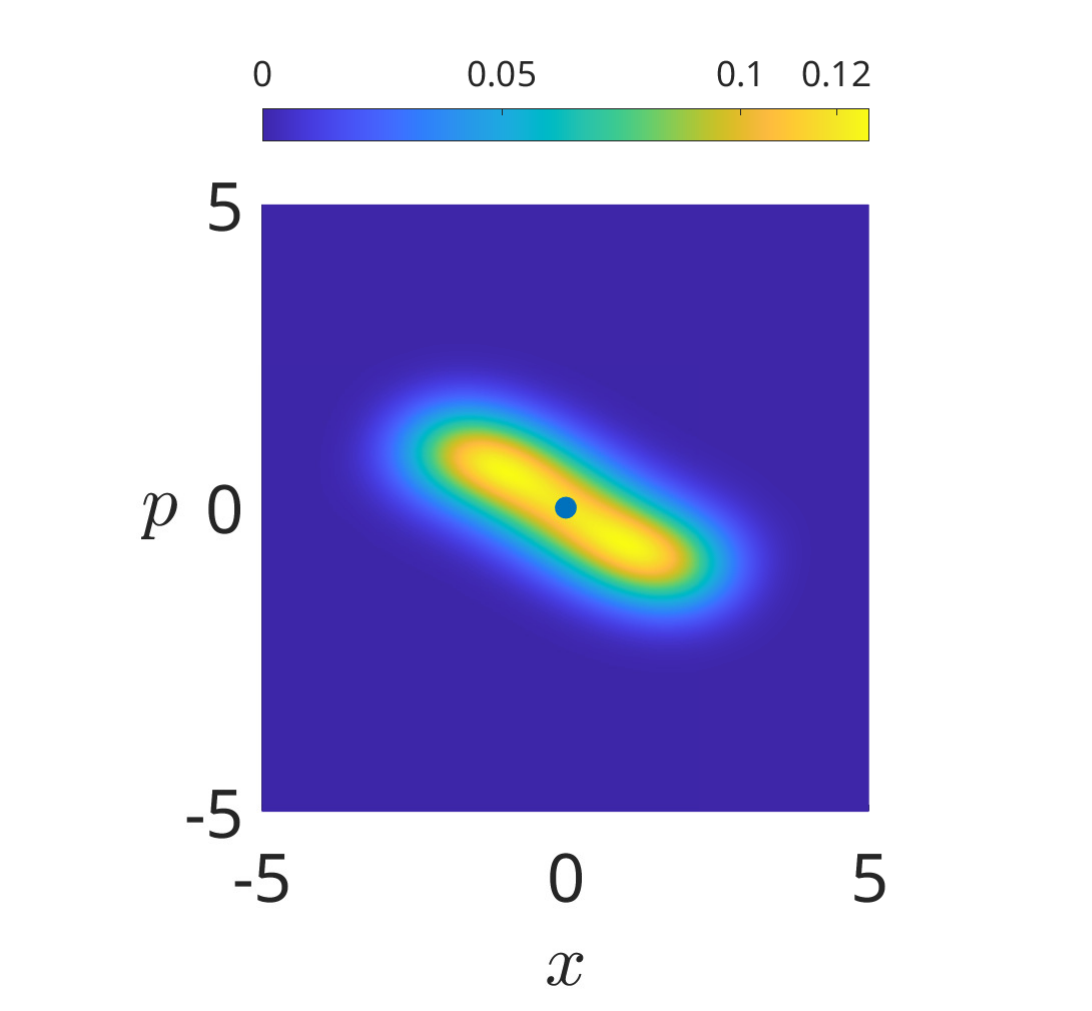}\label{fig:kappa008c}}	 \\
		\hspace*{-0.41cm}\subfigure[$\mathcal{P}(x, p)$, site 1]{\includegraphics[width=0.4\columnwidth]{ 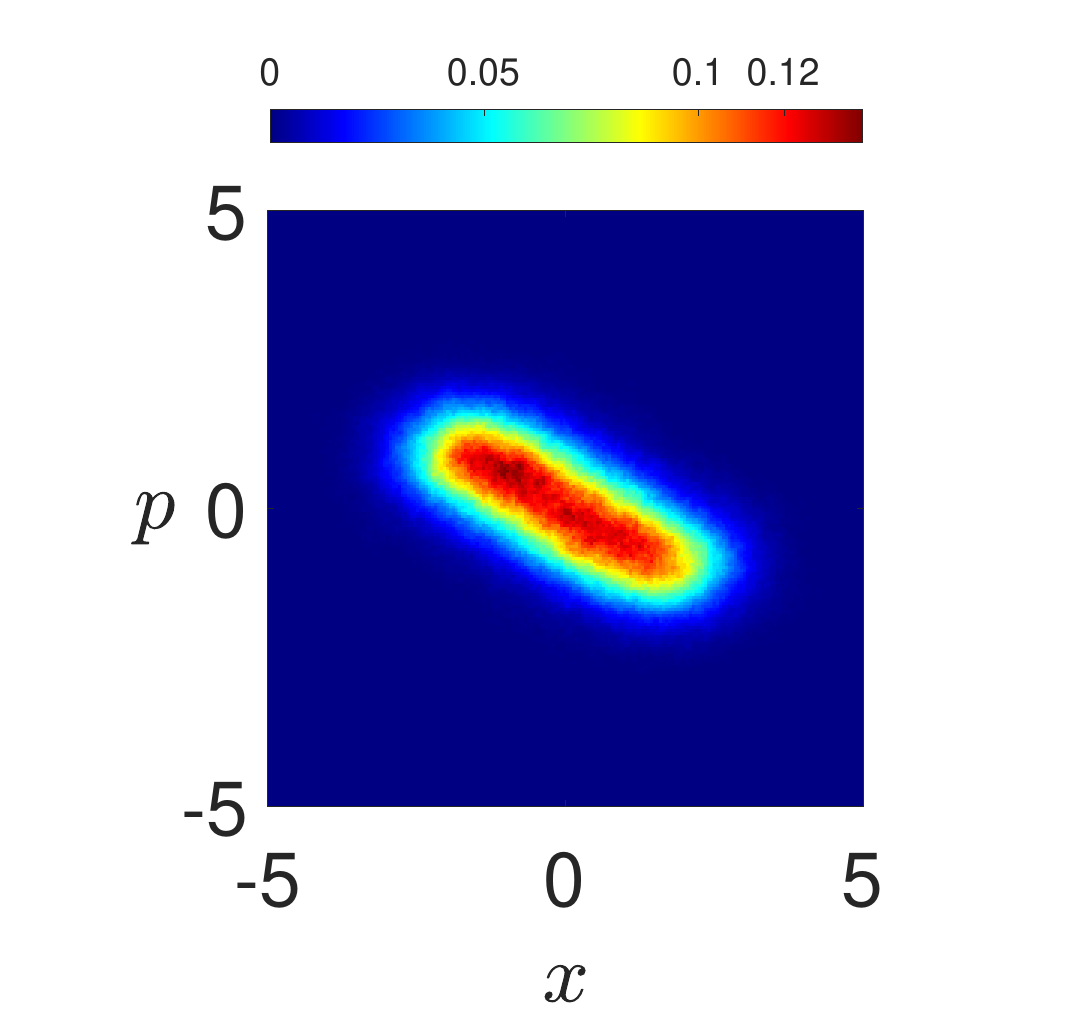}\label{fig:kappa008d}}
		\hspace*{-0.78cm}\subfigure[$\mathcal{P}(x, p)$, site 2]{\includegraphics[width=0.4\columnwidth]{ 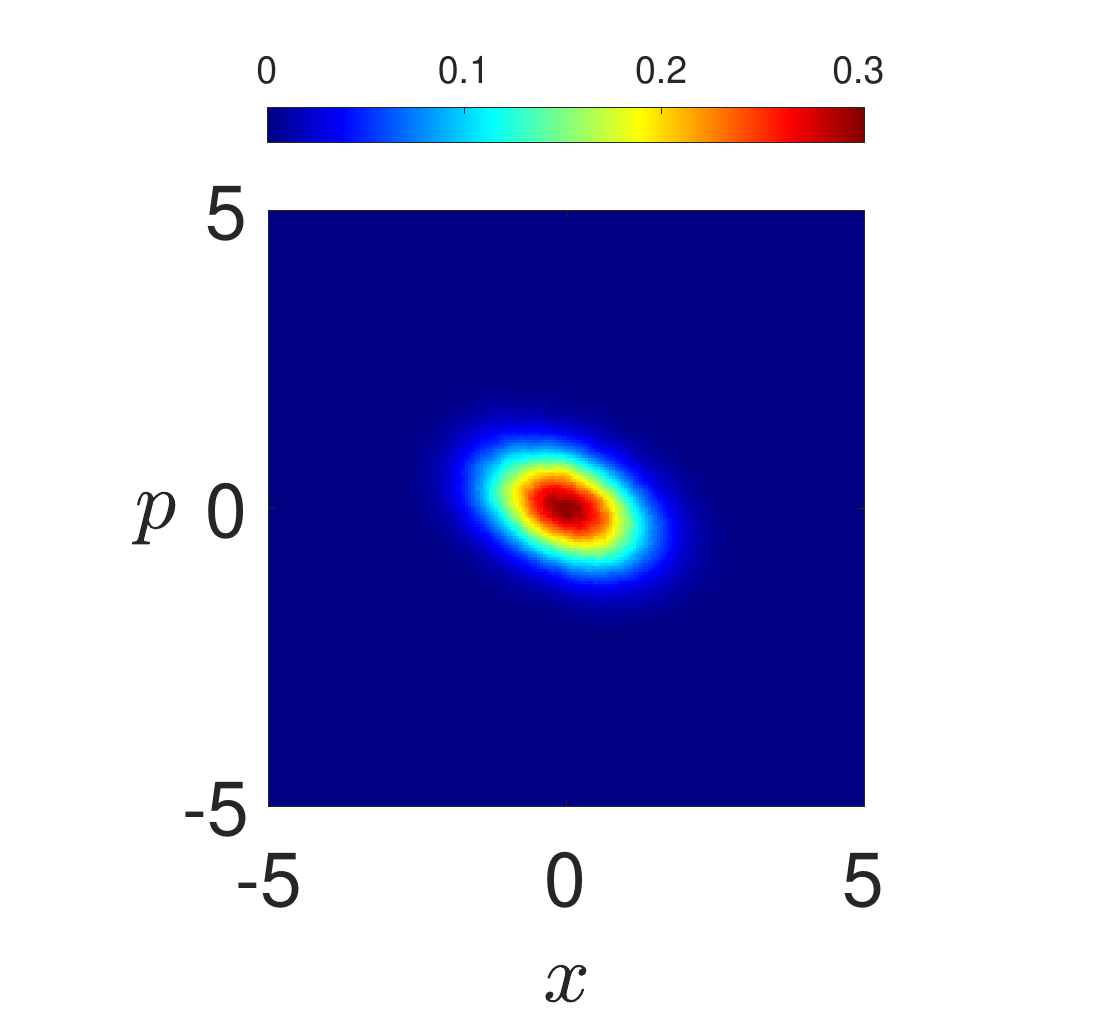}\label{fig:kappa008e}}
		\hspace*{-0.77cm}\subfigure[$\mathcal{P}(x, p)$, site 3]{\includegraphics[width=0.4\columnwidth]{ 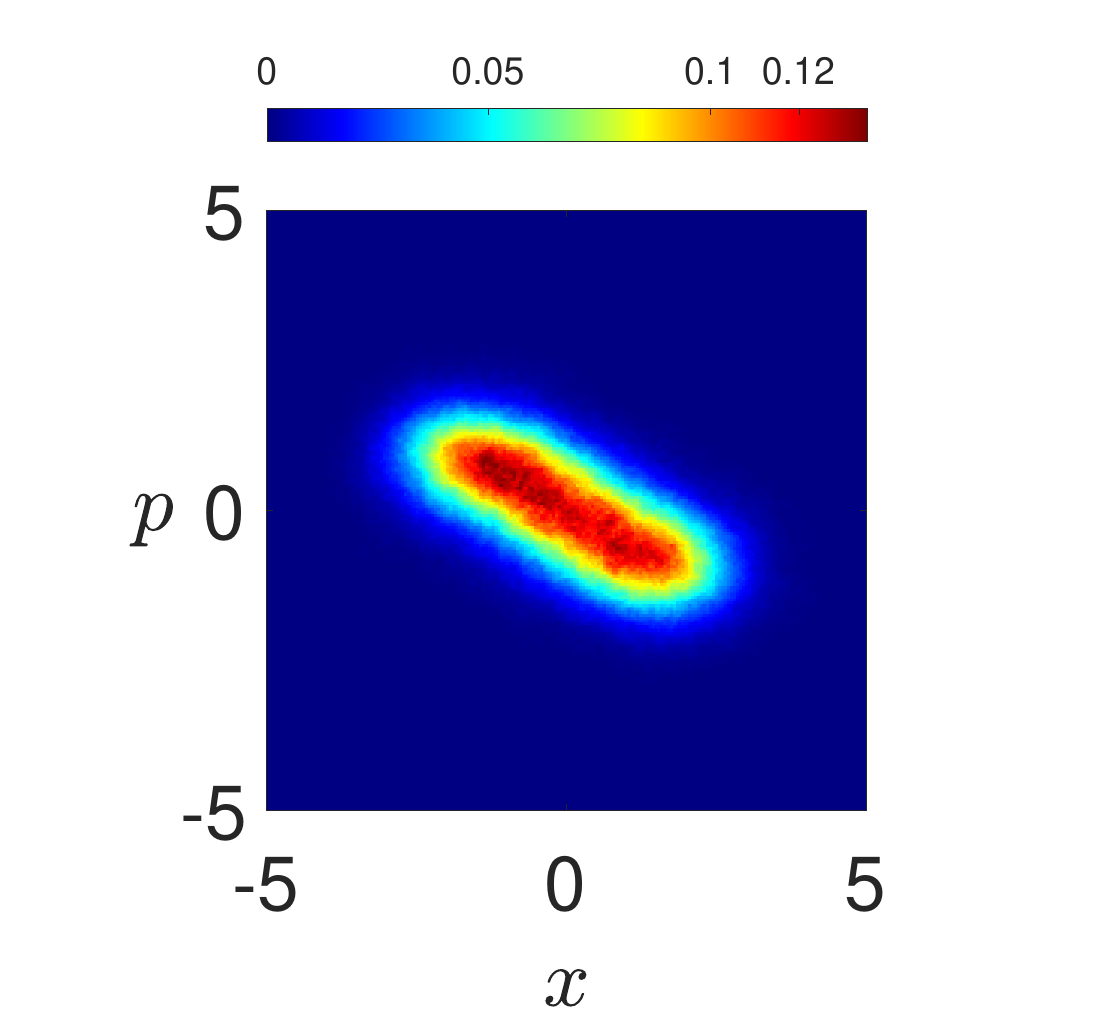}\label{fig:kappa008f}}
	\end{center}\caption{
		\textbf{(a)-(c)} Reduced Wigner functions for the three-site quantum system for the system parameters $\Delta=0.05$, $\gamma_1=0.03$, $\gamma_2=0.005$, $\theta=\pi,\eta=0.025,\lambda=0.08,\kappa=0.08$, corresponding to the stable homogeneous stationary branch of the deterministic bifurcation diagram in Fig.~\ref{fig:bifurcation_delta05}. 
		The distributions $W_1$ and $W_3$ almost coincide and exhibit a barely visible bimodal structure\, precursor of the nearby patterned solution obtained for lower values of $\kappa$.
		The distribution function $W_2$ remains unimodal and concentrated at $\alpha_2=0$. \textbf{(d)-(f)} Density distributions $\mathcal{P}(x, p)$ derived from the SDE for each site.
	}\label{fig:kappa008}
\end{figure}

By restoring the nonlocal interactions, we first consider a parameter set belonging to the stable branch of the homogeneous solution in the bifurcation diagram (the black branch), namely $\kappa = 0.08$, which strictly satisfies 
$\kappa > \kappa_1^{\mathrm{th}}>\kappa_2^{\mathrm{th}}$. 
The deterministic model predicts the stable steady homogeneous solution
$\alpha_1 =\alpha_2 =\alpha_3=0$. The reduced Wigner functions of the quantum system are shown in Fig.~\ref{fig:kappa008}, panels (a)-(c), at the steady state. 
We observe that $W_1$ and $W_3$ almost coincide and display again a squeezed Gaussian-like structure, now with two barely pronounced lateral peaks, while $W_2$ remains more concentrated around the origin. 
Therefore, although the deterministic model still predicts the stable homogeneous equilibrium $\alpha_1=\alpha_2=\alpha_3=0$ in this parameter region, the quantum steady state already carries a weak signature of the nearby patterned branch, although it is not fully developed. This mechanism is analogous to the notion of noisy precursors in classical pattern-forming systems, where fluctuations below the instability threshold already anticipate some properties of the pattern appearing above threshold (see Ref.~\cite{AgezEtAl2002}).

\begin{figure}[t!]
	\begin{center}
	\hspace*{-0.41cm}\subfigure[$W_1$]{\includegraphics[width=0.4\columnwidth]{ 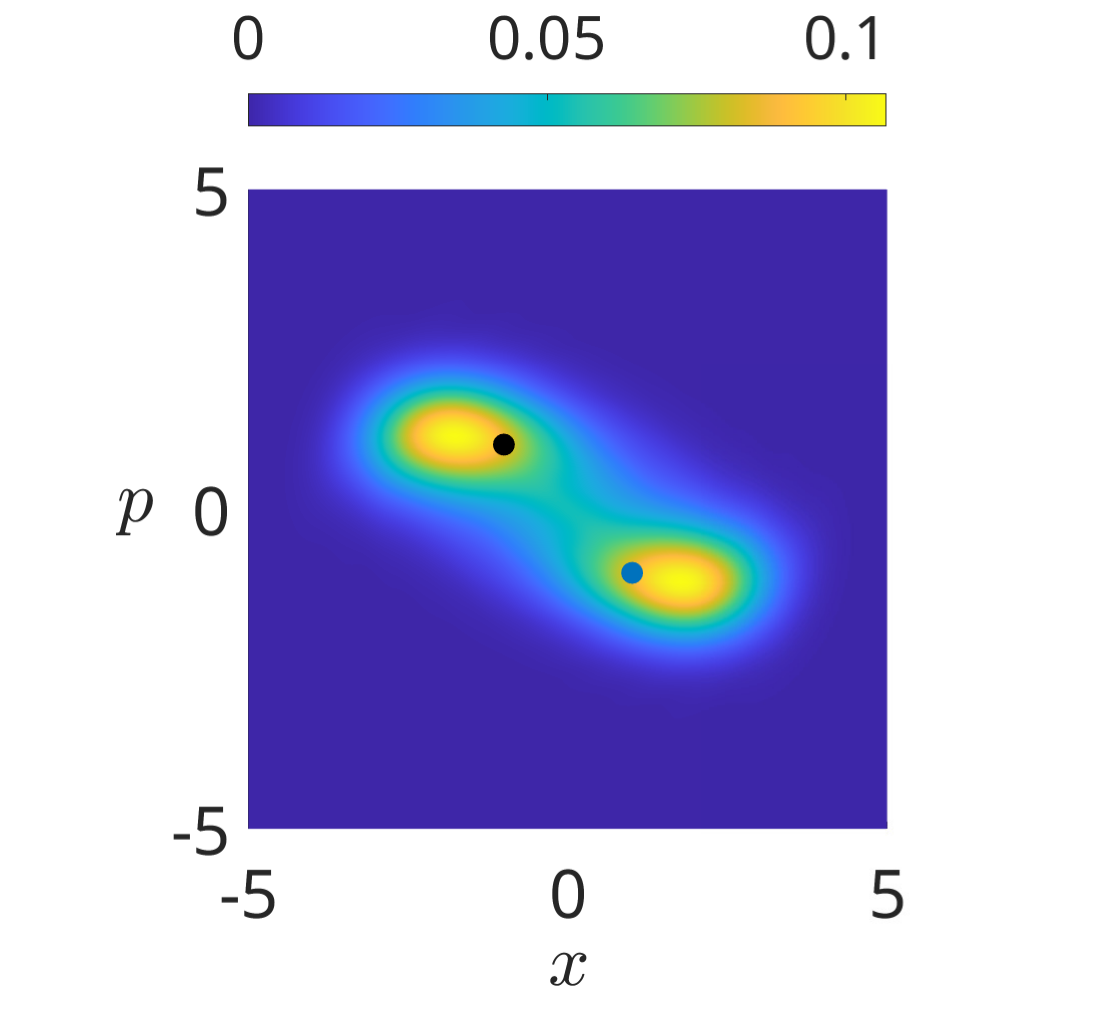}}
	\hspace*{-0.78cm}\subfigure[$W_2$]{\includegraphics[width=0.4\columnwidth]{ 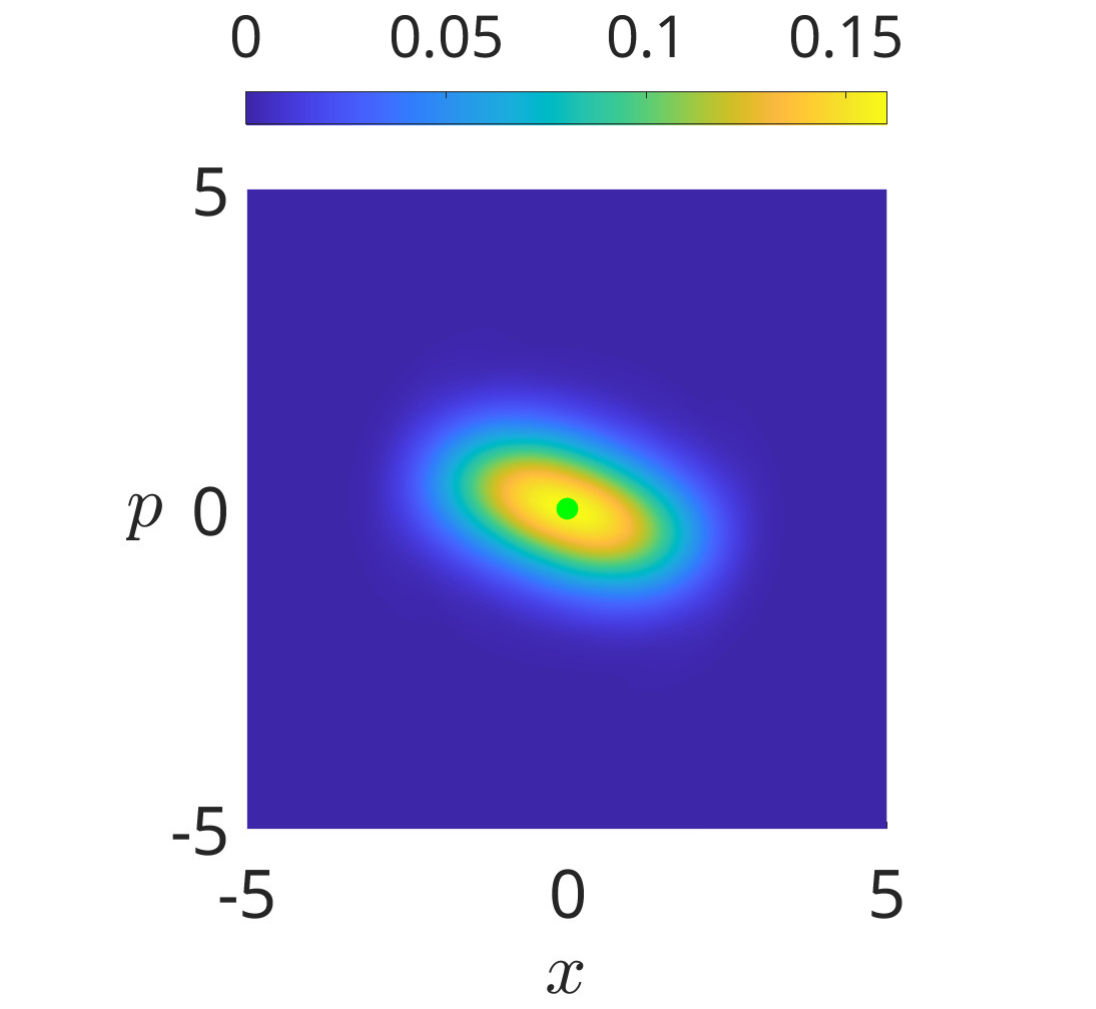}}
	\hspace*{-0.77cm}\subfigure[$W_3$]{\includegraphics[width=0.4\columnwidth]{ Tur_kappa004_mode1_3}\label{fig:kappa004c}}	
	\hspace*{-0.41cm}\subfigure[$\mathcal{P}(x, p)$, site 1]{\includegraphics[width=0.4\columnwidth]{ 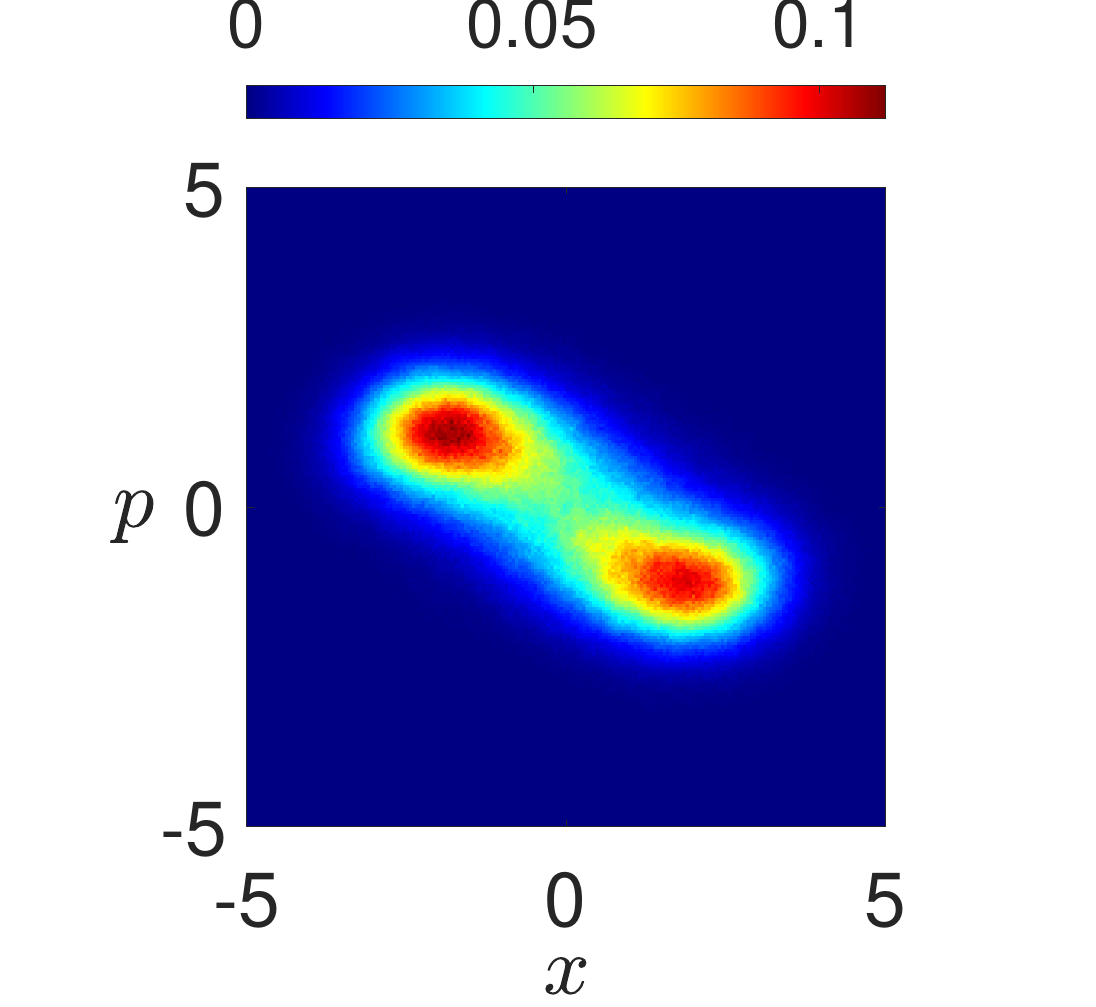}\label{fig:kappa004d}}
    \hspace*{-0.78cm}\subfigure[$\mathcal{P}(x, p)$, site 2]{\includegraphics[width=0.4\columnwidth]{ 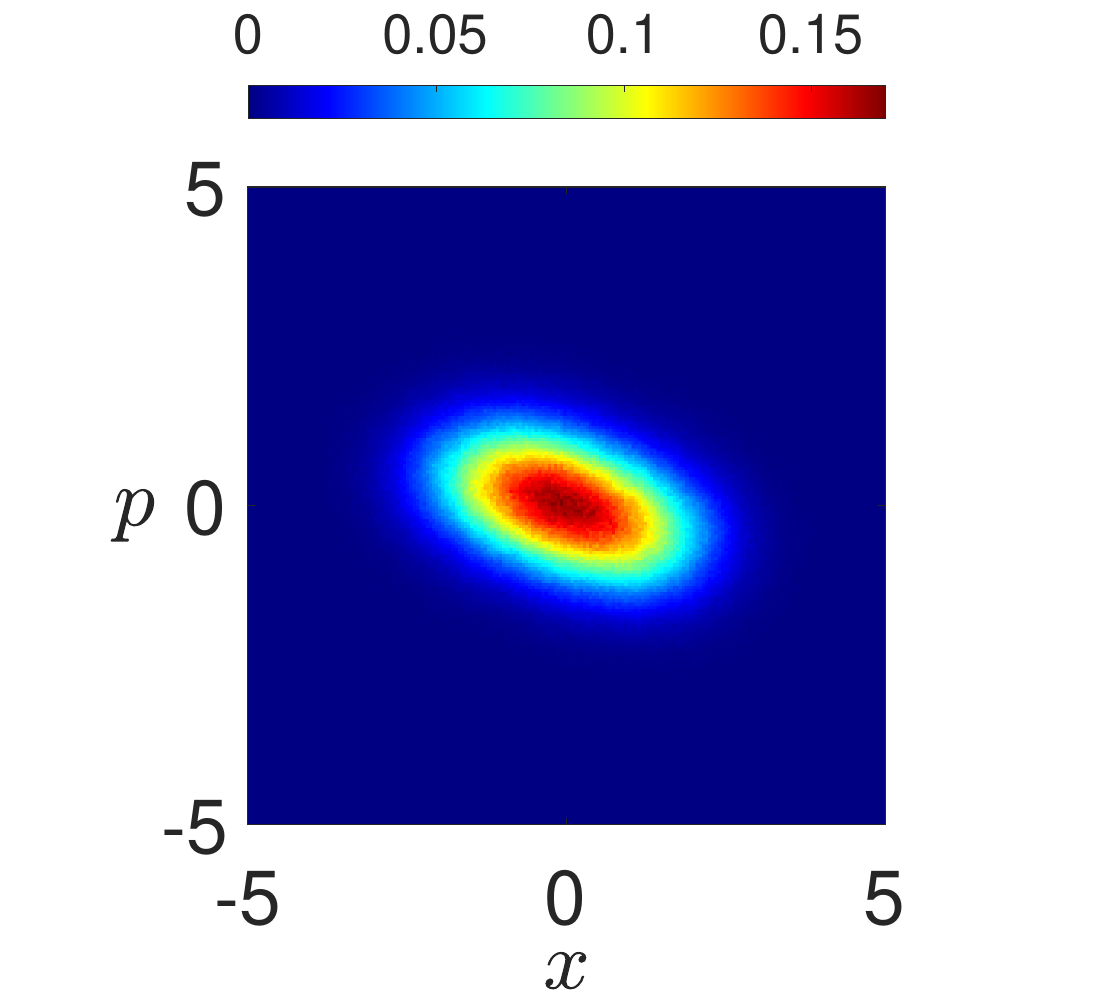}\label{fig:kappa004e}}
	\hspace*{-0.77cm}\subfigure[$\mathcal{P}(x, p)$, site 3]{\includegraphics[width=0.4\columnwidth]{ 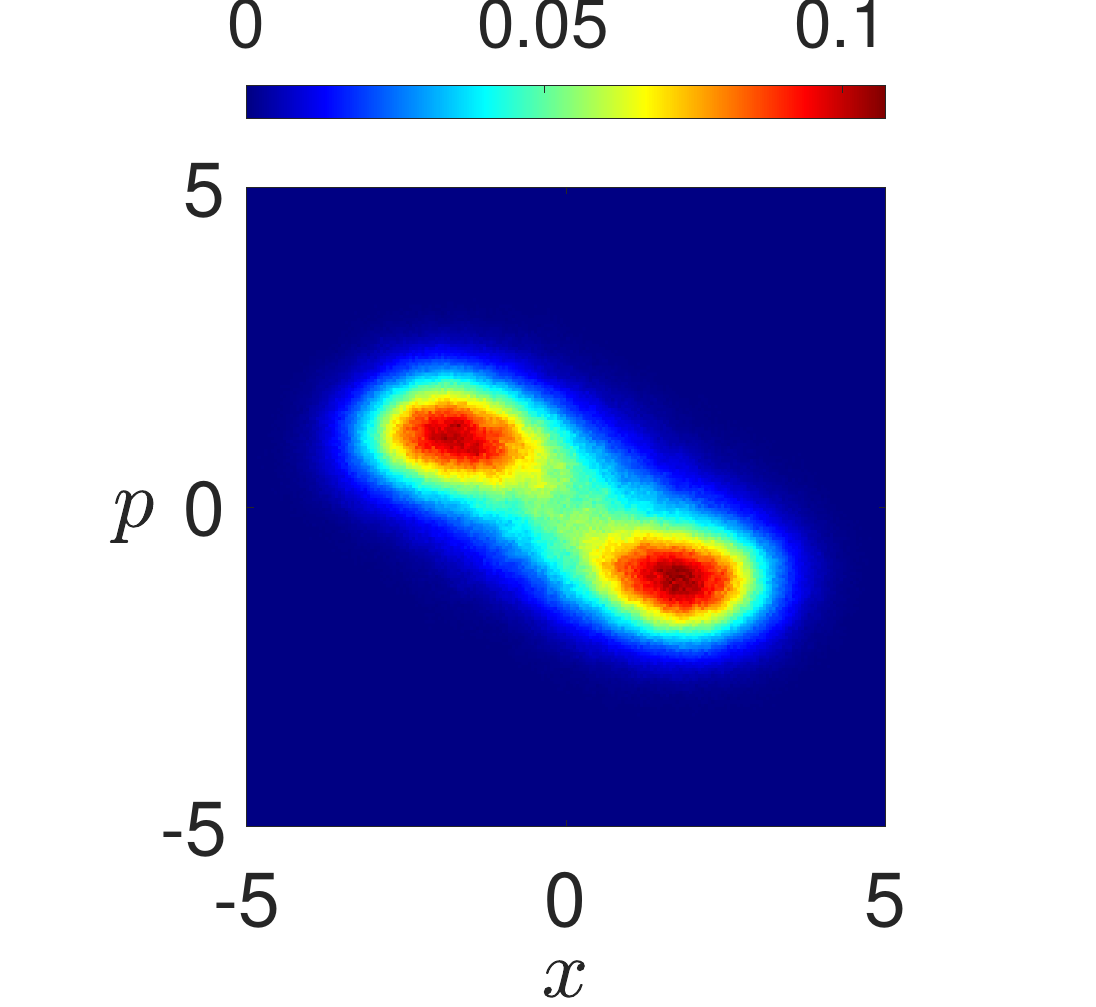}\label{fig:kappa004f}}
	\end{center}\caption{
	\textbf{(a)-(c)} Reduced Wigner functions for the three-site quantum system for the system parameters $\Delta=0.05$, $\gamma_1=0.03$, $\gamma_2=0.005$, $\theta=\pi, \eta=0.025, \lambda=0.08, \kappa=0.04$, corresponding to the first non-uniform stationary branch of the deterministic bifurcation diagram in Fig.~\ref{fig:bifurcation_delta05}. 
	The distributions $W_1$ and $W_3$ almost coincide and exhibit a bimodal structure with peaks located near the deterministic stationary values $\alpha_1=-\alpha_3=\pm(-1+1i)$, reflecting the underlying $\mathbb{Z}_2$ symmetry of the dynamics. 
	The reduced Wigner function $W_2$ remains unimodal and centered at $\alpha_2=0$. The deterministic values are represented by the black, green, and blue dots. \textbf{(d)-(f)} Density distributions $\mathcal{P}(x, p)$ derived from the SDE for each site.
}\label{fig:kappa004}
\end{figure}

We now consider a value of $\kappa$ in the first non-uniform branch of the bifurcation diagram, namely $\kappa = 0.04$, which strictly satisfies 
$\kappa_2^{\mathrm{th}} < \kappa < \kappa_1^{\mathrm{th}}$, showing that non-uniformity in the Wigner functions appears.
In this regime, only the first spatial mode is unstable, and the deterministic model predicts two symmetric stationary non-uniform configurations of the form
$\alpha_1 = -\alpha_3 \approx \pm (-1+1i)$ and $\alpha_2 = 0$, representing a spatial pattern dominated by the first eigenmode $\boldsymbol{v}_1=(a,0,-a)$ of $\nabla^{(2)}$.
The reduced Wigner functions are shown in Fig.~\ref{fig:kappa004}, panels (a)-(c), at the steady state. 
We observe that $W_1$ and $W_3$ coincide and now display a bimodal structure with two distinct peaks located near the classical stationary points $\pm \alpha_1=\mp \alpha_3$.
In contrast, the reduced Wigner function $W_2$ exhibits a unimodal distribution centered at $\alpha_2=0$, in agreement with the deterministic prediction. This patterned non-uniform behavior of the three Wigner functions is the clear signature of the emerging pattern selection.

We highlight here an important aspect: although the unstable eigenmode is $\boldsymbol{v}_1=(a,0,-a)$, the Wigner function preserves the $\mathbb{Z}_2$ symmetry, yielding $W_1=W_3$, because the quantum steady state remains a statistical mixture of the two equivalent patterns originating from the two symmetric patterns predicted by the deterministic model. For unimodal structures of $W_1$ and $W_3$ to emerge, peaked close to $\pm\alpha_1$ and $\pm\alpha_3$, one would require a measurement process, not considered in this work: this measurement breaks the symmetry and projects the state onto one of the two possible configurations, as clearly demonstrated in Refs.~\cite{Kato2022, JacobsSteck2006}.

We conclude the analysis of the first branch by giving a justification for adopting the deterministic evolution of the $\alpha_j$ to explain the pattern formation, which is well supported in the weak quantum regime for the selected parameters.

In fact, using $\gamma_2=0.005$, small compared to other system parameters, makes the semiclassical approximation reliable. To show this, we plot in Fig.~\ref{fig:kappa004}, panels (d)-(f), the density distributions determined by the SDE evolution and computed via a two-dimensional binning procedure in phase space. 
Specifically, the empirical probability density $\mathcal{P}(x, p)$ is reconstructed by discretizing the $(x, p)$ plane into a discrete grid with bin area $\delta_{x}\delta_{p}$, such that at each grid point:
\begin{equation}
	\mathcal{P}(x_i, p_j) = \frac{\Gamma(x_i, p_j)}{\delta_{x}\delta_{p} \sum_{i,j} \Gamma(x_i, p_j)},
\end{equation}
where $\Gamma(x_i, p_j)$ represents the number of sampled trajectory points falling into the 
corresponding bin, and the denominator ensures the distribution is 
properly normalized to unity over the entire domain. Specifically, we used $\delta_x=\delta_p=0.025$, and considered a total of $10^7$ sampling time stpdf after an initial transient. The density distributions compare well with the corresponding Wigner distributions, showing the good agreement between the SDE and the QME dynamics.

\begin{figure}[t!]
	\begin{center}
		\hspace*{-0.41cm}\subfigure[$W_1$]{\includegraphics[width=0.4\columnwidth]{ 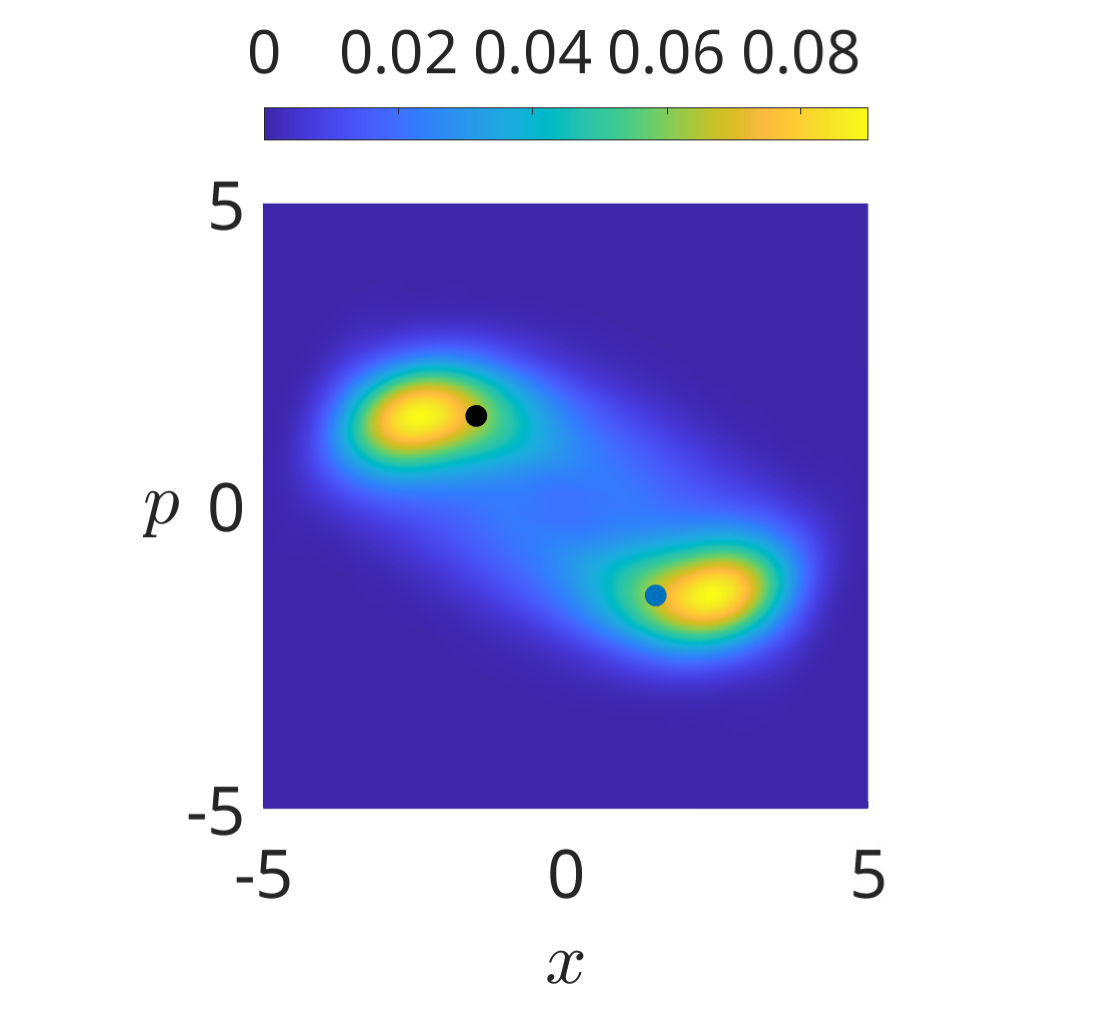}}
		\hspace*{-0.78cm}\subfigure[$W_2$]{\includegraphics[width=0.4\columnwidth]{ 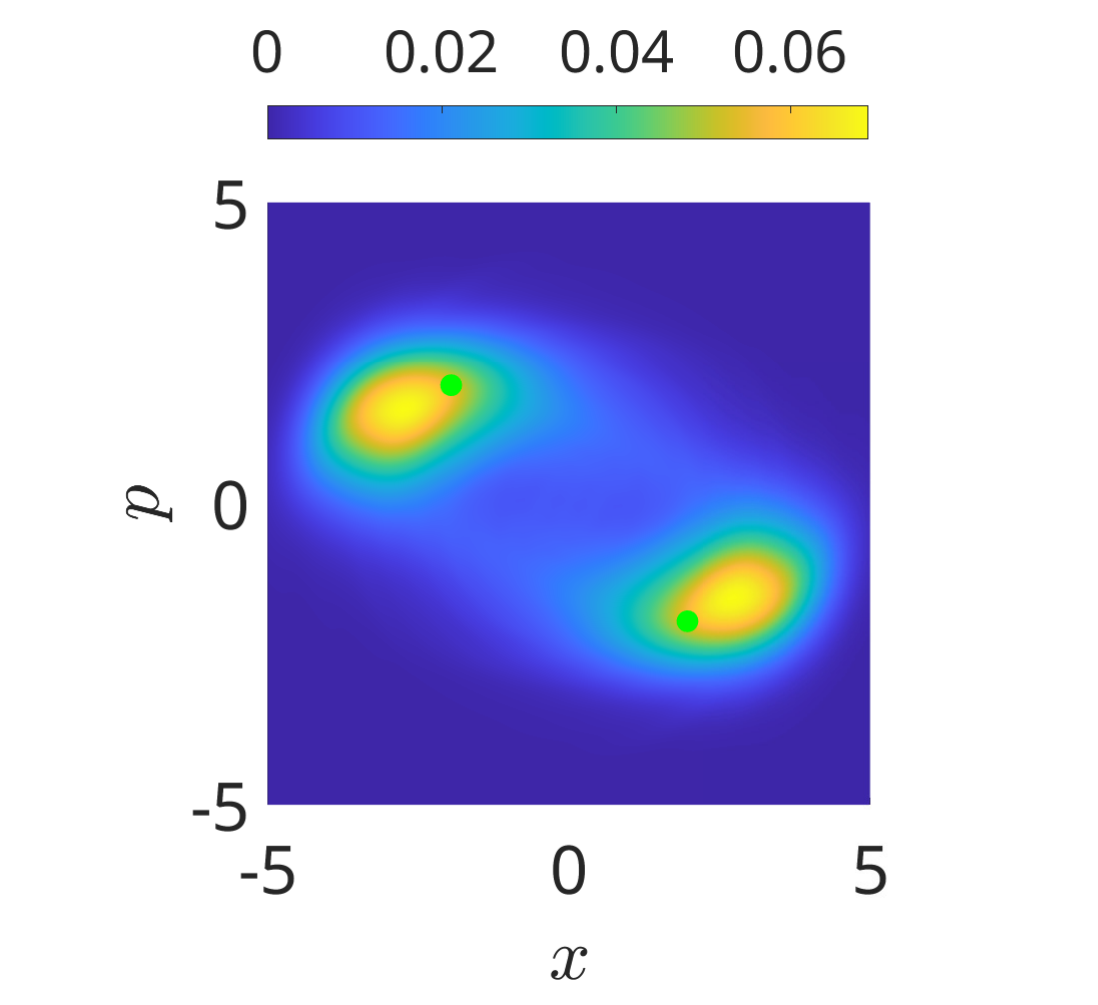}}
		\hspace*{-0.77cm}\subfigure[$W_3$]{\includegraphics[width=0.4\columnwidth]{ Tur_kappa00175_mode1_3}\label{fig:kappa00175c}}	
		\hspace*{-0.41cm}\subfigure[$\mathcal{P}(x, p)$, site 1]{\includegraphics[width=0.4\columnwidth]{ 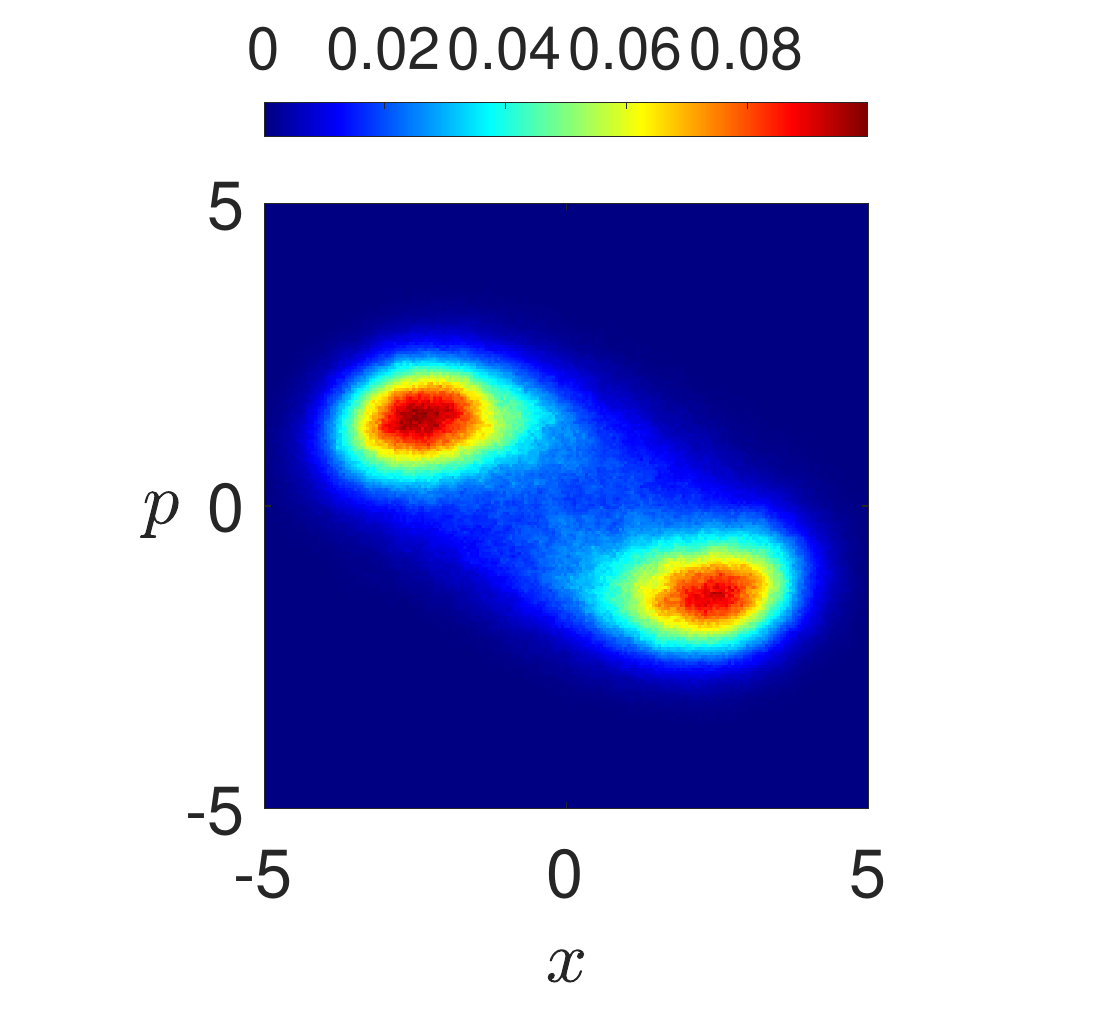}\label{fig:kappa00175d}}
		\hspace*{-0.78cm}\subfigure[$\mathcal{P}(x, p)$, site 2]{\includegraphics[width=0.4\columnwidth]{ 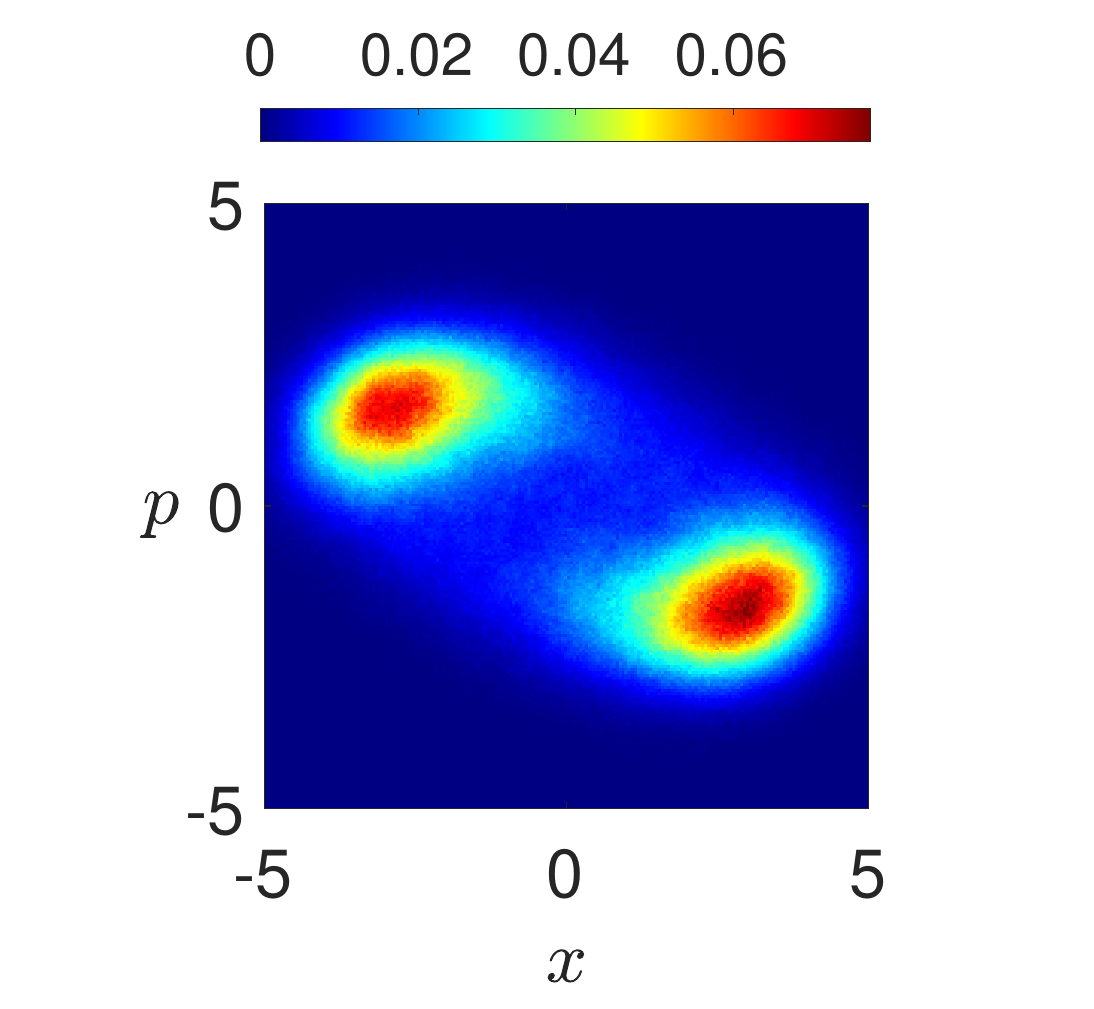}\label{fig:kappa00175e}}
		\hspace*{-0.77cm}\subfigure[$\mathcal{P}(x, p)$, site 3]{\includegraphics[width=0.4\columnwidth]{ 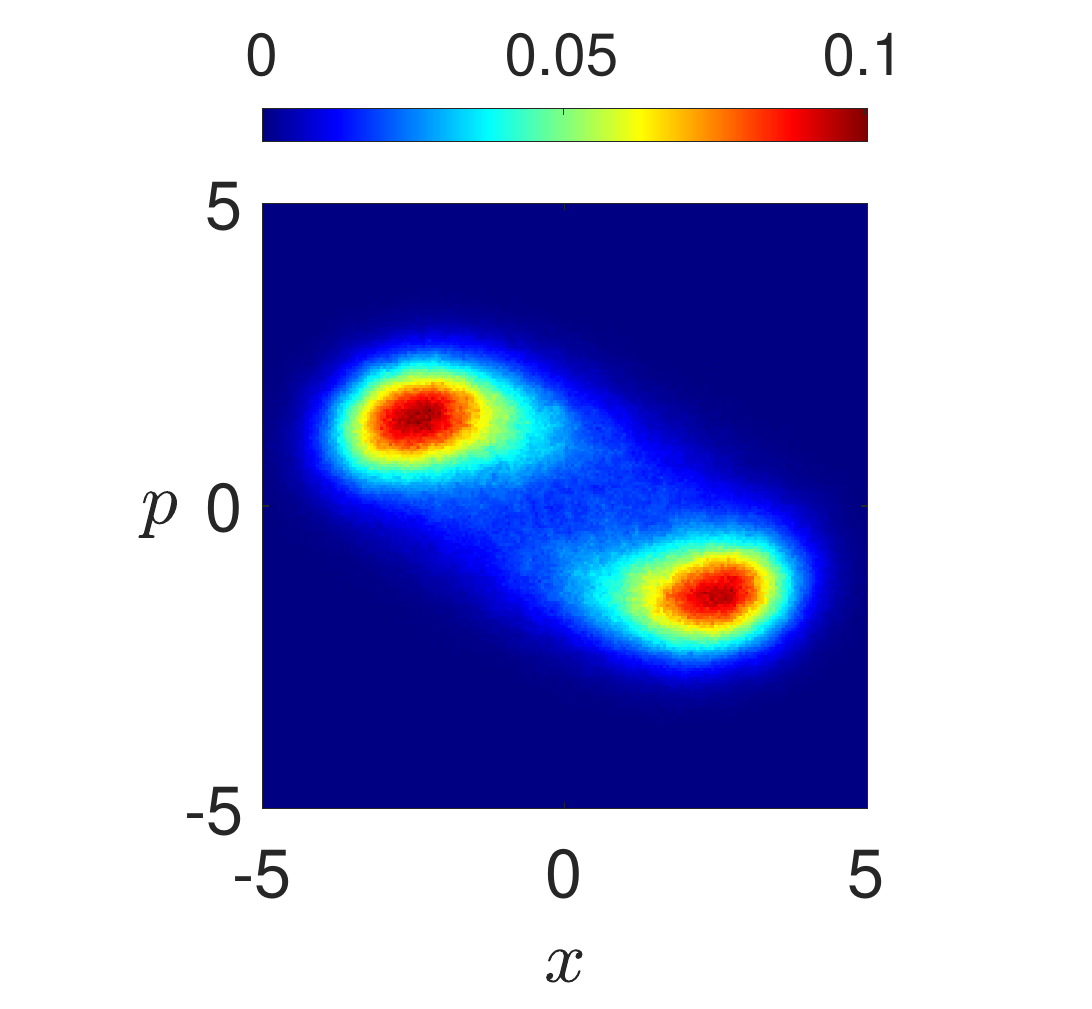}\label{fig:kappa00175f}}
	\end{center}\caption{
		\textbf{(a)-(c)} Reduced Wigner functions for the three-site quantum system for the system parameters $\Delta=0.05$, $\gamma_1=0.03$, $\gamma_2=0.005$, $\theta=\pi, \eta=0.025, \lambda=0.08, \kappa=0.0175$, corresponding to the second non-uniform stationary branch of the deterministic bifurcation diagram in Fig.~\ref{fig:bifurcation_delta05}. 
		The distributions $W_1$ and $W_3$ almost coincide and exhibit a bimodal structure with peaks located near the deterministic stationary values $\alpha_1=\alpha_3=\pm(-1.48+1.48i)$, reflecting the underlying $\mathbb{Z}_2$ symmetry of the dynamics. 
		Also, the reduced Wigner function $W_2$ has a bimodal behavior with peaks located near the deterministic stationary values {$\alpha_2=\pm(-1.96+1.96i)$}. The deterministic values are represented by the black, green, and blue dots. \textbf{(d)-(f)} Density distributions $\mathcal{P}(x, p)$ derived from the SDE for each site.
	}\label{fig:kappa00175}
\end{figure}

We now discuss the case for which  $\kappa$ falls below the second critical threshold $\kappa_2^{\mathrm{th}} \approx 0.0244$. A  clear bimodal behavior also appears in the second Wigner function $W_2$, as shown in Fig.~\ref{fig:kappa00175}, where the Wigner steady-state distributions are shown together with the densities $\mathcal{P}(x,p)$ reconstructed from the SDE trajectories.

In particular, $W_2$ has a bimodal structure with two peaks located near the deterministic stationary values $\alpha_2\approx\pm(-1.96+1.96i)$. From the point of
view of the deterministic model, this is explained by the second spatial non-uniform eigenmode $\boldsymbol{v}_2=(b,-2b,b)$ becoming dominant. As a consequence, the deterministic stationary pattern involves nonzero amplitudes on all three sites.

Also in this case, the SDE dynamics provides a consistent picture. The densities $\mathcal{P}(x,p)$ reconstructed from the SDE trajectories reproduce the same qualitative features of the Wigner functions, showing bimodal distributions for all sites and confirming that the semiclassical stochastic dynamics correctly captures the structure of the quantum steady state in this parameter regime.

\begin{figure}[!t]
	\begin{center}
		\hspace*{-0.38cm}\subfigure[]{\includegraphics[width=0.50\columnwidth]{ 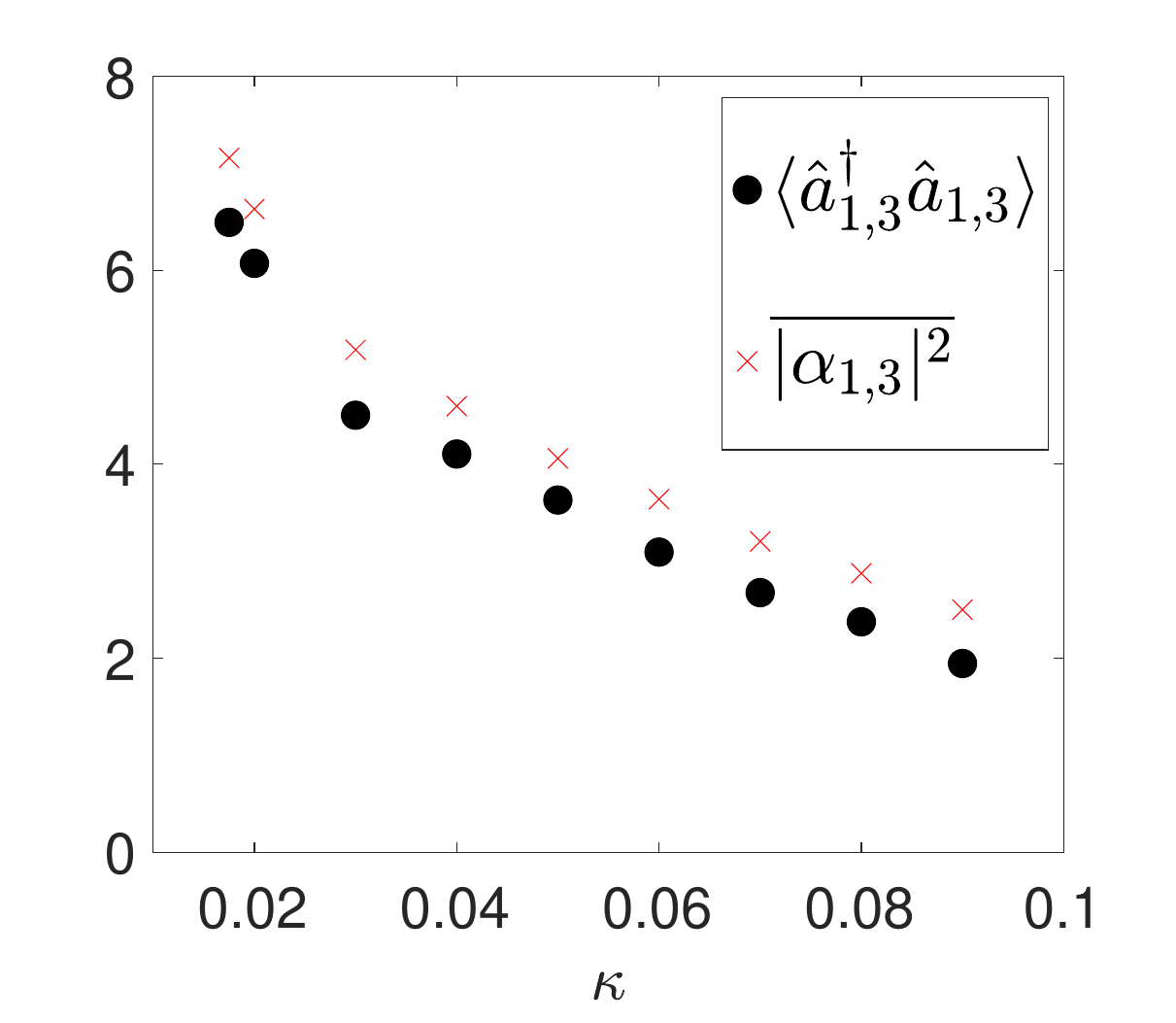}}
		\hspace*{0.25cm}\subfigure[]{\includegraphics[width=0.5\columnwidth]{ 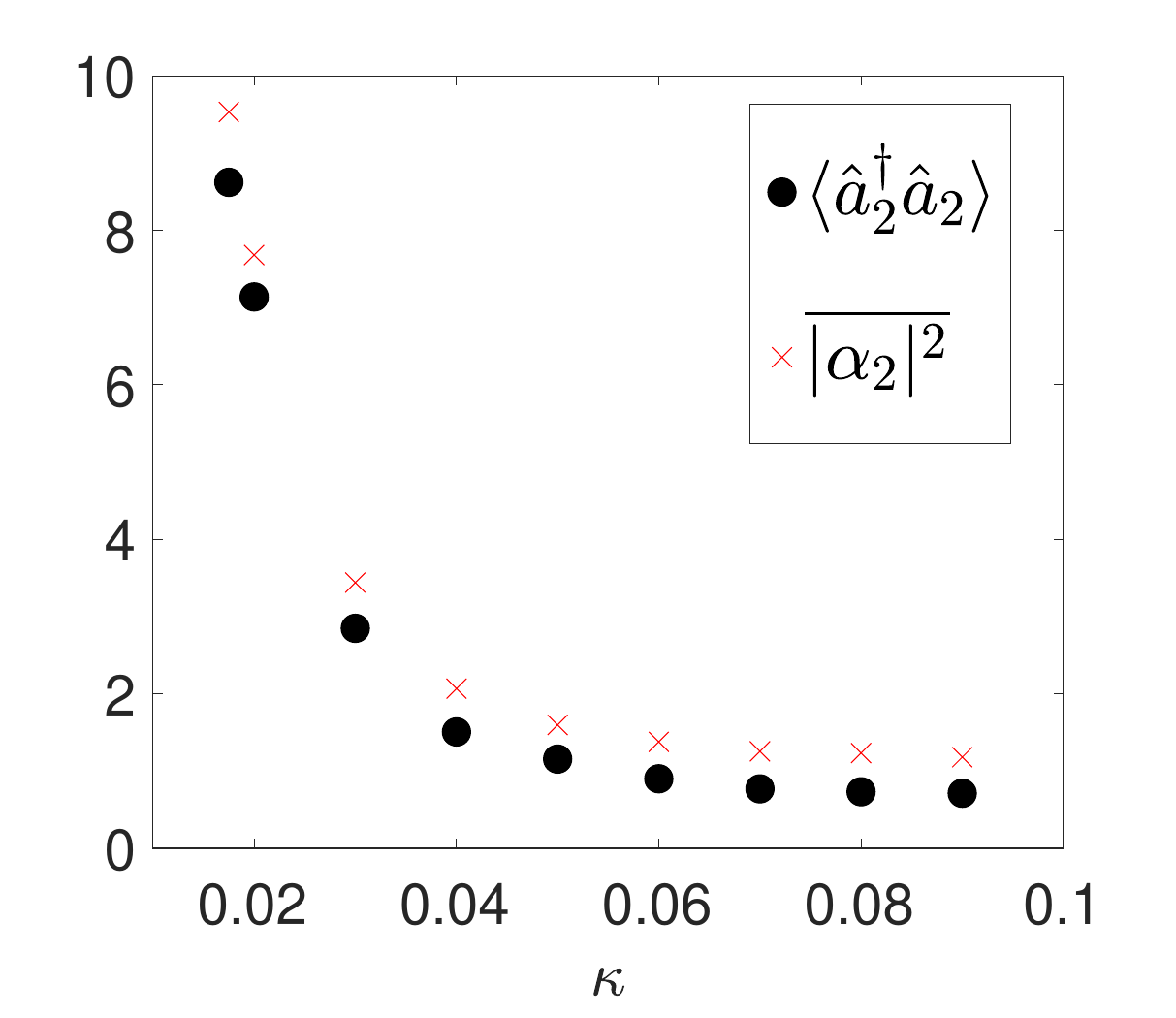}}
	\end{center}\caption{
		\textbf{(a)} The quantum mean phonon number for sites 1 and 3 (denoted by $\langle\hat{a}_{1,3}^\dagger \hat{a}_{1,3} \rangle$, black dots) plotted alongside the semiclassical averaged squared amplitude determined from the SDE ($\overline{|\alpha_{1,3}|^2}$, red crosses). \textbf{(b)} A similar comparison for site 2. In both cases, the numerical results show a constant shift of approximately $1/2$, confirming that the SDE provides a good approximation of the full QME in the weak quantum regime. The fixed parameters used for the simulations are $\Delta=0.05$, $\gamma_1=0.03$, $\gamma_2=0.005$, $\theta=\pi$, $\eta=0.025$, and $\lambda=0.08$, while $\kappa$ is varied.}
	\label{fig:SDE_QME_mode13}
\end{figure}

We conclude this part by showing the overall good agreement between the QME and SDE dynamics by analyzing the mean phonon
numbers $\langle\hat a_j^\dagger \hat a_j \rangle,\,j=1,2,3$ at each site from the quantum model, and the averaged amplitudes $\overline{|\alpha_j|^2},\,j=1,2,3,$ from the SDE model (notice that, in general, $\overline{|\alpha_j|^2}=\langle\hat a_j^\dagger \hat a_j \rangle+1/2$ because of the symmetric ordering \cite{Kato2022}). In Fig.~\ref{fig:SDE_QME_mode13}, we illustrate this comparison as a function of the parameter $\kappa$. Panel (a) illustrates the populations for sites 1 and 3 (they have the same values), while panel (b) displays the corresponding dynamics for site 2. As $\kappa$ is varied, the semiclassical predictions closely follow the exact quantum results, exhibiting the theoretically expected shift of approximately $1/2$. This confirms that the QME dynamics is captured by the SDE model in the weak quantum regime.

\paragraph{Modal characterization of quantum spatial structures.}
To further quantify the spatial structure of the quantum steady state, it is convenient to introduce a set of collective quadrature operators associated with the two non-uniform Laplacian eigenmodes of the three-site chain. 
More precisely, we define the modal quadratures
\bea
\hat X_{\mathrm{sym}}
=
\frac{\hat x_1-2\hat x_2+\hat x_3}{\sqrt{6}},
\qquad
\hat P_{\mathrm{sym}}
=
\frac{\hat p_1-2\hat p_2+\hat p_3}{\sqrt{6}},
\label{eq:XsymPsym}
\ena
and
\bea
\hat X_{\mathrm{asym}}
=
\frac{\hat x_1-\hat x_3}{\sqrt{2}},
\qquad
\hat P_{\mathrm{asym}}
=
\frac{\hat p_1-\hat p_3}{\sqrt{2}},
\label{eq:XasymPasym}
\ena
where, as usual, position and momentum operators are
\bea
\hat x_j=\frac{\hat a_j+\hat a_j^\dagger}{2},
\qquad
\hat p_j=\frac{\hat a_j-\hat a_j^\dagger}{2i},
\qquad j=1,2,3.
\ena
The operators $\hat X_{\mathrm{sym}},\hat P_{\mathrm{sym}}$ can be understood as the projection of the quantum state onto the symmetric non-uniform mode proportional to the eigenvector $\boldsymbol{v}_2$, whereas $\hat X_{\mathrm{asym}},\hat P_{\mathrm{asym}}$ probe the antisymmetric mode proportional to $\boldsymbol{v}_1$. They are introduced because they can reveal which spatial structure is dominant in each parameter regime. In analogy with the root-mean-square difference (RMSD) used to quantify spatial non-uniformity \cite{Kato2022}, here we characterize each mode via the corresponding root-mean-square modal quadrature amplitudes
\bea
\mathcal R_{X,r}
=
\sqrt{\left\langle
\left(
\hat X_r-\langle \hat X_r\rangle
\right)^2
\right\rangle},
\quad
\mathcal R_{P,r}
=
\sqrt{\left\langle
\left(
\hat P_r-\langle \hat P_r\rangle
\right)^2
\right\rangle},
\label{eq:RMS_modal}
\ena
with $r=\mathrm{sym},\mathrm{asym}$. Because of the underlying symmetry of the solution, the first moments $\langle \hat X_r\rangle$ and $\langle \hat P_r\rangle$ vanish or are numerically negligible, and hence
\bea
\mathcal R_{X,r}\simeq \sqrt{\langle \hat X_r^2\rangle},
\quad
\mathcal R_{P,r}\simeq \sqrt{\langle \hat P_r^2\rangle}.
\ena
For the representative values $\kappa=0.08,\,0.04,\,0.021,\,0.0175$, we obtained the steady-state amplitudes listed in Table~\ref{tabweak}.

\begin{table}[t!]
	\centering
	\begin{tabular}{c|cccc}
		\hline
		$\kappa$ & $\mathcal R_{X,\mathrm{sym}}$ & $\mathcal R_{P,\mathrm{sym}}$ & $\mathcal R_{X,\mathrm{asym}}$ & $\mathcal R_{P,\mathrm{asym}}$\\
		\hline
		$0.08$   & $0.76$   & $0.638$  & $1.81$  & $1.08$  \\
		$0.04$   & $1.25$   & $0.828$  & $2.31$  & $1.509$ \\
		$0.021$  & $2.82$   & $1.73$   & $2.23$  & $1.46$  \\
		$0.0175$ & $3.252$ & $2.148$ & $2.128$ & $1.34$  \\
		\hline
	\end{tabular}
	\caption{Steady-state root-mean-square values of the symmetric and antisymmetric modal quadratures. Other system parameters (squeezing-dominant regime, weak quantum regime): $\Delta=0.05$, $\gamma_1=0.03$, $\gamma_2=0.005$, $\theta=\pi,\eta=0.025,\lambda=0.08$.}
	\label{tabweak}
\end{table}

These values are fully consistent with the deterministic bifurcation diagram in Fig.~\ref{fig:bifurcation_delta05}, providing further support for adopting the same bifurcation scenario also in the quantum regime. 
For $\kappa=0.08$, namely above the first instability threshold $\kappa_1^{\mathrm{th}}\approx0.06$, the homogeneous solution is still stable, and correspondingly the symmetric RMSD modal amplitudes remain small, while the antisymmetric ones are already the largest among the two non-uniform projections. This indicates that the quantum steady state already displays enhanced fluctuations along the first patterned direction, namely, precursor signatures of the antisymmetric spatial structure. Although no true deterministic pattern is expected in this parameter region, quantum noise already enhances spatially structured fluctuations, similarly to the noisy pattern precursors described in classical noisy pattern-forming systems \cite{AgezEtAl2002}.

\begin{figure*}[t!]
	\includegraphics[width=0.75\textwidth]{ 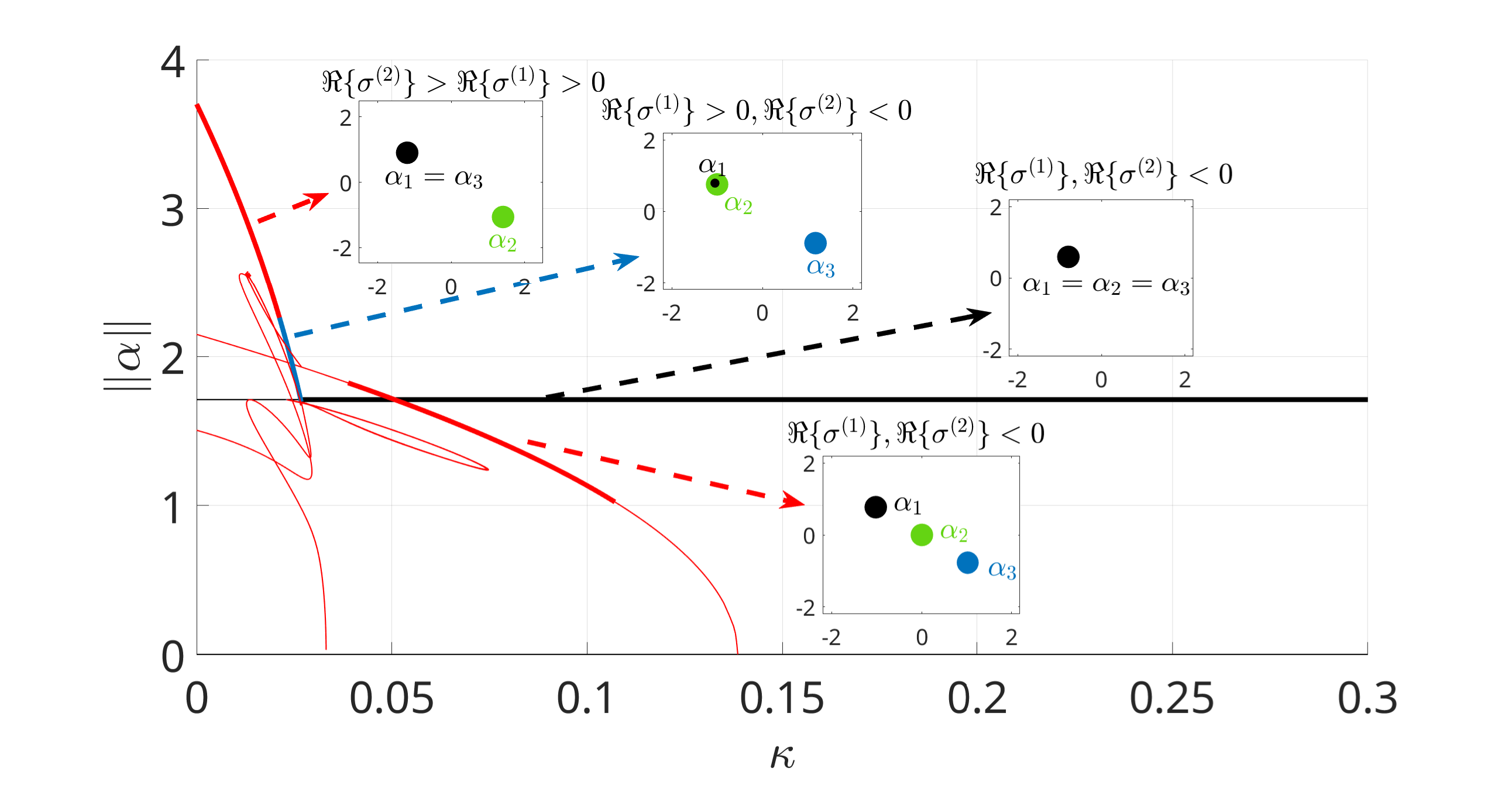}
	\caption{
		Bifurcation diagram of the deterministic model of Eq.~\eqref{alphaeq} for $N=3$ sites, obtained by numerical continuation in the parameter $\kappa$ in the squeezing-dominant regime $\eta>|\Delta|/2$ where a nonzero homogeneous stable solution exists. System parameters: $
		\Delta = 0.05,
		\gamma_1 = 0.03,  
		\gamma_2 = 0.03,
		\theta = \pi,  
		\eta = 0.026, 
		\lambda = 0.08.
		$
		The vertical axis represents the norm $\|\boldsymbol{\alpha}\|$ of the stationary solution. 
		Thick (thin) black lines denote stable (unstable) homogeneous equilibrium, whereas red curves represent stationary non–homogeneous solutions emerging via Turing instability. 
		The blue stable branch indicates the parameter region in which only one spatial mode is unstable.
		The inset shows representative spatial configurations $(\alpha_1,\alpha_2,\alpha_3)$ corresponding to selected points along the various branches.
	}\label{fig:bifurcation_nontrivial}
\end{figure*}

At $\kappa=0.04$, that is in the interval $\kappa_2^{\mathrm{th}}<\kappa<\kappa_1^{\mathrm{th}}$, the quantum state is clearly dominated by the antisymmetric mode. This is reflected in the fact that both $\mathcal R_{X,\mathrm{asym}}$ and $\mathcal R_{P,\mathrm{asym}}$ are substantially larger than their symmetric counterparts, in agreement with the deterministic prediction that the first non-uniform branch is organized by the antisymmetric spatial mode $\boldsymbol{v}_1=(a,0,-a)$.

A qualitatively different situation is found for $\kappa=0.021$, slightly below the second threshold $\kappa_2^{\mathrm{th}}\approx0.0244$. 
In this parameter region, both non-uniform modes are excited, consistently with the onset of the oscillatory branch in the bifurcation diagram. 
The values reported above indeed show that the antisymmetric contribution remains significant, but the symmetric modal amplitudes undergo a marked increase and become comparable to the antisymmetric ones, especially in the momentum-like quadratures.

Finally, for $\kappa=0.0175$, below the oscillatory window, the symmetric modal amplitudes become the dominant ones, with $\mathcal R_{X,\mathrm{sym}}=3.252$ and $\mathcal R_{P,\mathrm{sym}}=2.148$, while the antisymmetric contributions decrease with respect to the previous cases. 
This behavior is consistent with the leftmost stationary branch of Fig.~\ref{fig:bifurcation_delta05}, where the system selects a non-uniform stationary configuration shaped by the second eigenvector $\boldsymbol{v}_2=(b,-2b,b)$.

Hence, the root-mean-square amplitudes of the modal observables of Eqs.~\eqref{eq:XsymPsym}--\eqref{eq:XasymPasym} provide a direct quantitative bridge between the deterministic bifurcation structure and the quantum steady-state distributions characterized by the spatial structures of the Wigner functions, highlighting how the deterministic analysis can reveal important information about the fully quantum system.

\subsubsection{Deterministic roadmap for the nontrivial homogeneous state}

We now consider the deterministic bifurcation diagram for the nontrivial homogeneous equilibrium.
We consider the following set of system parameters
\bea
\Delta = 0.05,
\gamma_1 = 0.03,
\gamma_2 = 0.03,
\theta = \pi,
\eta = 0.026,
\lambda = 0.08,
\ena
so that a nontrivial homogeneous equilibrium exists when $r_+=0.987$ and $\phi=5.63691$ in Eq.~\eqref{condr2}, giving $\alpha_{+}^{\text{hom}}=0.7887-0.5950i$ in Eq.~\eqref{nontrivial}.
The instability thresholds of the stable nontrivial homogeneous branch are
\bea
\kappa_1^{\mathrm{th}}\simeq 0.02287,
\qquad
\kappa_2^{\mathrm{th}}\simeq 0.02032.
\ena

The corresponding bifurcation diagram is shown in Fig.~\ref{fig:bifurcation_nontrivial}. For sufficiently large values of
$\kappa$, the nontrivial homogeneous state is stable, as indicated by the thick black branch. As $\kappa$ decreases, this homogeneous state loses stability through a non-uniform instability, giving rise to stationary patterned solutions.
Since these two threshold values, $\kappa_1^{\mathrm{th}}$ and $\kappa_2^{\mathrm{th}}$, are very close, only one dominant non-uniform branch is visibly generated from the nontrivial homogeneous state.
It is also worth noting that a stable portion of a non-uniform branch connected with the trivial solution is still present.

\paragraph{Quantum counterpart}

Since $\gamma_2=0.03$ is comparable with the other system parameters, the dynamics is expected to belong to a moderate-to-strong quantum regime. In this case, the SDE approximation is no longer reliable.
Indeed, the diffusion matrix of Eq.~\eqref{diffmatrix}, used to construct the SDE in Appendix~\ref{APP:CNE}, is not always positive-semidefinite, especially for small values of $|\alpha|$. This follows from the relatively large value of $\gamma_2$, which makes the semiclassical truncation no longer justified. Moreover, the third-order derivative terms in the Wigner equation, \eqref{thirdorderderivWigner}, are expected to be non-negligible and may induce genuine quantum corrections. At the same time, the large value of $\gamma_2$ enhances the diffusive contribution in the Wigner dynamics.
Therefore, quantum fluctuations may smooth the phase-space distribution and mask or reduce the signatures of the Turing instability observed in the weak-quantum regime, such as bimodal Wigner functions with peaks close to the deterministic equilibria.

\begin{figure}[t!]
	\begin{center}
		\hspace*{-0.41cm}\subfigure[$W_1$]{\includegraphics[width=0.4\columnwidth]{ 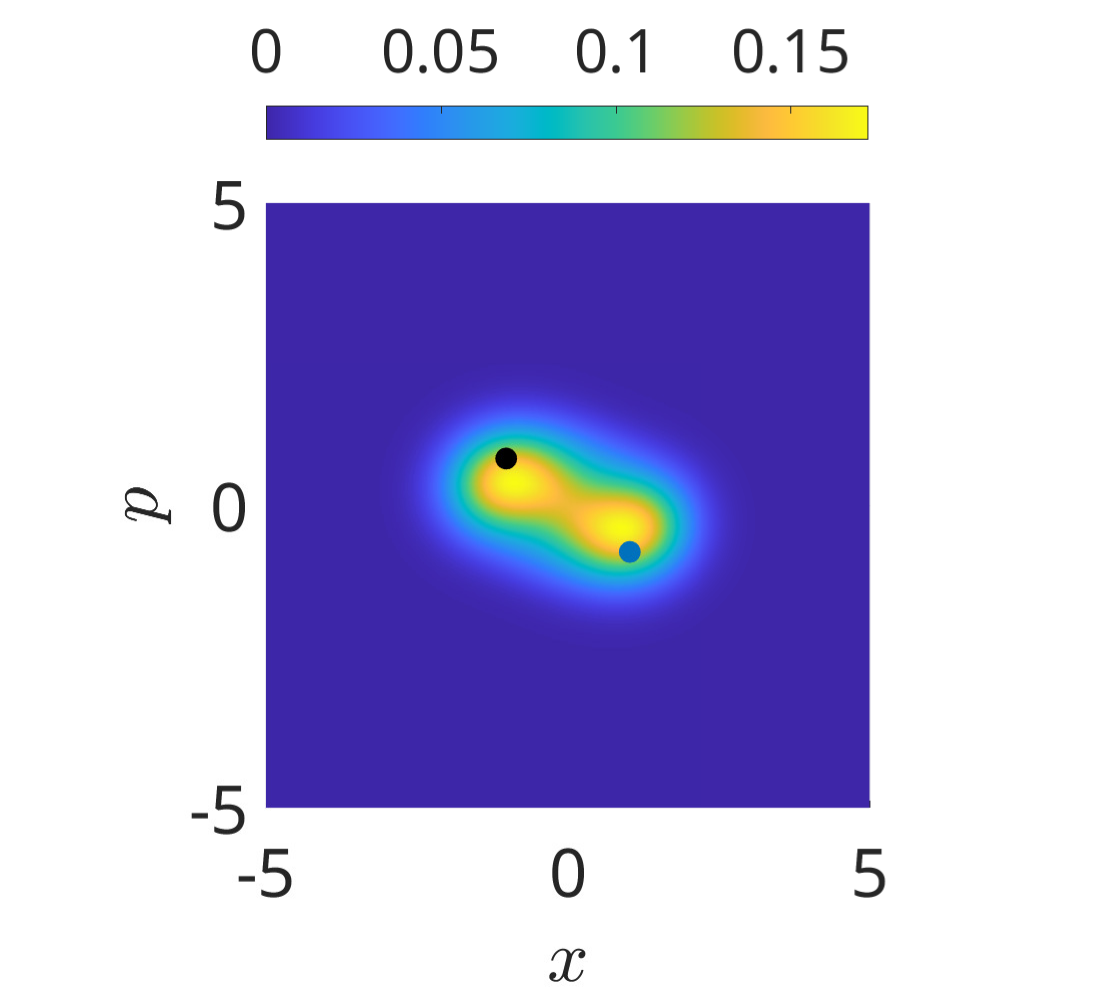}}
		\hspace*{-0.78cm}\subfigure[$W_2$]{\includegraphics[width=0.4\columnwidth]{ 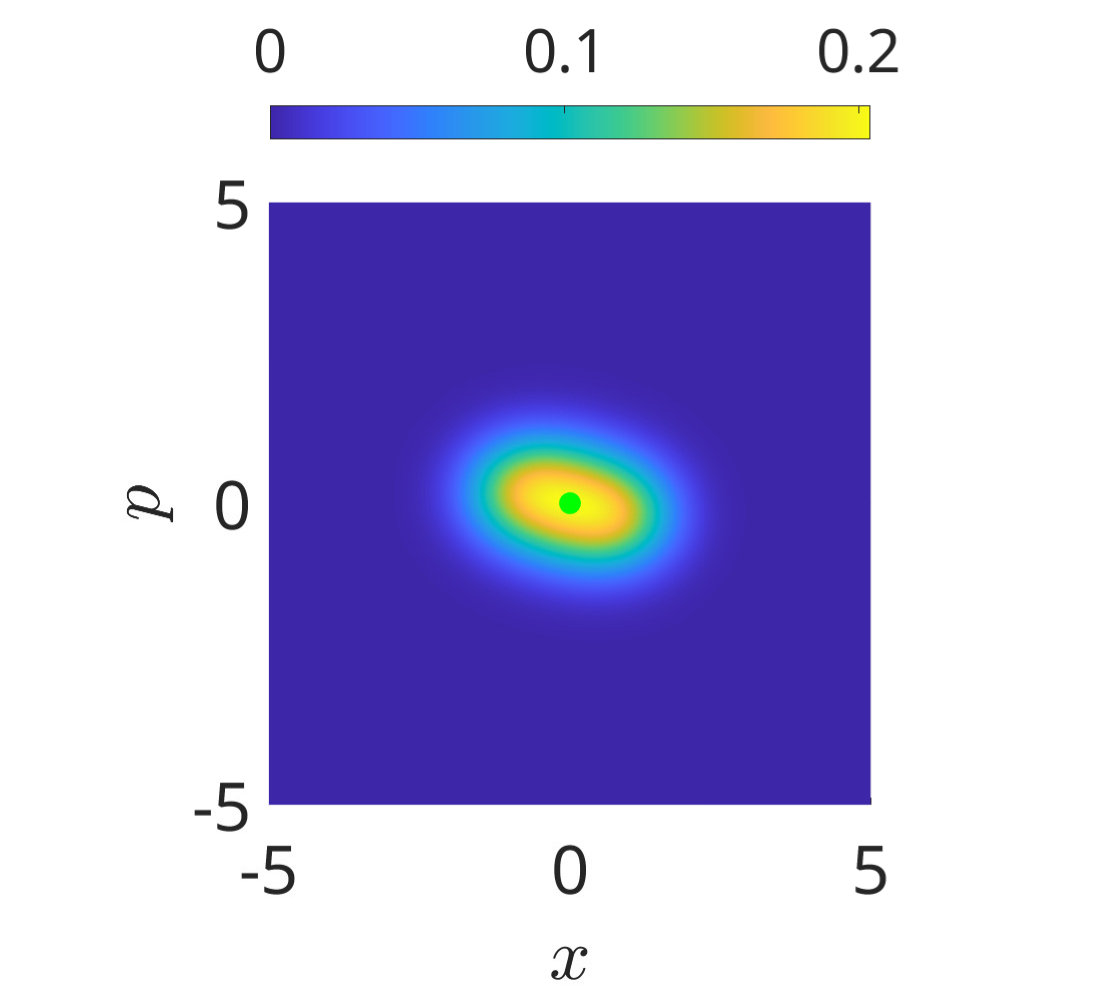}}
		\hspace*{-0.77cm}\subfigure[$W_3$]{\includegraphics[width=0.4\columnwidth]{ 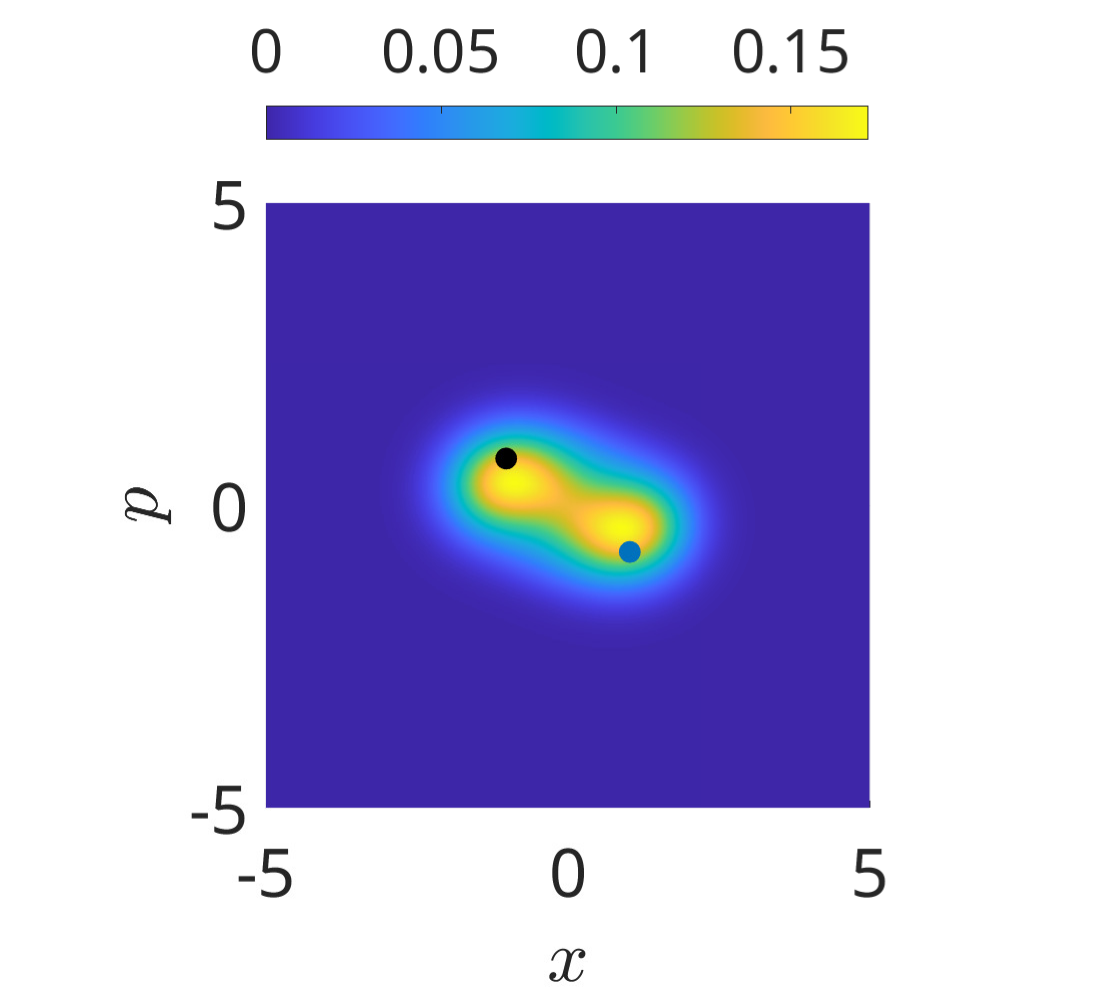}}	
	\end{center}
	\caption{
		\textbf{(a)--(c)} Reduced Wigner functions for the three-site quantum
	system in the squeezing-dominant regime for the parameters
	$\Delta=0.05$, $\gamma_1=0.03$, $\gamma_2=0.03$, $\theta=\pi$,
	$\eta=0.026$, $\lambda=0.08$, and $\kappa=0.04$. For these parameters,
	no Turing instability is activated for the nontrivial homogeneous
	solution, but a non-uniform branch connected with the null solution
	is still present, originating from the branch starting at
	$\kappa\simeq0.13$ in Fig.~\ref{fig:bifurcation_nontrivial}.
	The black, green, and blue dots mark the corresponding values
	$\alpha_1,\alpha_2,\alpha_3$ on this branch.
	}\label{fig:nontrivkappa004}
\end{figure}

\begin{figure}[t!]
	\begin{center}
		\hspace*{-0.41cm}\subfigure[$W_1$]{\includegraphics[width=0.4\columnwidth]{ 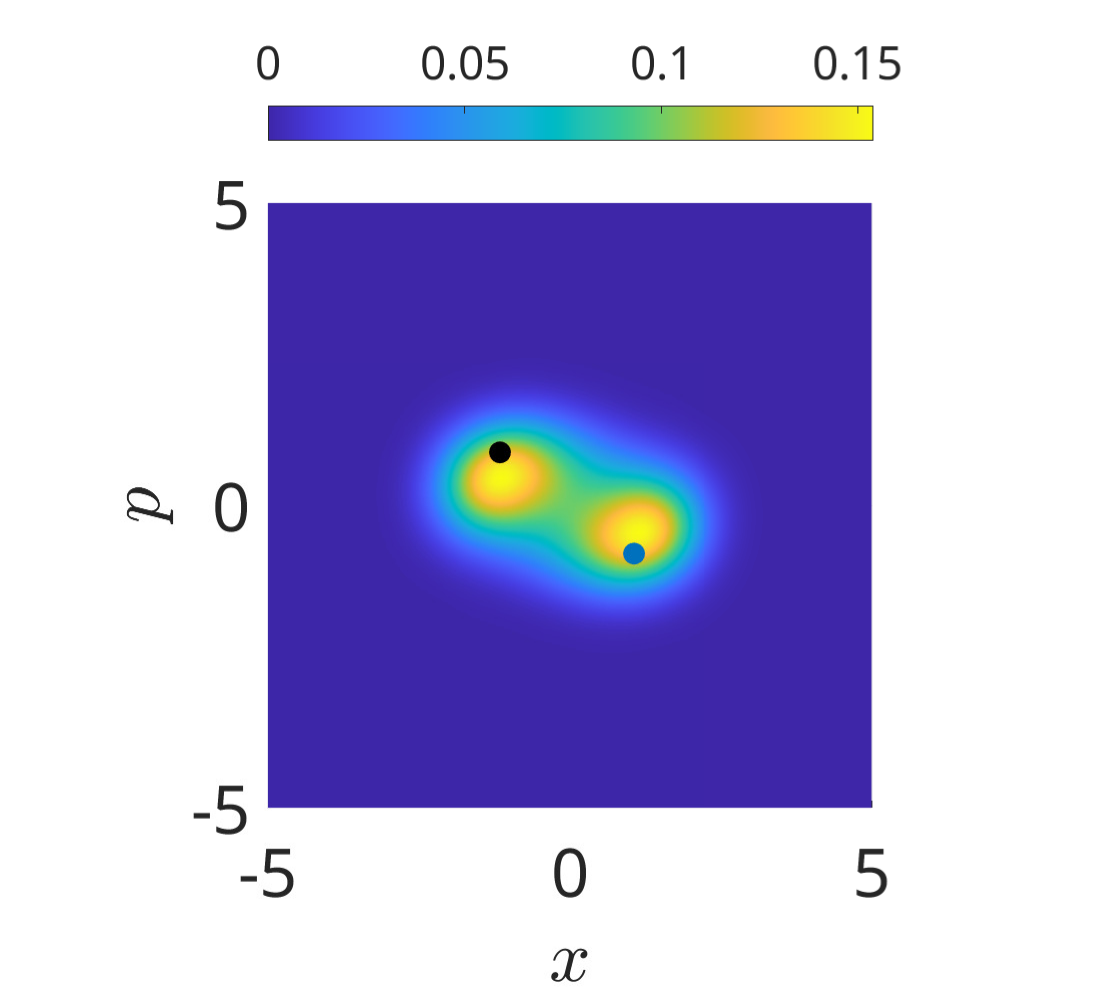}}
		\hspace*{-0.78cm}\subfigure[$W_2$]{\includegraphics[width=0.4\columnwidth]{ 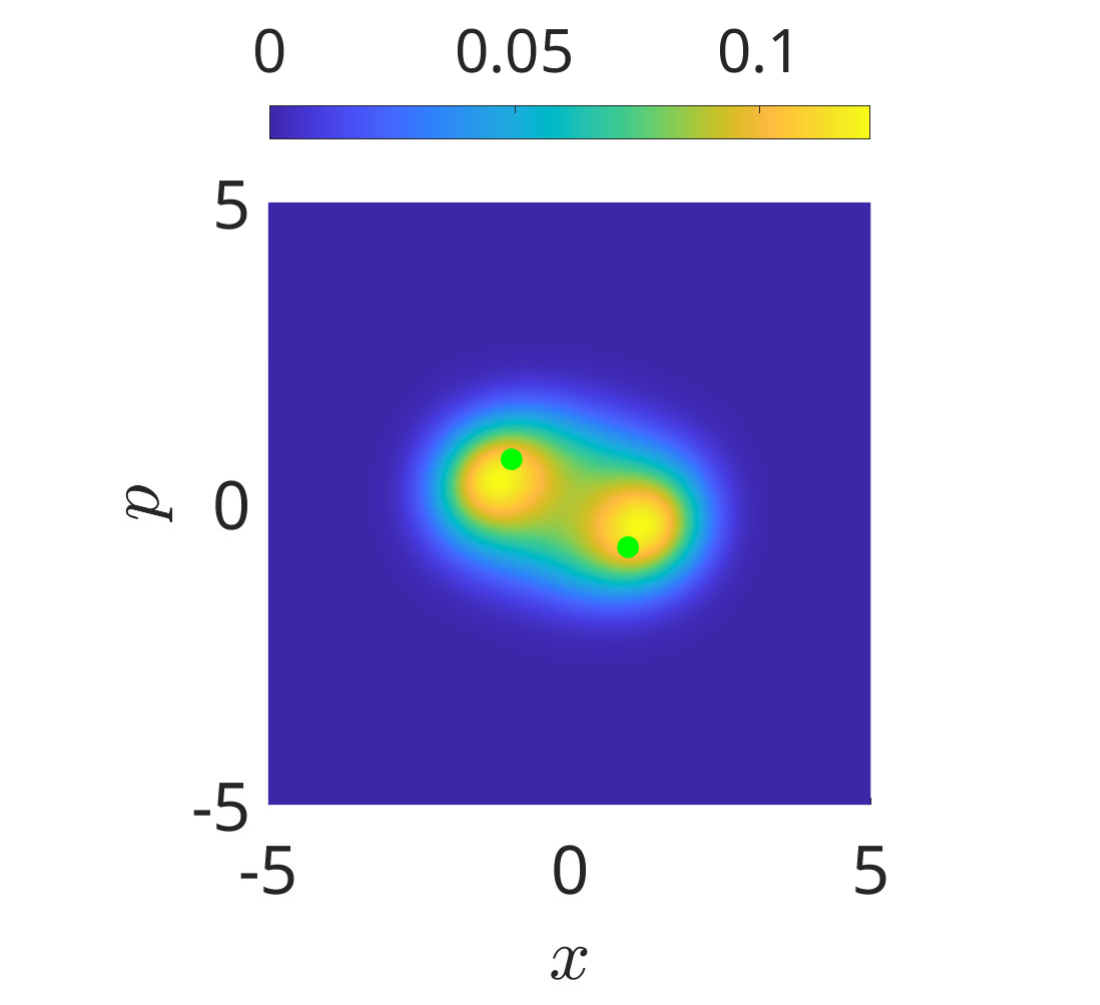}}
		\hspace*{-0.77cm}\subfigure[$W_3$]{\includegraphics[width=0.4\columnwidth]{ 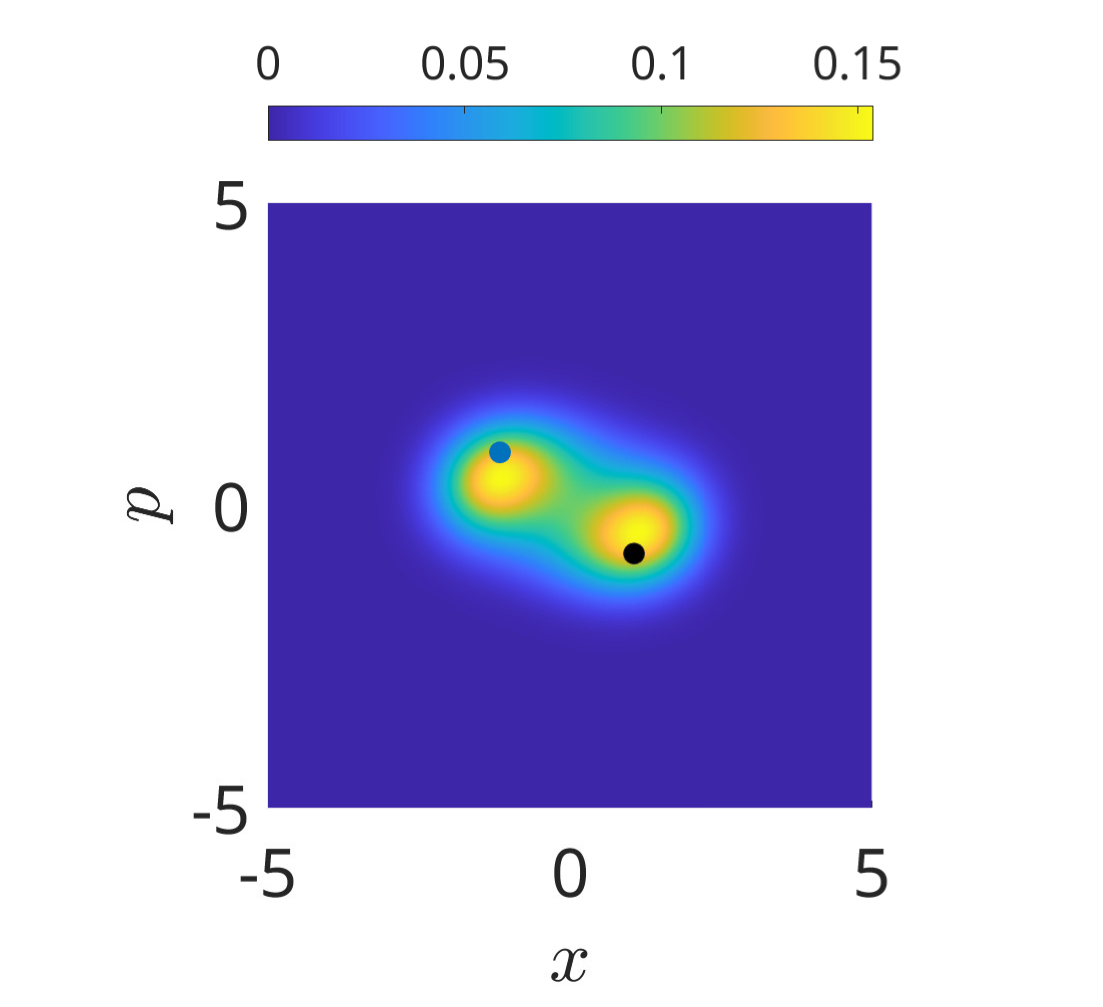}}	
	\end{center}
	\caption{
		\textbf{(a)--(c)} Reduced Wigner functions for the three-site quantum
system in the squeezing-dominant regime for the parameters
$\Delta=0.05$, $\gamma_1=0.03$, $\gamma_2=0.03$, $\theta=\pi$,
$\eta=0.026$, $\lambda=0.08$, and $\kappa=0.022$. This value of
$\kappa$ lies just below the first instability threshold of the
nontrivial homogeneous equilibrium, so that only one non-uniform spatial mode is unstable. The black, green, and blue dots mark the deterministic values $\alpha_1,\alpha_2,\alpha_3$ of the two symmetry-related non-uniform stationary solutions.
	}\label{fig:nontrivkappa0022}
\end{figure}

\begin{figure}[t!]
	\begin{center}
		\hspace*{-0.41cm}\subfigure[$W_1$]{\includegraphics[width=0.4\columnwidth]{ 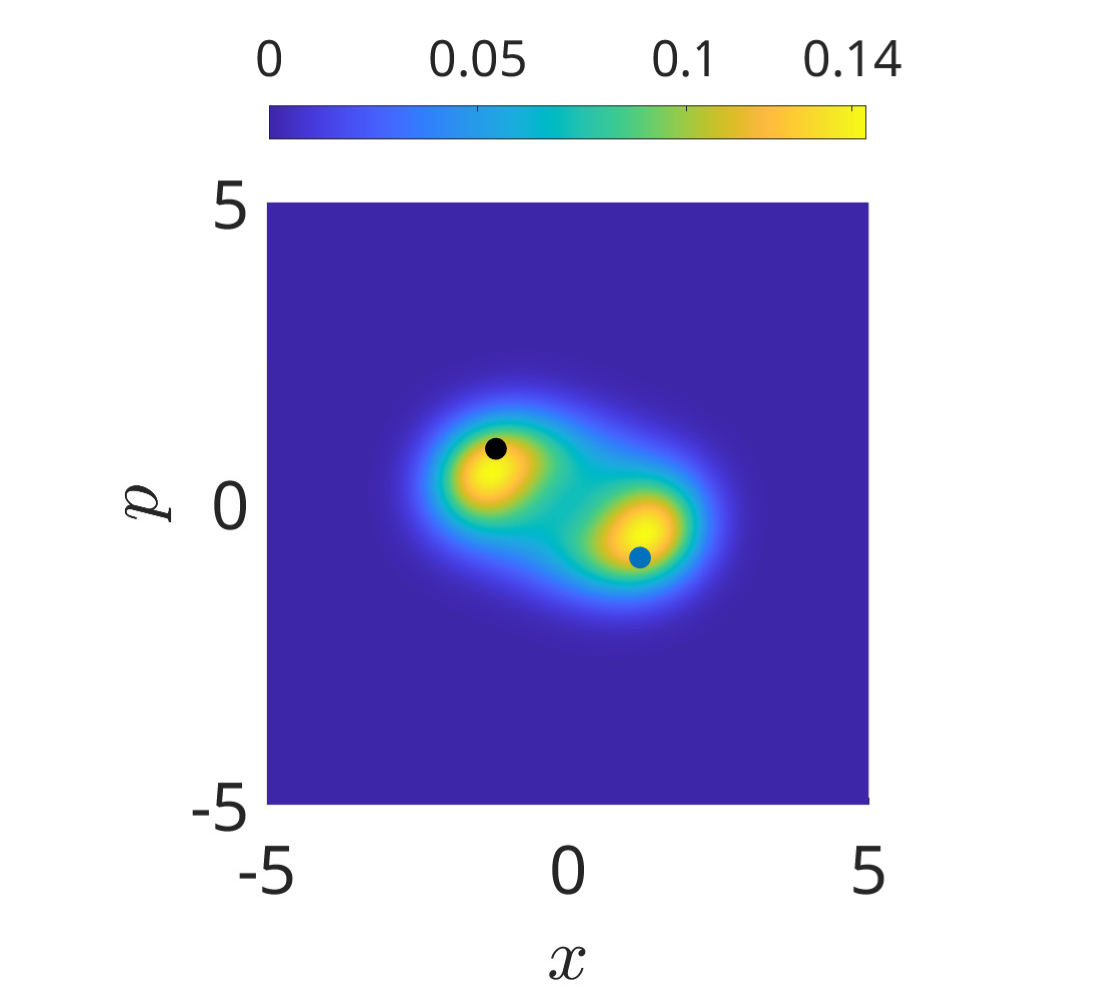}}
		\hspace*{-0.78cm}\subfigure[$W_2$]{\includegraphics[width=0.4\columnwidth]{ 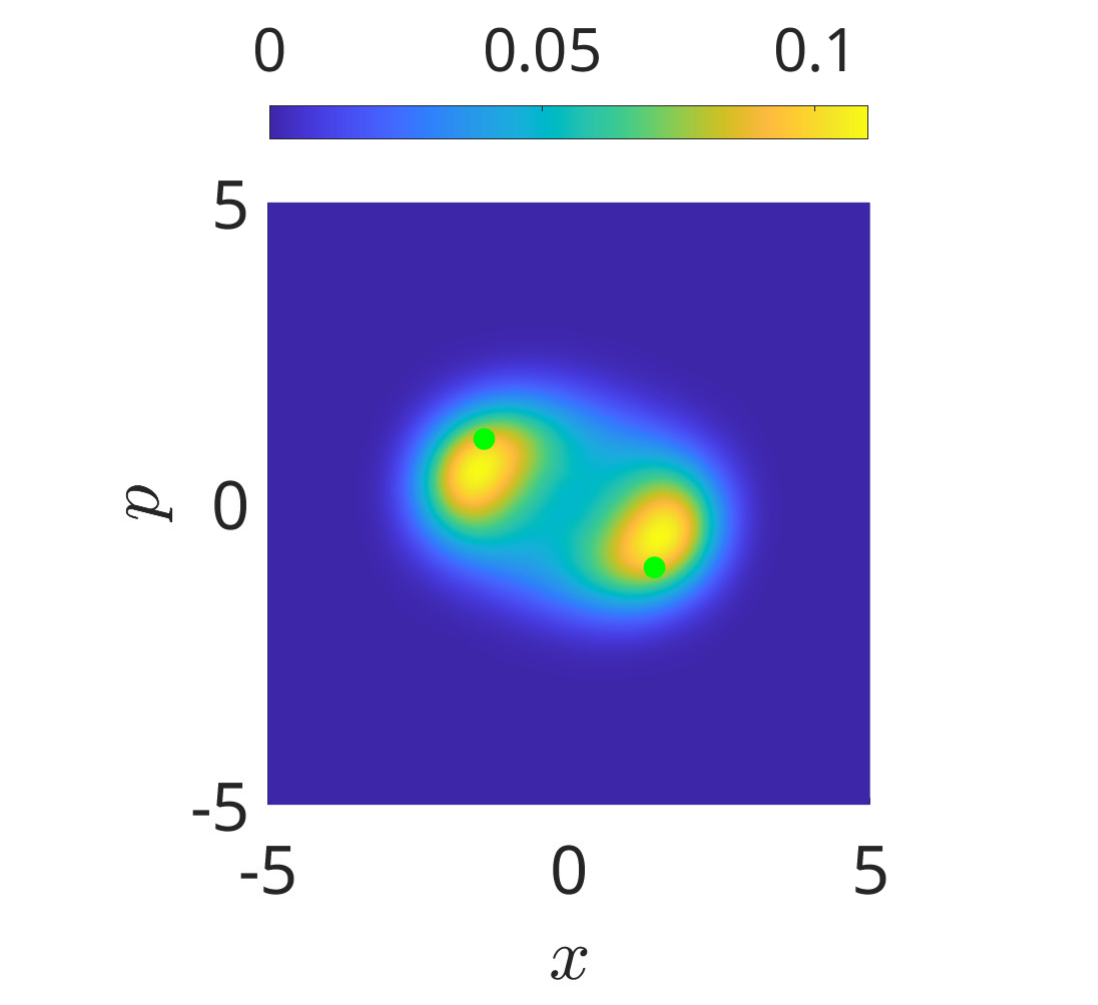}}
		\hspace*{-0.77cm}\subfigure[$W_3$]{\includegraphics[width=0.4\columnwidth]{ Tur_nontriv_zerobranch_kappa0015_mode1_3.pdf}}	
	\end{center}
	\caption{
		\textbf{(a)--(c)} Reduced Wigner functions for the three-site quantum
	system in the squeezing-dominant regime for the parameters
	$\Delta=0.05$, $\gamma_1=0.03$, $\gamma_2=0.03$, $\theta=\pi$,
	$\eta=0.026$, $\lambda=0.08$, and $\kappa=0.015$. Two non-uniform
	spatial modes are unstable. The black, green, and blue dots mark the deterministic values $\alpha_1,\alpha_2,\alpha_3$ of the two
	symmetry-related non-uniform stationary solutions.
	}\label{fig:nontrivkappa0015}
\end{figure}

The Wigner distributions are shown in Figs.~\ref{fig:nontrivkappa004}--\ref{fig:nontrivkappa0015} for three representative values: $\kappa=0.04$, where no Turing instability of the nontrivial homogeneous state is present; $\kappa=0.022$, where only one spatial mode is unstable; and $\kappa=0.015$, where both non-uniform modes are unstable. We stress that, for $\kappa=0.04$, although the nontrivial homogeneous state is stable, a non-uniform branch connected with the null solution is still present, originating from the branch starting at $\kappa\simeq0.13$ in Fig.~\ref{fig:bifurcation_nontrivial}. The overall results still display a bimodal, non-uniform structure among the sites, as in the previous case. However, the peaks are now smoother and closer to the origin of phase space, reflecting the stronger quantum noise as well as the strong nonlinear damping induced by the larger value of $\gamma_2$. This can be more clearly seen, for instance, by comparing the case $\kappa=0.04$ in Fig.~\ref{fig:nontrivkappa004}, obtained in the present moderate-to-strong quantum regime, with Fig.~\ref{fig:kappa004}, corresponding to the weak quantum regime. Despite the slightly different values of $\eta$, the main differences arise from the different nonlinear damping rate: in the weak-quantum case $\gamma_2=0.005$, the Wigner peaks are sharper and less homogenized compared to the case $\gamma_2=0.03$.

%\FloatBarrier

We also report the outcomes related to the modal quadratures defined in Eqs.~\eqref{eq:XsymPsym}--\eqref{eq:XasymPasym} to quantify the spatial organization of the quantum steady state for the parameter set of Figs.~\ref{fig:nontrivkappa004}--\ref{fig:nontrivkappa0015}, varying only $\kappa$. The resulting RMSD values, for selected $\kappa$, are reported in Table~\ref{tabstrong}.
	\begin{table}
		\begin{center}
	\begin{tabular}{c|cccc}
		\hline
		$\kappa$ 
		& $\mathcal R_{X,\mathrm{sym}}$ 
		& $\mathcal R_{P,\mathrm{sym}}$ 
		& $\mathcal R_{X,\mathrm{asym}}$ 
		& $\mathcal R_{P,\mathrm{asym}}$\\
		\hline
		$0.125$ & $0.64$ & $0.58$ & $1.15$ & $0.69$\\
		$0.04$  & $1.04$ & $0.76$ & $1.32$ & $0.81$\\
		$0.022$ & $1.51$ & $0.98$ & $1.29$ & $0.81$\\
		$0.015$ & $1.82$ & $1.17$ & $1.22$ & $0.77$\\
		\hline
\end{tabular}
	\caption{Steady-state root-mean-square values of the symmetric and antisymmetric modal quadratures. Other system parameters (squeezing-dominant regime, strong quantum regime): $\Delta=0.05$, $\gamma_1=0.03$, $\gamma_2=0.03$, $\theta=\pi,\eta=0.026,\lambda=0.08$.}
\label{tabstrong}
\end{center}
\end{table}
The trend is consistent with the deterministic bifurcation diagram, but the modal separation is much less pronounced than in the weak-quantum case reported in Table \ref{tabweak}. For large $\kappa$, the antisymmetric projection is dominant, indicating that the
quantum state already carries the signature of the first non-uniform structure. As $\kappa$ is decreased, the symmetric mode progressively grows and eventually becomes dominant, in agreement with the activation of the second non-uniform spatial mode. The main difference with respect to the weak-quantum regime is quantitative and physical. Here, the larger value of $\gamma_2$ enhances quantum noise and smooths the Wigner distributions. Consequently, the modal amplitudes are not sharply separated, and the transition between the two spatial structures is less evident: this effect was already reported in a pair of coupled quantum oscillators \cite{Paul2024}. However, we stress that the deterministic analysis is still a useful tool in organizing the quantum steady state, retaining the signatures of the underlying non-uniform organization, although they are partially washed out by the stronger quantum noise.

\subsection{Detuning-dominant regime $\eta<|\Delta|/2$}

We now consider the detuning-dominant regime $\eta<|\Delta|/2$. In this parameter region, the deterministic model admits only the trivial null equilibrium $\boldsymbol{\alpha}=(0,0,0)$, since the non-null homogeneous stationary states exist only in the squeezing-dominant regime. Moreover, the linear stability analysis of Section~\ref{sec_Turing} shows that the eigenvalues associated with the homogeneous equilibrium form complex-conjugate pairs. Therefore, when the homogeneous mode remains linearly stable, 
$
\mathrm{Re}\{\sigma^{(0)}\}=s<0,
$ 
and there exists at least one non-uniform mode $m\geq 1$ such that
$
\mathrm{Re}\{\sigma^{(m)}\}>0,
$
the instability corresponds to a discrete version of wave instability. In contrast to the squeezing-dominant case, the destabilization of the homogeneous state then leads to oscillatory spatial patterns rather than stationary ones.

\begin{figure*}[t!]
	\includegraphics[width=0.75\textwidth]{ 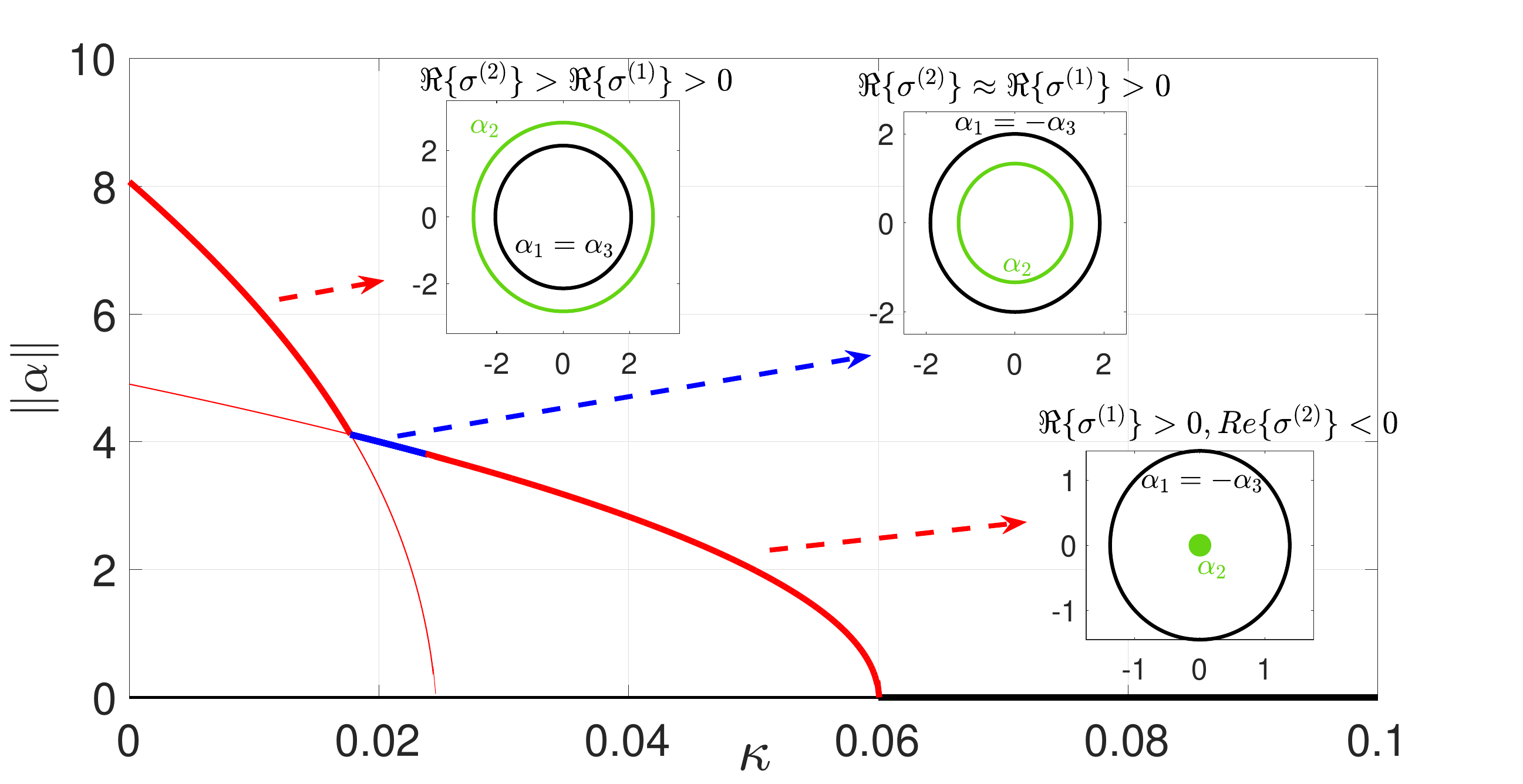}
	\caption{
		Bifurcation diagram of the deterministic model of Eq.~\eqref{alphaeq} for $N=3$ sites, obtained by numerical continuation in the parameter $\kappa$ in the detuning-dominant regime $\eta<|\Delta|/2$. System parameters: $
		\Delta = 1,
		\gamma_1 = 0.03,  
		\gamma_2 = 0.005,
		\theta = \pi,  
		\eta = 0.025, 
		\lambda = 0.08.
		$
		The vertical axis represents the norm $\|\boldsymbol{\alpha}\|$ of the solution. 
		Thick (thin) lines denote stable (unstable) branches. 
		Black curves correspond to the homogeneous equilibrium, while the blue branches represent time-periodic non-uniform solutions generated by the discrete wave instability. 
		The insets show representative spatial configurations $(\alpha_1,\alpha_2,\alpha_3)$ along the different oscillatory branches.
	}\label{fig:bifurcation_HOPF2}
\end{figure*}

To illustrate this scenario, we consider the deterministic model of Eq.~\eqref{alphaeq} again with the same parameters given in Eq.~\eqref{param1}, except that the detuning is now chosen as
$\Delta = 1.$
The corresponding bifurcation diagram is shown in Fig.~\ref{fig:bifurcation_HOPF2}. 
For this choice of parameters, the thresholds for the activation of the non-uniform modes remain the same as in the squeezing-dominant regime for the null homogeneous solution,
\bea
\kappa_1^{\mathrm{th}} \approx 0.06,
\qquad
\kappa_2^{\mathrm{th}} \approx 0.0244,
\ena
because for the selected parameters, the real parts of the growth rates coincide.
For $\kappa>\kappa_1^{\mathrm{th}}$, the null homogeneous equilibrium is linearly stable. 
When $\kappa$ crosses the first threshold, $\kappa_1^{\mathrm{th}}$, the first spatial mode becomes unstable, and the system develops a time-periodic state mainly associated with the antisymmetric configuration $\alpha_1=-\alpha_3$, while $\alpha_2$ remains null. 
As $\kappa$ is further decreased and approaches $\kappa_2^{\mathrm{th}}$, the first and second non-uniform modes become simultaneously relevant (the blue branch in the bifurcation diagram), with
\bea
\mathrm{Re}\{\sigma^{(1)}\}\approx \mathrm{Re}\{\sigma^{(2)}\}>0,
\ena
and the oscillatory solution acquires a mixed spatial structure where  $\alpha_2$ is also nonzero.
For smaller values of $\kappa$, the second mode becomes dominant, and the oscillatory pattern increasingly reflects the geometry of the second Laplacian eigenmode, with $\alpha_1$ and $\alpha_3$ oscillating synchronously while $\alpha_2$ exhibits larger excursions in phase space.

\begin{figure}[t!]
	\begin{center}
		\hspace*{-0.41cm}\subfigure[$W_1$]{\includegraphics[width=0.4\columnwidth]{ 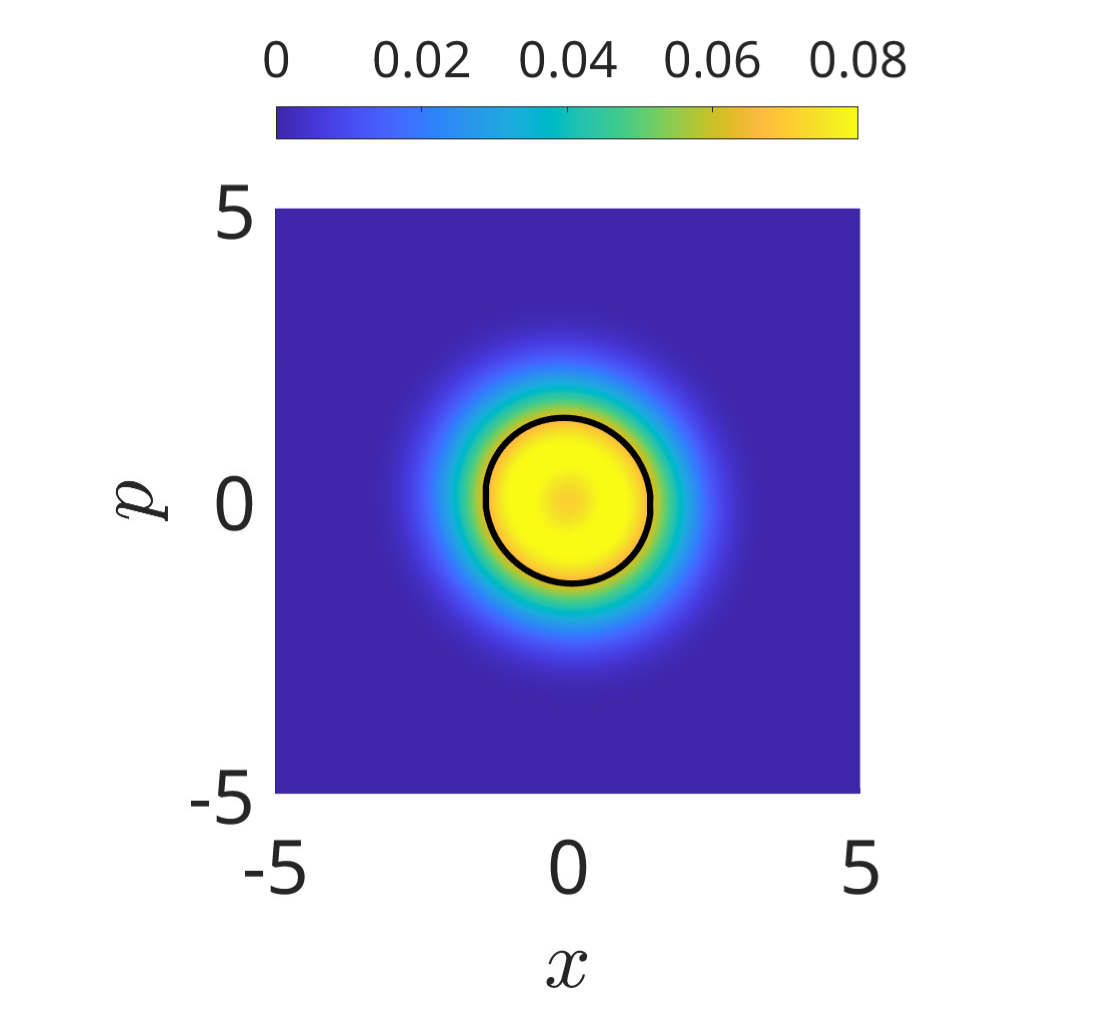}}
		\hspace*{-0.78cm}\subfigure[$W_2$]{\includegraphics[width=0.4\columnwidth]{ 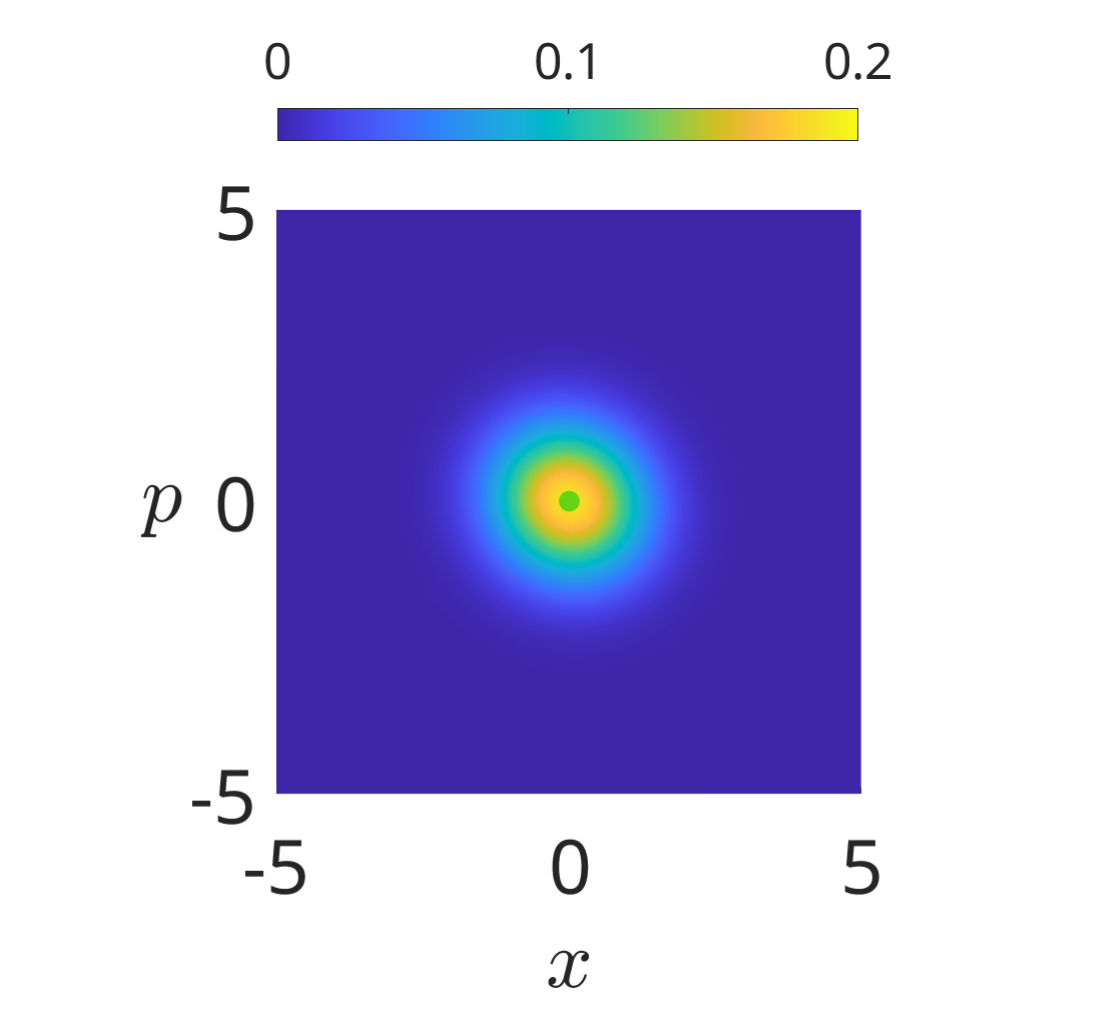}}
		\hspace*{-0.77cm}\subfigure[$W_3$]{\includegraphics[width=0.4\columnwidth]{ 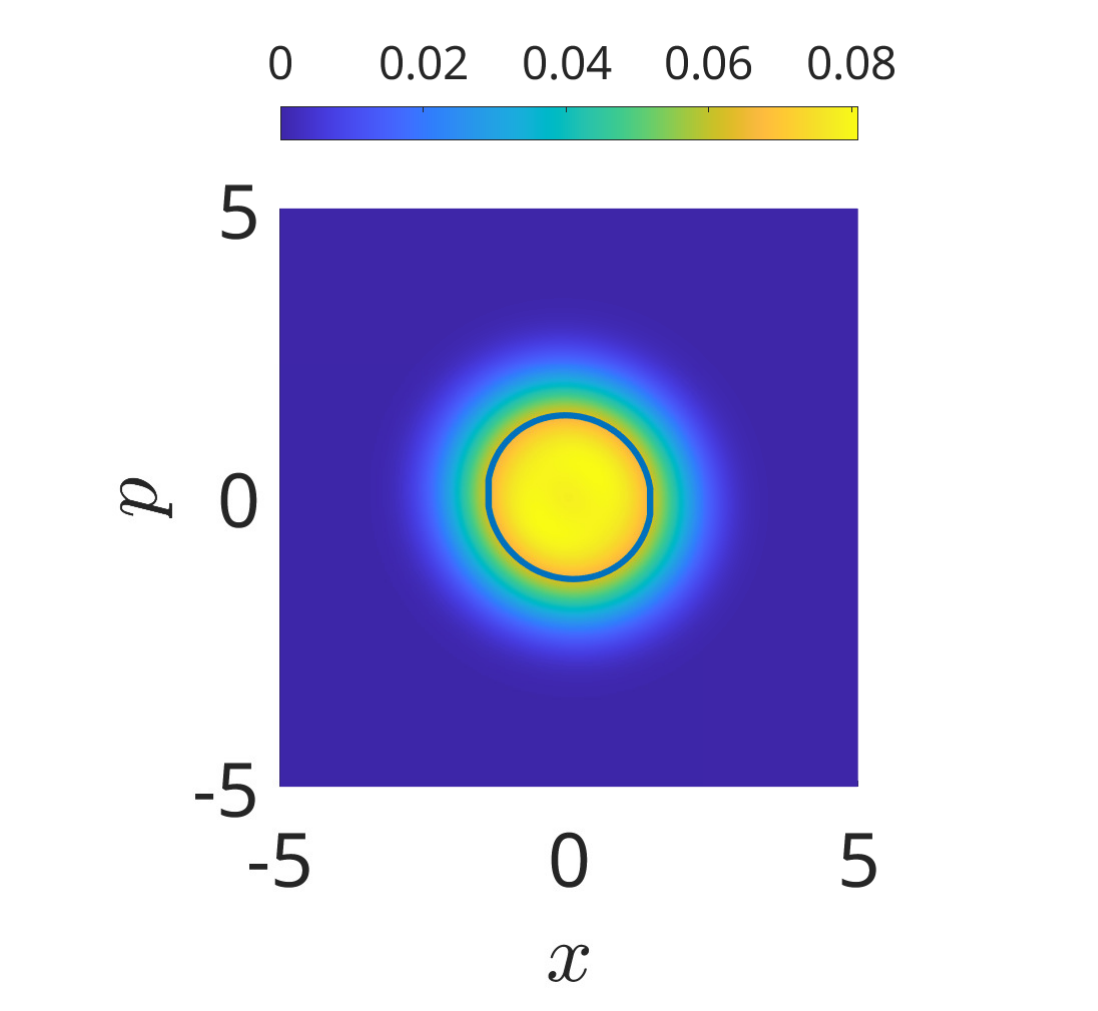}\label{fig:hopfkappa004c}}	
		\hspace*{-0.41cm}\subfigure[$\mathcal{P}(x, p)$, site 1]{\includegraphics[width=0.4\columnwidth]{ 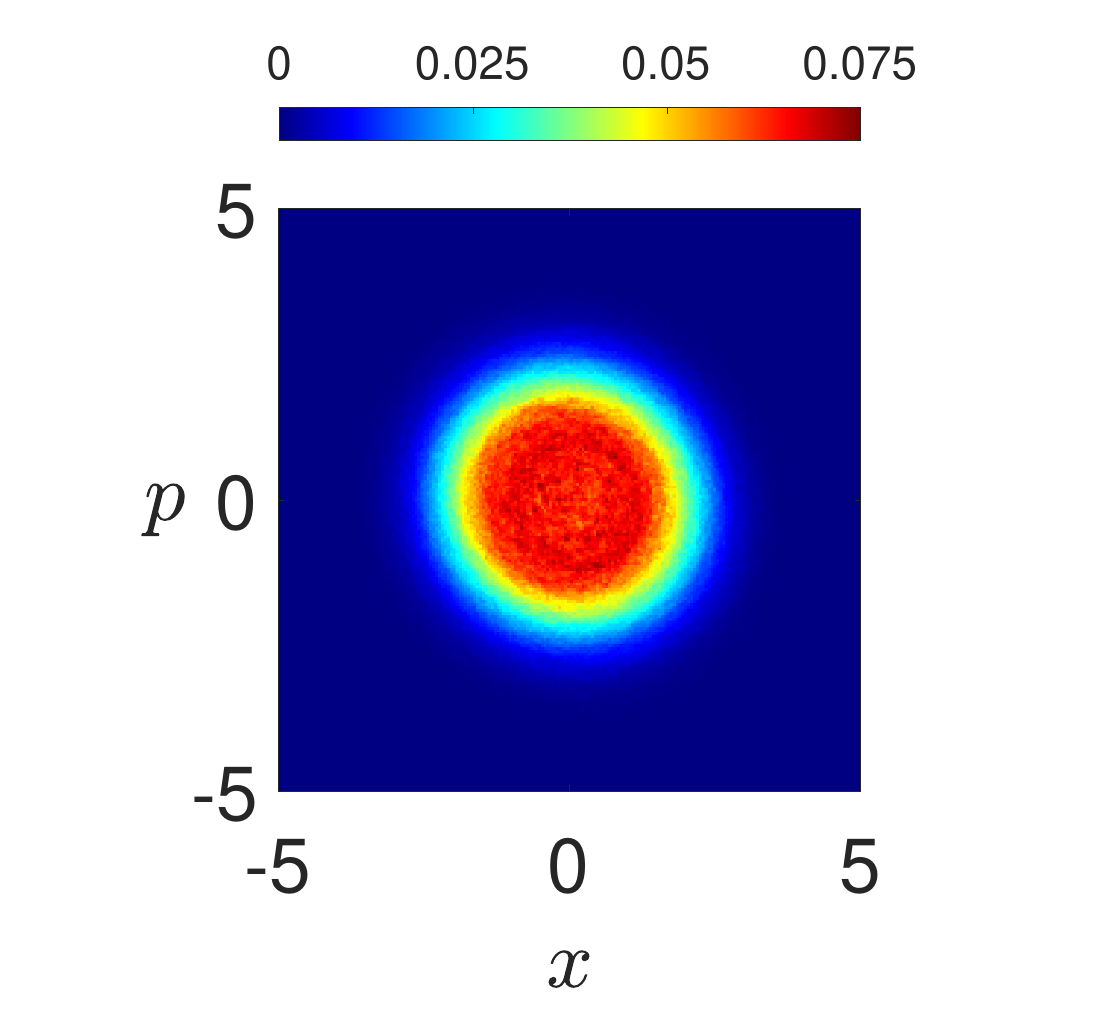}\label{fig:hopfkappa004d}}
		\hspace*{-0.78cm}\subfigure[$\mathcal{P}(x, p)$, site 2]{\includegraphics[width=0.4\columnwidth]{ 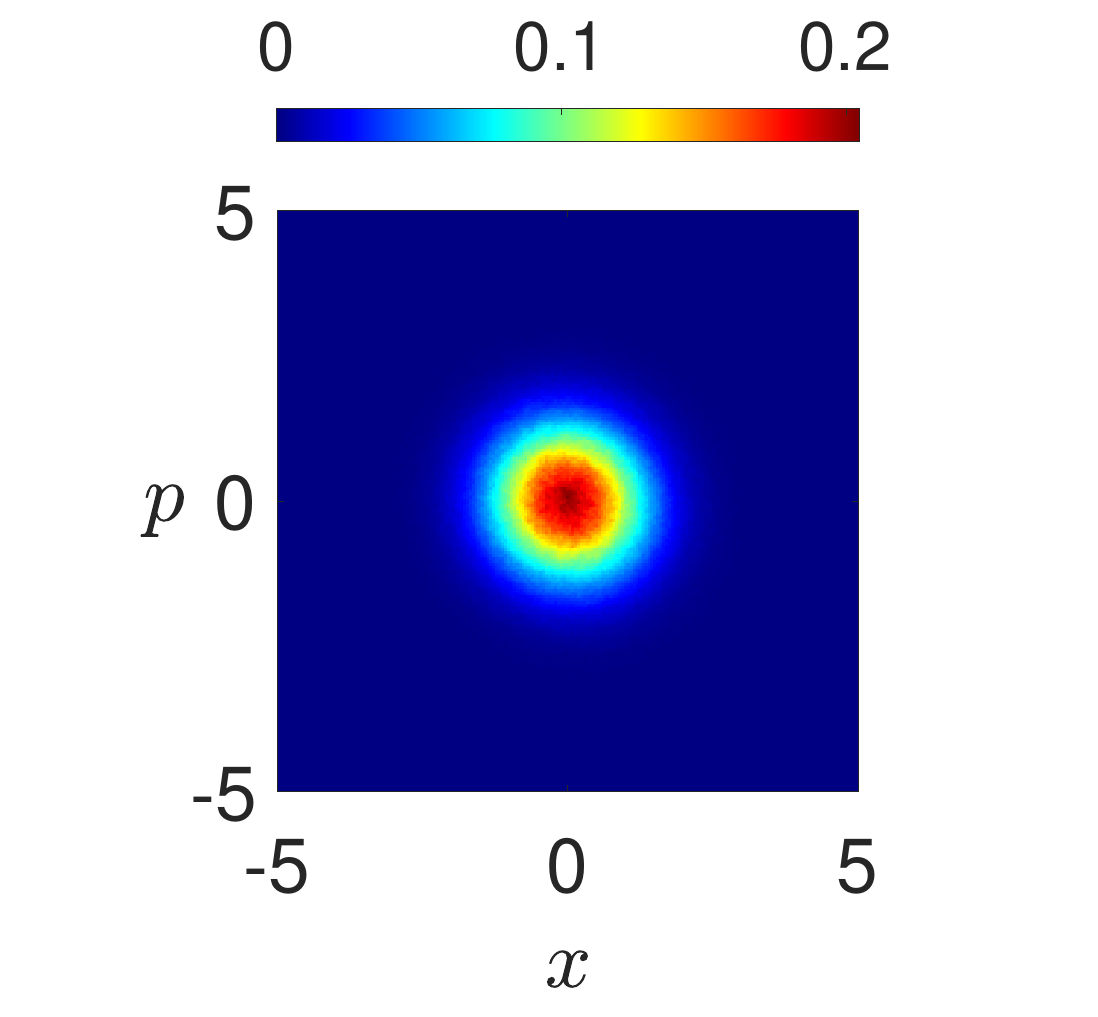}\label{fig:hopfkappa004e}}
		\hspace*{-0.77cm}\subfigure[$\mathcal{P}(x, p)$, site 3]{\includegraphics[width=0.4\columnwidth]{ 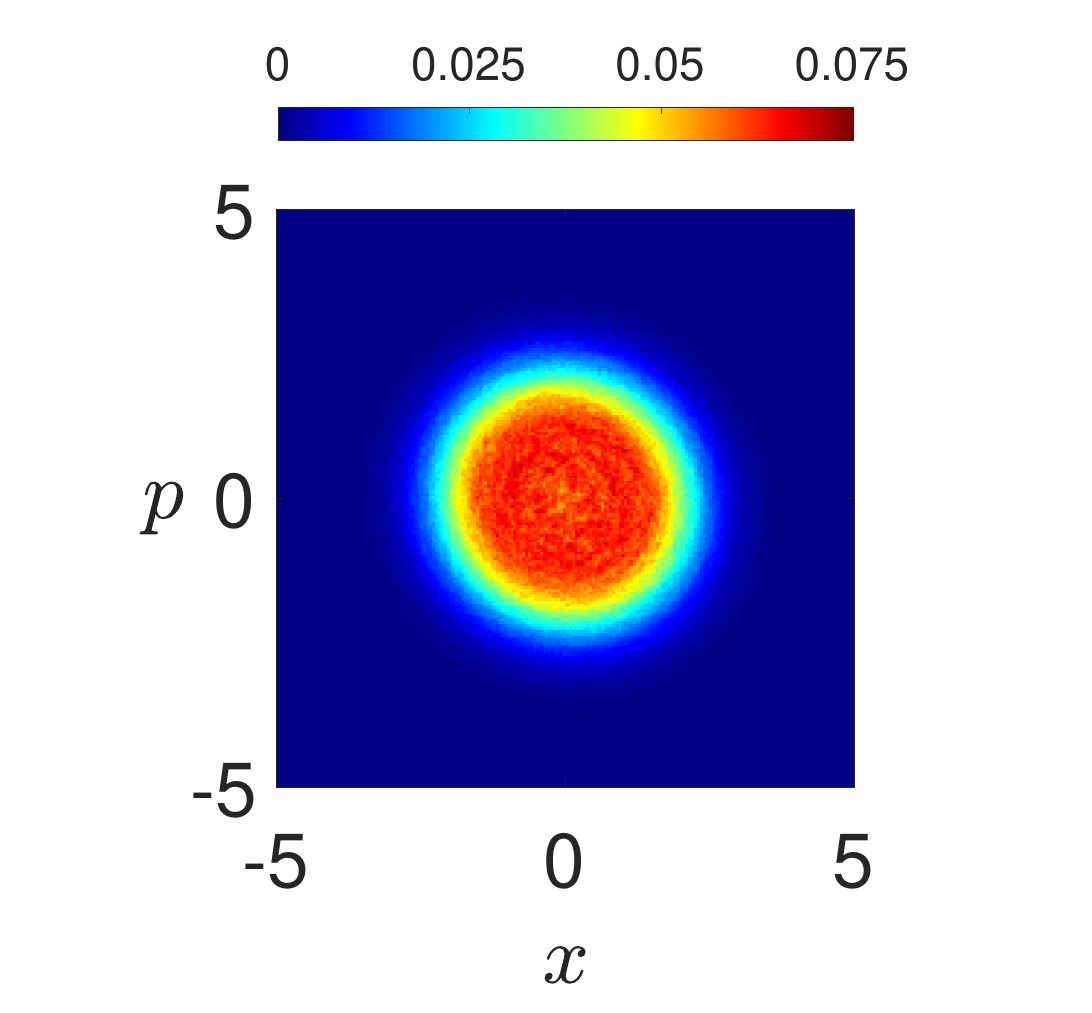}\label{fig:hopfkappa004f}}
	\end{center}
	\caption{
		\textbf{(a)-(c)} Reduced Wigner functions for the three-site quantum system in the detuning-dominant regime for the parameters $\Delta=1$, $\gamma_1=0.03$, $\gamma_2=0.005$, $\theta=\pi$, $\eta=0.025$, $\lambda=0.08$, and $\kappa=0.04$, corresponding to the first oscillatory branch of the deterministic bifurcation diagram in Fig.~\ref{fig:bifurcation_HOPF2}. Black and blue limit cycles of the deterministic solution for the first and third sites are shown in \textbf{(a),(c)}, as well as the zero green point in \textbf{(b)} for the second site. 
		The distributions $W_1$ and $W_3$ coincide and exhibit a deformed ring-like structure, reflecting the underlying time-periodic motion in phase space, while $W_2$ remains unimodal and localized near the origin. 
		\textbf{(d)-(f)} Empirical densities $\mathcal{P}(x,p)$ reconstructed from the SDE trajectories for the three sites. The qualitative agreement with the corresponding Wigner functions confirms that the semiclassical stochastic dynamics captures the structure of the quantum steady state in this oscillatory regime.
	}\label{fig:hopfkappa004}
\end{figure}

\paragraph{Quantum counterpart.} We now consider the quantum counterpart of the full three-site system. 
As in the previous subsection, the local quantum state at each site is characterized through the reduced Wigner functions.
We first select $\kappa=0.04$, corresponding to the first oscillatory branch in Fig.~\ref{fig:bifurcation_HOPF2}. 
The reduced Wigner functions are shown in Fig.~\ref{fig:hopfkappa004}, panels (a)-(c), together with the densities $\mathcal{P}(x,p)$ reconstructed from the SDE trajectories in panels (d)-(f). 
In contrast to the squeezing-dominant regime, here the reduced Wigner functions of sites 1 and 3 display a deformed ring-like structure, reflecting the underlying time-periodic motion in the phase space of $\alpha_1$ and $\alpha_3$, while $W_2$ remains unimodal and concentrated around the origin, compatible with the deterministic prediction $\alpha_2=0$.
The SDE densities $\mathcal{P}(x,p)$ reproduce the same qualitative features, showing that the semiclassical stochastic dynamics captures the annular phase-space structure of the quantum steady state also in the detuning-dominant regime. In this case, we observe that the noise-induced diffusion is strong enough to significantly smear out the deterministic limit-cycle trajectory. As a consequence, the ring-like structure appears only weakly defined, with a substantial spreading of the probability density towards the inner part of the ring. In this setting, we also notice that the SDE approximation degrades, at least compared with the previous squeezing-dominant setting. In fact, the resulting mean values from the QME dynamics are $\langle\hat a_{1,3}^\dagger \hat a_{1,3} \rangle\approx2.15$ nearly identical for the first and third sites), $\langle\hat a_{2}^\dagger \hat a_{2} \rangle\approx1.03$, while those from the SDE dynamics are $\overline{|\alpha_{1,3}|^2}\approx3$ (again the same value for the first and third sites), $\overline{|\alpha_{2}|^2}\approx1.59$, showing differences slightly larger than the empirical 1/2 shift.

\begin{figure}[t!]
	\begin{center}
		\hspace*{-0.41cm}\subfigure[$W_1$]{\includegraphics[width=0.4\columnwidth]{ 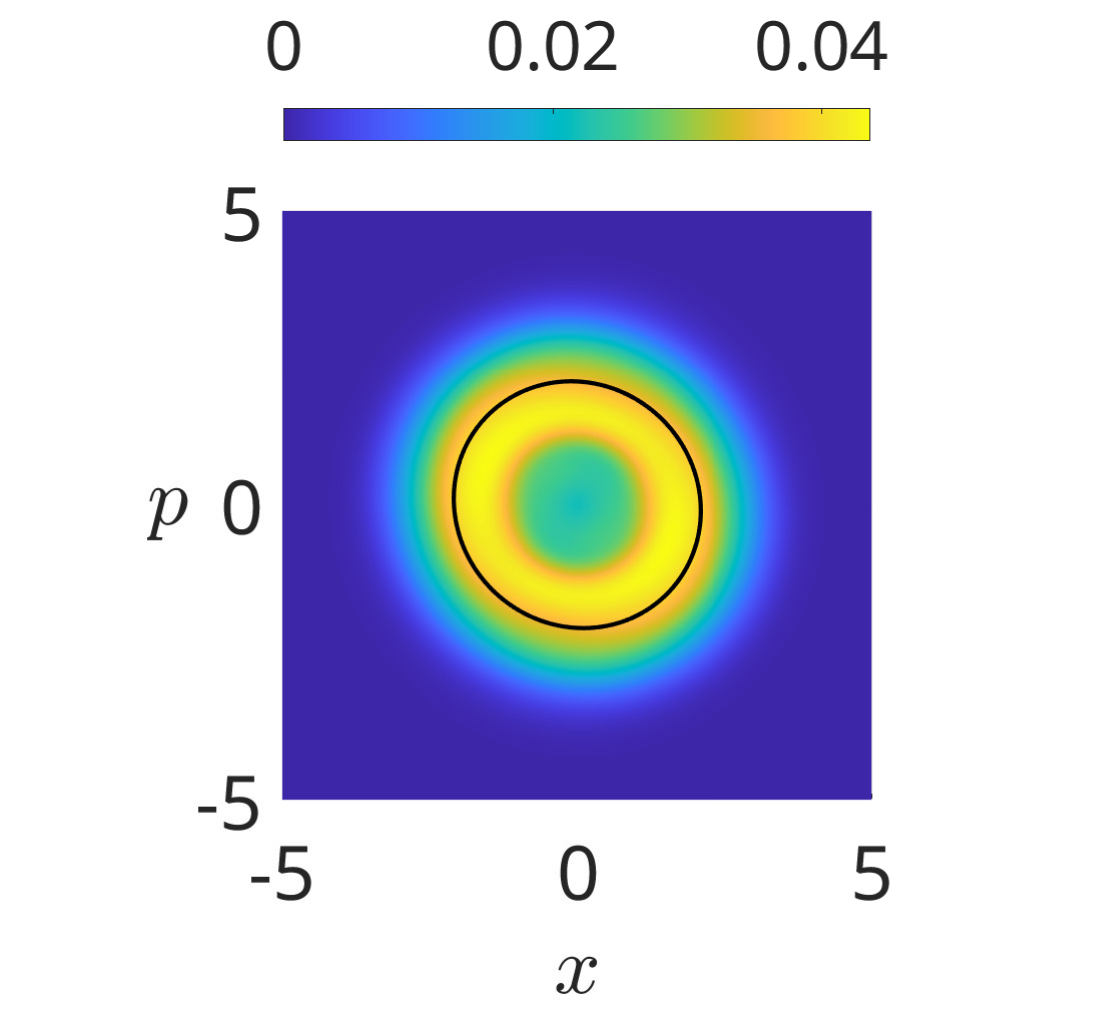}}
		\hspace*{-0.78cm}\subfigure[$W_2$]{\includegraphics[width=0.4\columnwidth]{ 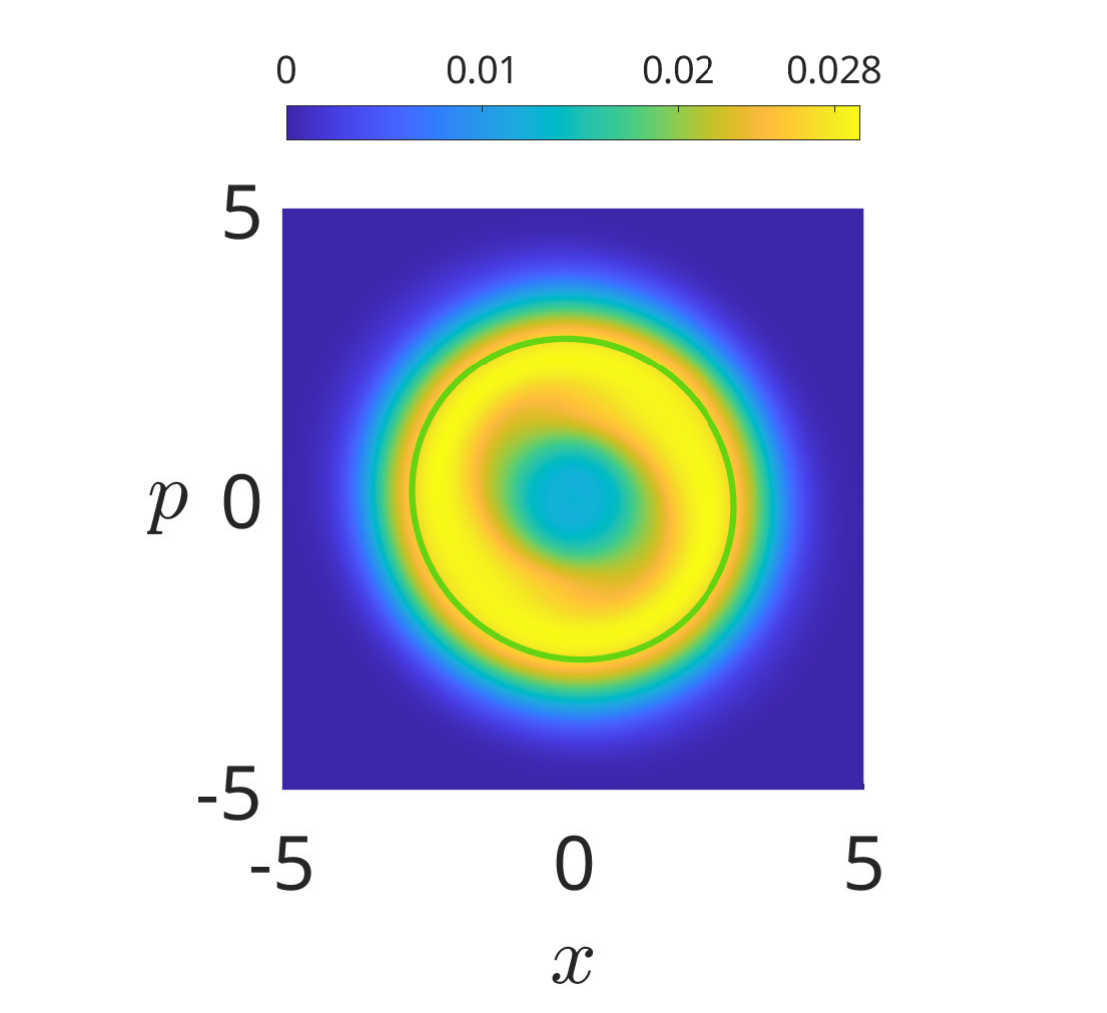}}
		\hspace*{-0.77cm}\subfigure[$W_3$]{\includegraphics[width=0.4\columnwidth]{ 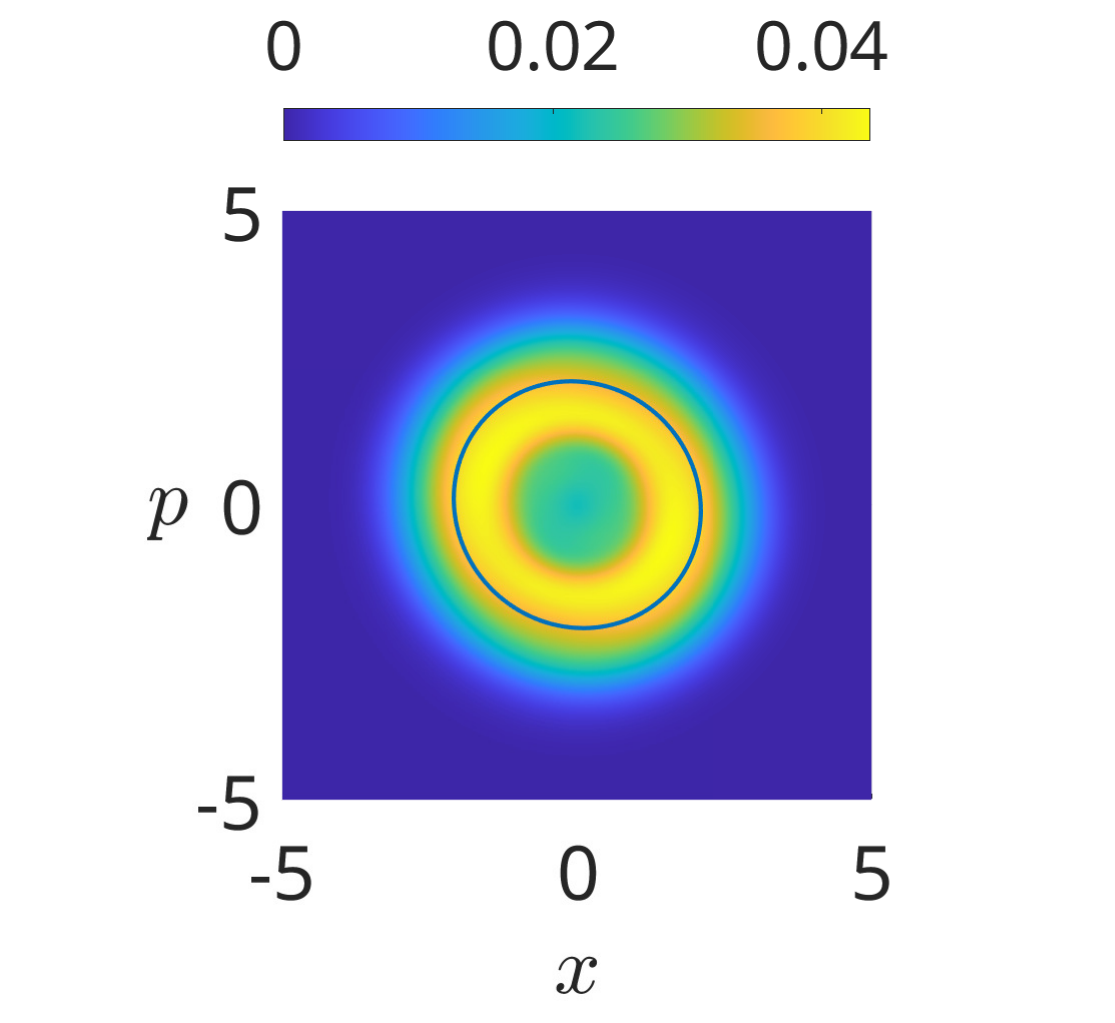}\label{fig:hopfkappa00175c}}	
		\hspace*{-0.41cm}\subfigure[$\mathcal{P}(x, p)$, site 1]{\includegraphics[width=0.4\columnwidth]{ 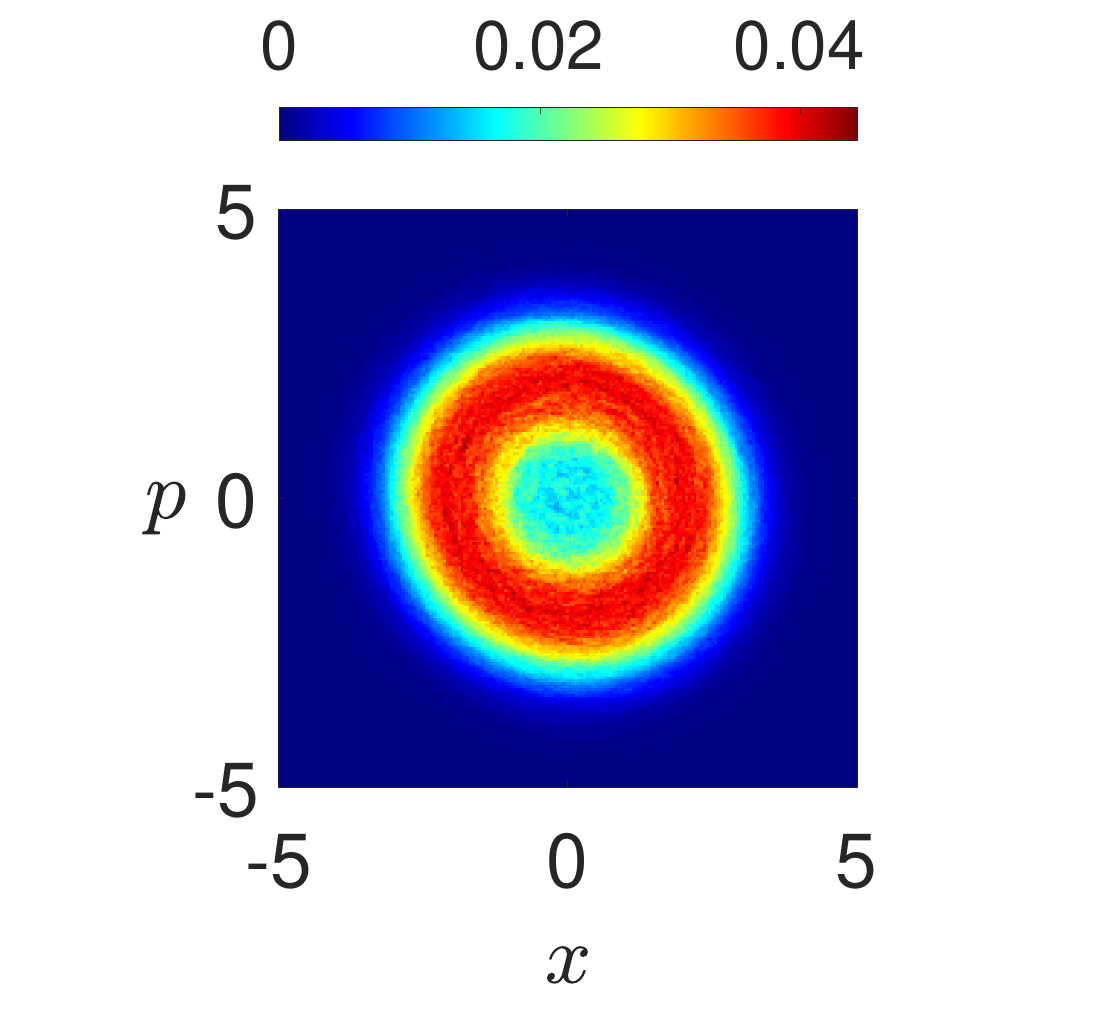}\label{fig:hopfkappa00175d}}
		\hspace*{-0.78cm}\subfigure[$\mathcal{P}(x, p)$, site 2]{\includegraphics[width=0.4\columnwidth]{ 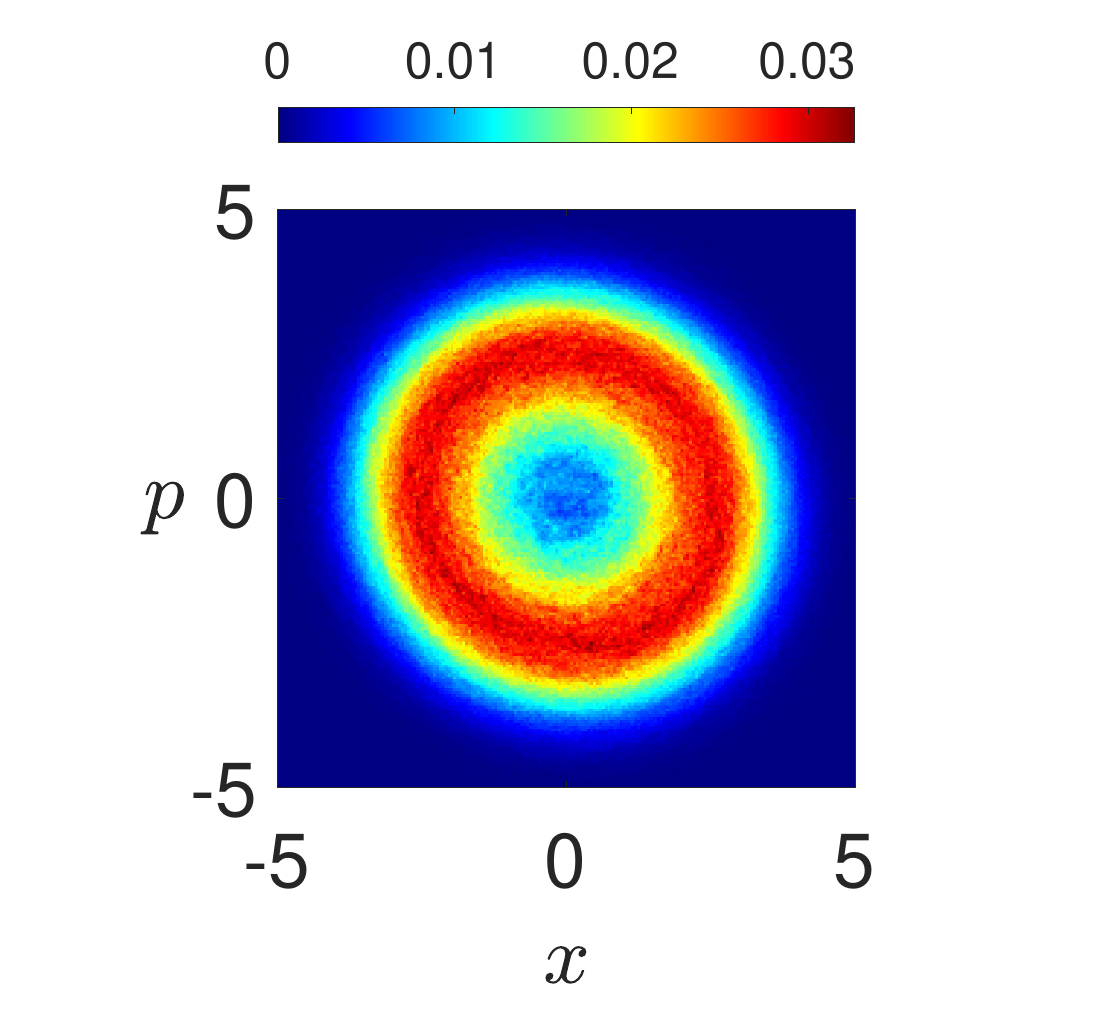}\label{fig:hopfkappa00175e}}
		\hspace*{-0.77cm}\subfigure[$\mathcal{P}(x, p)$, site 3]{\includegraphics[width=0.4\columnwidth]{ 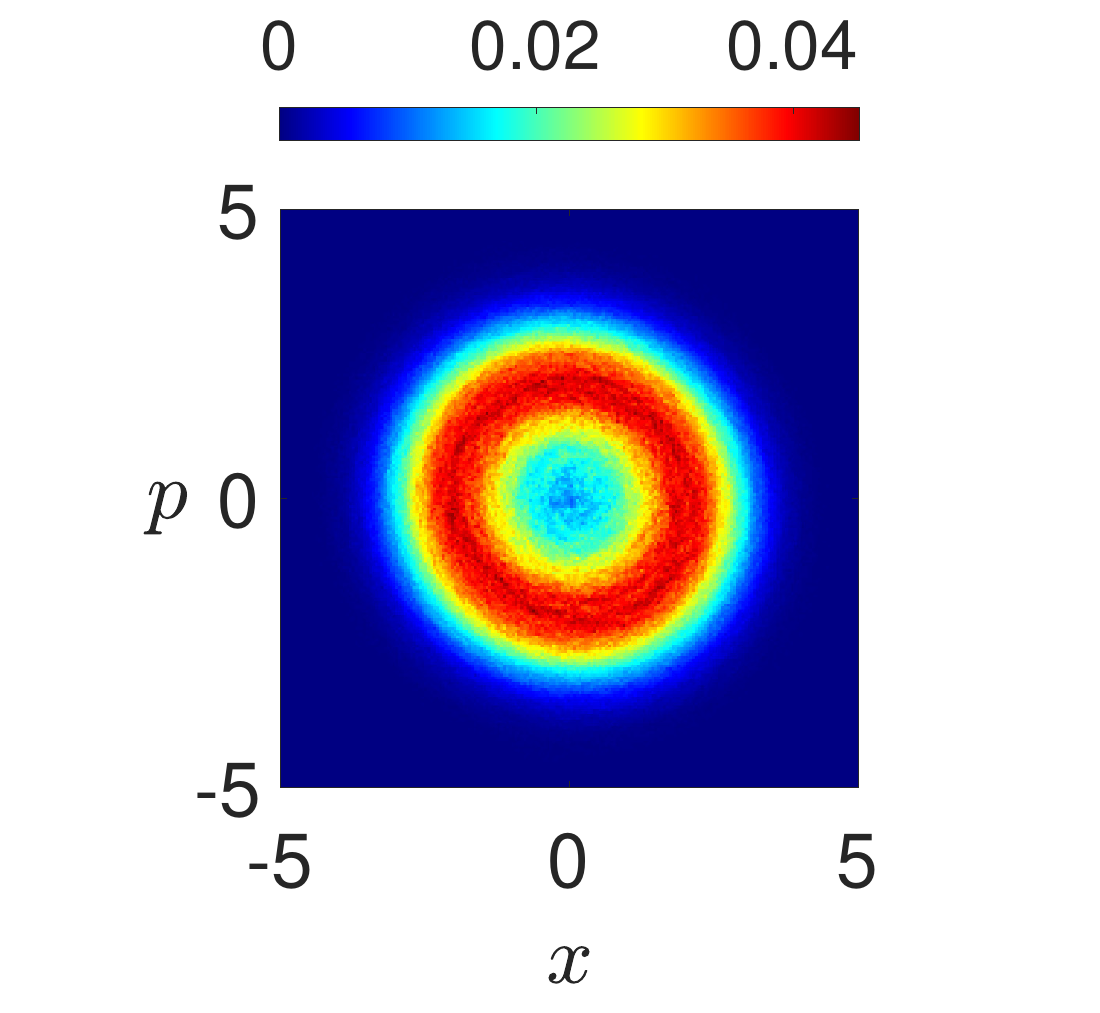}\label{fig:hopfkappa00175f}}
	\end{center}
	\caption{
		\textbf{(a)-(c)} Reduced Wigner functions for the three-site quantum system in the detuning-dominant regime for the parameters $\Delta=1$, $\gamma_1=0.03$, $\gamma_2=0.005$, $\theta=\pi$, $\eta=0.025$, $\lambda=0.08$, and $\kappa=0.0175$, corresponding to the second oscillatory branch of the deterministic bifurcation diagram in Fig.~\ref{fig:bifurcation_HOPF2}. Black, green, and blue limit cycles of the deterministic solution are shown in the corresponding sites. 
		All the distributions $W_1, W_2$ and $W_3$, with $W_1$ and $W_3$ almost coinciding, exhibit a clear ring-like structure, reflecting the underlying time-periodic motion in phase space
		\textbf{(d)-(f)}. Empirical densities $\mathcal{P}(x,p)$ reconstructed from the SDE trajectories for the three sites. The qualitative agreement with the corresponding Wigner functions confirms that the semiclassical stochastic dynamics captures the structure of the quantum steady state in this oscillatory regime.
	}\label{fig:hopfkappa00175}
\end{figure}

Selecting  $\kappa=0.0175$, corresponding to the second oscillatory branch in Fig.~\ref{fig:bifurcation_HOPF2} originating at $\kappa_2^{\text{th}}$,  
the reduced Wigner functions are shown in Fig.~\ref{fig:hopfkappa00175}, panels (a)-(c), together with the  densities $\mathcal{P}(x,p)$ reconstructed from the SDE trajectories in panels (d)-(f). 
In all cases, all three reduced Wigner functions exhibit a ring-like structure, reflecting the underlying time-periodic motion in the phase space of $\alpha_1,\alpha_2,\alpha_3$ as shown in the leftmost limit cycles in the inset of Fig.~\ref{fig:bifurcation_HOPF2}. Compared to the first branch case $\kappa=0.04$ in Fig.~\ref{fig:hopfkappa004}, here the ring structures are more evident, consistent with the weaker noise-induced diffusion at the smaller value of $\kappa$. Although the phase-space prediction is better resolved at smaller $\kappa$, the quantitative agreement between the QME and SDE estimates of mean phonon numbers does not improve accordingly. In fact, the resulting mean values from the QME dynamics slightly differ from the corresponding values computed from the SDE (up to the factor 1/2): $\langle\hat a_{1,3}^\dagger \hat a_{1,3} \rangle\approx4.3$, $\langle\hat a_{2}^\dagger \hat a_{2} \rangle\approx6.6$, while those from the SDE dynamics are $\overline{|\alpha_{1,3}|^2}\approx5.5$, $\overline{|\alpha_{2}|^2}\approx7.8$.

\section{Conclusions}\label{sec:conclusions}

In this work, we have introduced and analyzed a multimode open quantum system in which Turing-type mechanisms arise from the microscopic GKSL dynamics of a finite chain of bosonic sites. The model combines local coherent dynamics with single-photon losses, nonlinear two-photon damping, nearest-neighbor incoherent pumping, and a longer-range dissipative channel that acts as an effective hyper-diffusive mechanism. In this way, the system implements, at the level of engineered quantum dissipation, the basic ingredients of the classical short-range enhancement and long-range suppression paradigm. At the same time, it remains directly connected to phase-space descriptions and to a deterministic reaction-diffusion-like limit. Starting from the GKSL master equation, we derived the mean-field equations for the complex field amplitudes and showed that they allow for a finite-dimensional analogue of Turing instability analysis, where homogeneous configurations lose stability against selected non-uniform spatial modes.

In the squeezing-dominant regime, we identified two deterministic routes to stationary spatial organization. In the first one, the null homogeneous state becomes unstable and gives rise to non-uniform stationary branches. In the second one, when a non-null homogeneous equilibrium exists, this state can also lose stability through a Turing-type mechanism. These scenarios are naturally connected with the classical transition from homogeneous to inhomogeneous steady states \cite{KoseskaVolkovKurths2013}, as well as with its quantum extensions in coupled oscillator systems \cite{Ban20,BandyopadhyayKhatunBanerjee2021}. In both cases, the deterministic bifurcation diagrams provide an organizing framework for interpreting the full quantum dynamics: they identify the relevant spatial modes, clarify the parameter regions where non-uniformity is expected, and reveal the competition between different patterned configurations.

The comparison between the quantum dynamics and the semiclassical stochastic description further clarifies the role of quantum fluctuations. In the weak quantum regime, where the nonlinear damping associated with two-photon loss is small, the semiclassical stochastic equations reproduce the main features of the quantum steady state. The reduced Wigner functions and the empirical phase-space densities obtained from the stochastic dynamics show compatible structures, while the local occupations agree up to the expected symmetric-ordering shift. This agreement confirms that, in this regime, the deterministic and stochastic phase-space descriptions capture the essential mechanisms behind the emergence of spatial organization. When the nonlinear damping is increased, the system enters a stronger quantum regime in which fluctuations become more relevant, and the deterministic pattern structure is progressively smoothed. Nevertheless, the reduced Wigner functions still retain identifiable signatures of the underlying non-uniform organization. Thus, although the semiclassical description ceases to be quantitatively complete in this regime, the deterministic bifurcation analysis remains useful as a qualitative guide to the quantum steady state.

We also investigated the detuning-dominant regime, where the instability of the homogeneous state is oscillatory rather than stationary. In this case, the deterministic dynamics predicts wave-like time-periodic patterns, and the corresponding reduced Wigner functions develop ring-like phase-space structures. This behavior shows that the same dissipative architecture can support not only stationary pattern selection but also oscillatory non-uniform states. It also emphasizes that bifurcation mechanisms and quantum noise may combine in nontrivial ways in driven-dissipative quantum systems, consistently with related studies on nonlinear quantum oscillators and noise-induced dynamical transitions \cite{Paul2024,Chia2023}.

Overall, our results show that multimode bosonic lattices with suitably engineered dissipative couplings can support spatial quantum counterparts of Turing mechanisms. The model displays stationary pattern formation, competition between different non-uniform modes, and oscillatory wave-like states. Importantly, these effects already appear in the minimal three-site chain considered here. This minimal setting is sufficient to go beyond one- and two-mode descriptions, because it allows distinct spatial modes to coexist, compete, and determine the selected quantum pattern.

Several extensions naturally follow from this work. A first direction is the study of larger lattices, where wavelength selection, boundary effects, and multimode competition can be explored by continuum limit constructions  \cite{Fagotti2020,Gerbino2025}. Adapting these approaches to the competing gradient-pumping and curvature-damping channels considered in our work could provide an effective open quantum field-theoretical description of pattern selection. A second direction concerns the role of measurement-induced symmetry breaking in selecting one of the possible patterned configurations, as well as the possible generation of multipartite entanglement. In this context, entanglement measures and quantum coherence quantifiers commonly used in quantum information theory may provide valuable tools to characterize the emerging states \cite{HorodeckiQEntanglement,Streltsov2017,Sun2022,LoFranco2018,LoFranco2016,Aolita_2015,Kobra2024}. An additional quantum-information perspective would be to characterize the nonclassical resources generated by the driven-dissipative dynamics not only at the level of steady states, but also at the level of quantum processes \cite{Wang_2019}. Moreover, it would be interesting to relate the spatially organized states generated by dissipative pattern-forming mechanisms to other nonequilibrium quantum many-body settings exhibiting emergent spatial self-organization, such as quenched superconducting systems displaying pattern formation and spatial inhomogeneities \cite{SciPostPhysFan}, as well as to engineered bosonic platforms where controlled light–matter interactions can produce multiphoton entanglement and nonclassical photon-emission processes \cite{SciPostRLF}. 
Finally, it would be interesting to compare the present GKSL-based approach with quantum-like operatorial formulations of reaction-diffusion systems, where ladder and number operators have been used to describe spatially distributed populations and pattern-forming mechanisms \cite{Bagarello18,Khrennikov23,Bagarello23,Gorgone2025}. Incorporating Lindblad dynamics into such frameworks may open further possibilities for describing richer forms of spatiotemporal organization.

\vspace*{0.5cm}
\noindent
%\textbf{Author Contributions:}
%F.G. and R.L.F. conceptualized the work; G.C. and F.G. conducted the numerical experiments and the formal analysis. All authors wrote/reviewed/edited the manuscript.

\appendix
\renewcommand{\thesection}{A\arabic{section}}
\renewcommand{\thesubsection}{\thesection.\arabic{subsection}}
\renewcommand{\theequation}{\thesection.\arabic{equation}}
\counterwithin*{equation}{section}

\section{Wigner equation}\label{APP:WIGNER}

Using the standard phase-space representation,
we introduce the Wigner distribution $W(\boldsymbol{\alpha},\boldsymbol{\alpha}^\ast,t)$ associated with the density operator $\hat\rho(t)$.
For a system of $N$ bosonic modes, the Wigner function is defined as
\bea
W(\boldsymbol{\alpha},\boldsymbol{\alpha}^\ast,t)
=
\frac{1}{\pi^{2N}}
\int d^{2N}\boldsymbol{\zeta}\;
\exp\!\left(
\boldsymbol{\zeta}^\ast\!\cdot\!\boldsymbol{\alpha}
-
\boldsymbol{\zeta}\!\cdot\!\boldsymbol{\alpha}^\ast
\right)
\,
\Tr\!\left[
\hat\rho(t)\,
\exp\!\left(
\boldsymbol{\zeta}\!\cdot\!\hat{\boldsymbol a}^\dagger
-
\boldsymbol{\zeta}^\ast\!\cdot\!\hat{\boldsymbol a}
\right)
\right].\label{Wigner}
\ena
Here
$\boldsymbol{\alpha}=(\alpha_1,\dots,\alpha_N)$, with
$\alpha_j=x_j+i p_j$,
and
$d^{2N}\boldsymbol{\zeta}=\prod_{j=1}^N d\zeta_j\,d\zeta_j^\ast$.
This definition corresponds to Weyl-symmetric (Cahill-Glauber)
ordering and provides a real quasi-probability distribution on phase
space.

The mapping from the master equation to a phase-space PDE relies on the symmetric-ordering (Wigner) correspondences \cite{GardinerQuantumNoise,Carmichael1999}, valid mode by mode:
\bea
\hat a_j\hat\rho \ \longleftrightarrow\ 
\Bigl(\alpha_j+\frac12\frac{\partial}{\partial\alpha_j^\ast}\Bigr)W,
\quad
\hat\rho \hat a_j \ \longleftrightarrow\ 
\Bigl(\alpha_j-\frac12\frac{\partial}{\partial\alpha_j^\ast}\Bigr)W,
\ena
\bea
\hat a_j^\dagger\hat\rho \ \longleftrightarrow\ 
\Bigl(\alpha_j^\ast-\frac12\frac{\partial}{\partial\alpha_j}\Bigr)W,
\quad
\hat\rho \hat a_j^\dagger \ \longleftrightarrow\ 
\Bigl(\alpha_j^\ast+\frac12\frac{\partial}{\partial\alpha_j}\Bigr)W.
\ena
Products of operators are represented by compositions of the corresponding differential operators, in the same left/right order in which they act on $\hat\rho$.
This calculus yields a partial differential equation for $W$ containing first-order derivatives (drift), mixed second-order derivatives $\partial_{\alpha_j}\partial_{\alpha_k^\ast}$ (diffusion), and higher-order derivatives generated by nonlinear jump operators such as $\hat a_j^2$.

We write the resulting Wigner equation in conservative (Fokker-Planck) form as
\bea
\frac{\partial W}{\partial t}
=
-\sum_{j=1}^N\left[
\frac{\partial}{\partial\alpha_j}\big(A_j W\big)
+
\frac{\partial}{\partial\alpha_j^\ast}\big(A_j^\ast W\big)
\right]
+\frac12\sum_{j,k=1}^N
\frac{\partial^2}{\partial\alpha_j\,\partial\alpha_k^\ast}
\big(D_{jk} W\big)
+Q_{\gamma_2}[W].\nonumber\\
\ena
In this model, no second-order terms of the type
$\partial_{\alpha_j}\partial_{\alpha_k}$ (anomalous diffusion) appear, because none of the linear nonlocal jump operators mixes $\hat a$ and $\hat a^\dagger$ within the same Lindblad operator.
Hence, the diffusion sector is entirely encoded in the normal diffusion matrix $D_{jk}$.
The operator $Q_{\gamma_2}[W]$ collects the third-order derivative corrections produced by the nonlinear dissipators
$\gamma_2\mathcal D[\hat a_j^2]\hat\rho$.

To obtain the drift terms $A_j$ and the diffusion coefficients $D_{jk}$, we separately consider the contributions coming from the Hamiltonian and the Lindblad operators.
We first decompose the local Hamiltonian into a detuning part and a parametric squeezing part,
\bea
\hat H_{\mathrm{loc}}=\hat H_\Delta+\hat H_\eta,
\qquad
\hat H_\Delta=\sum_{j=1}^N \Delta\,\hat a_j^\dagger \hat a_j,
\qquad
\hat H_\eta=\sum_{j=1}^N
i\eta\left(
\hat a_j^2 e^{-i\theta}
-
\hat a_j^{\dagger 2} e^{i\theta}
\right).\nonumber
\\
\ena
Using the Wigner correspondences to evaluate the commutator contribution $-i[\hat H_{\mathrm{loc}},\hat\rho]$, one obtains a purely first-order (drift) contribution,
\bea
A_j^{(H)} = A_j^{(\Delta)} + A_j^{(\eta)},
\qquad
A_j^{(\Delta)} = -i\Delta\,\alpha_j,
\qquad
A_j^{(\eta)} = -2\eta e^{i\theta}\,\alpha_j^\ast.
\ena
The term $A_j^{(\Delta)}$ generates a local phase rotation at frequency $\Delta$, while $A_j^{(\eta)}$ couples $\alpha_j$ to its complex conjugate $\alpha_j^\ast$, implementing a local squeezing mechanism in phase space.

For the local linear loss channels, the Wigner mapping closes at second order and yields
\bea
A_j^{(\gamma_1)} = -\frac{\gamma_1}{2}\,\alpha_j,
\qquad
D_{jk}^{(\gamma_1)} = \gamma_1\,\delta_{jk}.
\ena
This contribution produces linear damping in the drift and additive, state-independent diffusion.

For the local nonlinear two-photon loss channels, the Wigner correspondence generates three distinct contributions:
a cubic drift term, a state-dependent diffusion, and a purely quantum correction involving third-order derivatives.
Explicitly,
\bea
A_j^{(\gamma_2)} =
\gamma_2\,\alpha_j
-
\gamma_2\,\alpha_j^\ast \alpha_j^2,
\ena
while the associated diffusion matrix reads
\bea
D_{jk}^{(\gamma_2)} =
4\gamma_2
\left(
|\alpha_j|^2 - \frac12
\right)
\delta_{jk}.
\ena
The higher-order contribution is
\bea
Q_{\gamma_2}[W]
=
\sum_{j=1}^N
\left[
\frac{\gamma_2}{4}\,
\frac{\partial^3}{\partial\alpha_j^2\,\partial\alpha_j^\ast}
\Big(\alpha_j W\Big)
+
\frac{\gamma_2}{4}\,
\frac{\partial^3}{\partial\alpha_j^{\ast 2}\,\partial\alpha_j}
\Big(\alpha_j^\ast W\Big)
\right].\label{thirdorderderivWigner}
\ena
In semiclassical regimes, where $\gamma_2$ is small, or the occupations are
large, the third-order term $Q_{\gamma_2}[W]$ is often neglected,
yielding a second-order complex Fokker-Planck approximation.

Finally, we compute the contributions coming from the nonlocal dissipators.
For the operators $\hat L_j^{(\lambda)}$, we obtain in compact form
\bea
\boldsymbol{A}^{(\lambda)} = \frac{\lambda}{2}\,\nabla^{(2)}\boldsymbol{\alpha}^{\mathrm T},
\qquad
D^{(\lambda)} = \lambda\,\nabla^{(2)},
\ena
where the discrete (minus) Laplacian $\nabla^{(2)}$ has entries
\bea
(\nabla^{(2)})_{jk} =
\begin{cases}
	1, & j=k=1 \text{ or } j=k=N,\\
	2, & j=k=2,\dots,N-1,\\
	-1, & |j-k|=1,\\
	0, & \text{otherwise}.
\end{cases}\label{matD2}
\ena
This matrix corresponds to the standard centered finite-difference
representation of the minus second-order spatial derivative on a
uniform one-dimensional lattice, with no-flux boundary conditions, that is the Discrete Cosine Transform-2 matrix, see \cite{strang99}.

For the nonlocal dissipators $\hat L_j^{(\kappa)}$, being linear in
$\hat a$, the Wigner mapping produces only drift and normal diffusion.
In compact form,
\bea
\boldsymbol{A}^{(\kappa)} = -\frac{\kappa}{2}\,\nabla^{(4)}\boldsymbol{\alpha}^{\mathrm T},
\qquad
D^{(\kappa)} = \kappa\,\nabla^{(4)},
\ena
where the induced bi-Laplacian is
\bea
\nabla^{(4)} = (\nabla^{(2)})^2.\label{matD4}
\ena

\noindent Collecting all drift contributions, we obtain
\bea
A_j
=
\Bigl(\gamma_2-\frac{\gamma_1}{2}-i\Delta\Bigr)\alpha_j
-\gamma_2\,\alpha_j^\ast \alpha_j^2
-2\eta e^{i\theta}\alpha_j^\ast
+\Bigl(\frac{\lambda}{2} \nabla^{(2)}\boldsymbol{\alpha}^{\mathrm T}\Bigr)_j
-\Bigl(\frac{\kappa}{2} \nabla^{(4)}\boldsymbol{\alpha}^{\mathrm T}\Bigr)_j,\nonumber \label{driftentries}
\\
\ena
while the total diffusion matrix reads
\bea\label{diffmatrix}
D_{jk}
=
\Biggl[
\gamma_1
+
4\gamma_2
\left(
|\alpha_j|^2 - \frac12
\right)
\Biggr]\delta_{jk}
+
\lambda(\nabla^{(2)})_{jk}
+
\kappa(\nabla^{(4)})_{jk}.
\ena
The diagonal contribution arises from local dissipation and is
multiplicative, while the minus Laplacian and bi-Laplacian terms encode spatially correlated noise induced by the nonlocal Lindblad channels.
Hence, the noise is intrinsically multiplicative, as the diffusion
matrix depends on the instantaneous amplitudes $|\alpha_j|^2$.

Substituting these expressions into the general structure yields the full
Wigner evolution equation,
\bea
\frac{\partial W}{\partial t}
=
-\sum_{j=1}^N
\left[
\frac{\partial}{\partial\alpha_j}\big(A_j W\big)
+
\frac{\partial}{\partial\alpha_j^\ast}\big(A_j^\ast W\big)
\right]
+\frac12\sum_{j,k=1}^N
\frac{\partial^2}{\partial\alpha_j\,\partial\alpha_k^\ast}
\big(D_{jk} W\big)
\nonumber\\
+
\sum_{j=1}^N
\left[
\frac{\gamma_2}{4}\,
\frac{\partial^3}{\partial\alpha_j^2\,\partial\alpha_j^\ast}
\Big(\alpha_j W\Big)
+
\frac{\gamma_2}{4}\,
\frac{\partial^3}{\partial\alpha_j^{\ast 2}\,\partial\alpha_j}
\Big(\alpha_j^\ast W\Big)
\right].
\ena

This equation provides a phase-space representation of the full quantum
dynamics. A semiclassical Fokker-Planck approximation is obtained by
neglecting the third-order derivative terms, while retaining the
state-dependent normal diffusion matrix $D_{jk}$, which encodes quantum fluctuations around the deterministic drift.

Some comments on this final form follow.
By construction, the anomalous diffusion sector is absent in this model,
since no second-order terms of the form
$\partial_{\alpha_j}\partial_{\alpha_k}$ arise.
The diagonal diffusion coefficients of $D$ may become negative for
sufficiently large $\gamma_2$ and small $|\alpha|$, violating the positivity
requirement of a genuine Fokker-Planck equation.
In the quantum context, this signals the breakdown of the truncated
Fokker-Planck approximation and the onset of non-classical fluctuations.
Restoring the higher-order derivative terms yields the full Wigner equation, which remains well defined as an evolution equation for a quasi-probability distribution but cannot be associated with a standard Langevin process driven by Gaussian noise.

\section{Mean-field equation}\label{APP:MFE}
Here we show how to obtain the mean field equations for $\alpha_j=\langle \hat a_j\rangle=\mathrm{Tr}(\hat a_j \hat \rho)$, obtained from the general relation
$\frac{d}{dt}\langle \hat a_j\rangle=\mathrm{Tr}(\hat a_j\dot{\hat\rho})$, where the dynamics of $\hat \rho$ is governed by the QME \eqref{QME}. 

\noindent The coherent contribution generated by the local Hamiltonian yields
\bea
-i\langle[\hat a_j, \hat H_{\mathrm{loc}}]\rangle=(-i\Delta)\alpha_j-2\eta e^{i\theta}\alpha_j^*.
\ena
Each linear local dissipator $\hat L_j^{(sl)}$ contributes as
$\gamma_1\langle\mathcal D[\hat a_j]^\dagger \hat a_j\rangle=-\frac{\gamma_1}{2}\alpha_j$,
producing a standard linear damping.
\noindent
Each nonlinear dissipator $\hat L_j^{(dl)}$  yields (using the symmetric ordering):
\bea
\langle\gamma_2\mathcal D[\hat a_j^2]^\dagger \hat a_j\rangle
= -\gamma_2 \langle \hat a_j^\dagger \hat a_j^2 \rangle
= -\gamma_2 \langle \hat a_j \hat a_j^\dagger \hat a_j - \hat a_j \rangle
= -\gamma_2 \langle \hat a_j \hat N_j \rangle + \gamma_2 \langle \hat a_j \rangle.
\ena
Applying the mean-field approximation $\langle \hat a_j \hat N_j \rangle \approx |\alpha_j|^2\alpha_j$, $\langle \hat a_j \rangle \approx \alpha_j$, this channel introduces both a cubic saturation term
and a positive linear correction:
\bea
\dot{\alpha}_j\Big|_{\gamma_2} = -\gamma_2 |\alpha_j|^2\alpha_j + \gamma_2 \alpha_j.
\ena
Collecting all local contributions, one obtains the local reaction dynamics:
\bea
\dot\alpha_j=(s-i\Delta)\alpha_j-\gamma_2|\alpha_j|^2\alpha_j-2\eta e^{i\theta}\alpha_j^*,
\ena
where we define the effective linear gain-loss parameter
\bea
s = \frac{2\gamma_2 - \gamma_1}{2}.
\ena

The nonlocal dissipators introduce spatial couplings that, in compact form, can be written as 
\bea
\dot{\boldsymbol{\alpha}}\Big|_{\text{non-loc}} =
\frac{\lambda}{2} \nabla^{(2)} \boldsymbol{\alpha}^\text{T}
-\frac{\kappa}{2} \nabla^{(4)} \boldsymbol{\alpha}^\text{T},
\ena
where the (minus) discrete Laplacian $\nabla^{(2)}$ and the discrete bi-Laplacian $\nabla^{(4)}$ are defined in \eqref{matD2} and \eqref{matD4}, so that for each node $j$ we obtain
\bea
\dot{\alpha}_j = (s-i\Delta)\alpha_j - \gamma_2|\alpha_j|^2\alpha_j - 2\eta e^{i\theta}\alpha_j^*
+ \frac{\lambda}{2} \sum_k (\nabla^{(2)})_{jk} \alpha_k
- \frac{\kappa}{2} \sum_k (\nabla^{(4)})_{jk} \alpha_k. \\
\label{MFE}
\ena
These represent a discrete reaction-diffusion system in which compete the effective gain $s$, the anti-diffusive Laplacian term of strength $\lambda$, and the stabilizing bi-Laplacian with intensity $\kappa$.
Notice that this equation is exactly $\dot \alpha_j=A_j$, where $A_j$ was derived in \eqref{driftentries} as the drift part of the Fokker-Planck equation.

\section{Classical noisy equations}\label{APP:CNE}
In the weak quantum regime, where the third-order correction
$Q_{\gamma_2}[W]$ can be neglected because $\gamma_2$ is small, the Wigner equation reduces to a
second-order (truncated-Wigner) Fokker-Planck equation for the
$N$ complex phase-space variables
\bea
\alpha_j = x_j + i p_j, \qquad j=1,\ldots,N.
\ena
This equation admits an equivalent representation in terms of an
It\^o stochastic differential equation (SDE).
In compact notation, the SDE reads
\bea
d\boldsymbol{\alpha}(t)
=
\boldsymbol{A}(\boldsymbol{\alpha}(t),\boldsymbol{\alpha}^\ast(t))\,dt
+
B(\boldsymbol{\alpha}(t),\boldsymbol{\alpha}^\ast(t))\,d\boldsymbol{W}(t),\label{SDE}
\ena
where $\boldsymbol{\alpha}=(\alpha_1,\ldots,\alpha_N)^T\in\mathbb{C}^N$,
$\boldsymbol{A}\in\mathbb{C}^N$ is the drift vector whose entries are defined according to \eqref{driftentries} and the right-hand side of \eqref{MFE},
$B\in\mathbb{C}^{N\times N}$ is a complex noise-amplitude matrix, and
$d\boldsymbol{W}\in\mathbb{C}^N$ is a vector of complex Wiener
increments.
Specifically, to correctly match the diffusion terms, which appear with a prefactor of $1/2$ in the Wigner Fokker-Planck equation, $B$ must satisfy
\bea
B(\boldsymbol{\alpha},\boldsymbol{\alpha}^\ast)
B(\boldsymbol{\alpha},\boldsymbol{\alpha}^\ast)^\dagger
=
\frac{1}{2} D(\boldsymbol{\alpha},\boldsymbol{\alpha}^\ast),
\ena
where $^\dagger$ denotes the conjugate transpose (Hermitian adjoint), and $D$ is the diffusion matrix.
Since $B$ depends on $\boldsymbol{\alpha}$, it must be computed at each time step.
Following Refs.~\cite{Ban20,Kato2022,Paul2024}, we use the spectral factorization
\bea
D
=
U\,\Lambda\,U^\dagger,
\qquad
\Lambda=\mathrm{diag}(\mu_1,\ldots,\mu_N),
\ena
where, taking advantage of the real and symmetric nature of $D$, $U$ is actually a real orthogonal matrix ($U^T=U^\dagger, U^T U = I$) and $\mu_r$ are the non-negative real eigenvalues of $D$.
Then \bea
B
=
\frac{1}{\sqrt{2}} \, U\,\Lambda^{1/2}.
\ena
The complex Wiener noise $d\boldsymbol{W}(t)=(dW_1(t),\ldots,dW_N(t))^T$ is defined as a complex Gaussian process, according to the requirements
\bea
\mathbb{E}\,\!\big[d\boldsymbol{W}(t)\big]=0,
\quad
\mathbb{E}\,\!\big[d\boldsymbol{W}(t)\,d\boldsymbol{W}(t)^\dagger\big]
=
I\,dt,
\quad
\mathbb{E}\,\!\big[d\boldsymbol{W}(t)\,d\boldsymbol{W}(t)^T\big]=0,
\ena
where $I$ denotes the $N\times N$ identity matrix.
A convenient construction uses two independent real standard Wiener vectors
$d\boldsymbol{W}_r,d\boldsymbol{W}_i\in\mathbb{R}^N$,
\bea
d\boldsymbol{W}
=
\frac{1}{\sqrt{2}}
\bigl(d\boldsymbol{W}_r+i\,d\boldsymbol{W}_i\bigr),
\ena
with
\bea
\mathbb{E}\,\!\big[d\boldsymbol{W}_r d\boldsymbol{W}_r^T\big]
=
\mathbb{E}\,\!\big[d\boldsymbol{W}_i d\boldsymbol{W}_i^T\big]
=
I\,dt,
\qquad
\mathbb{E}\,\!\big[d\boldsymbol{W}_r d\boldsymbol{W}_i^T\big]=0.
\ena
The prefactor $1/\sqrt{2}$ ensures the correct normalization for the complex variance,
\bea
\mathbb{E}\!\big[d\boldsymbol{W}\,d\boldsymbol{W}^\dagger\big]
=
I\,dt.
\ena
The above conditions define complex Wiener increments that are Gaussian, have zero mean, are temporally and spatially uncorrelated, have equal variance in their real and imaginary parts, exhibit no cross-correlations between different components, and are fully isotropic in the complex phase space.

%\bibliographystyle{apsrev4-2}
%\bibliography{biblio}


\end{document}